\documentclass[%
 reprint,
superscriptaddress,
nofootinbib,
 amsmath,amssymb,
 aps,
]{revtex4-1}

\usepackage{graphicx}% Include figure files
\usepackage{dcolumn}% Align table columns on decimal point
\usepackage{bm}% bold math
\usepackage{lipsum}
\usepackage{amsmath,amssymb}
\usepackage{dsfont}
\usepackage{float}
\usepackage{array}
\makeatletter
\let\newfloat\newfloat@ltx
\makeatother
\usepackage{algorithm}
\usepackage{algpseudocode}
\usepackage{multirow}
\usepackage{colortbl}
\usepackage{booktabs}
\usepackage{stmaryrd}
\newcolumntype{C}{>{$}c<{$}}

\AtBeginDocument{%
  \heavyrulewidth=.08em
  \lightrulewidth=.05em
  \cmidrulewidth=.03em
  \belowrulesep=.65ex
  \belowbottomsep=0pt
  \aboverulesep=.4ex
  \abovetopsep=0pt
  \cmidrulesep=\doublerulesep
  \cmidrulekern=.5em
  \defaultaddspace=.5em
}

\usepackage{verbatim}
\usepackage[left=2cm, right=2cm, top=2cm]{geometry}

\usepackage[normalem]{ulem}
\usepackage[utf8]{inputenc}
\usepackage{epigraph}
\usepackage{siunitx}
\usepackage[english]{babel}
\usepackage{listings}
\usepackage{graphicx}
\usepackage[utf8]{inputenc}
\usepackage{listings}
\usepackage{matlab-prettifier}
\usepackage{xcolor}
\usepackage{amsfonts}
\usepackage{amsmath}
\usepackage{amssymb}
\usepackage{amsthm}
\newcolumntype{P}[1]{>{\centering\arraybackslash}p{#1}}
\newcolumntype{M}[1]{>{\centering\arraybackslash}m{#1}}
\usepackage{braket}
\usepackage{xcolor}
\usepackage[colorlinks = true,
            linkcolor = blue,
            urlcolor  = blue,
            citecolor = blue,
            anchorcolor = blue]{hyperref}
\usepackage{comment}
\usepackage{hyperref}
\usepackage{tcolorbox} % For fancy boxes
\usepackage{color}

\newcommand{\be}{\begin{equation}}
\newcommand{\ee}{\end{equation}}

\begin{document}

\preprint{APS/123-QED}

\title{Mind the gap: addressing data gaps and assessing noise mismodeling in LISA}

\author{Ollie Burke}
\email[]{ollie.burke@glasgow.ac.uk}
\affiliation{School of Physics and Astronomy, University of Glasgow, University Avenue, Glasgow, G12 8QQ, UK}
\affiliation{Laboratoire des 2 Infinis - Toulouse (L2IT-IN2P3), Université de Toulouse, CNRS, UPS, F-31062 Toulouse Cedex 9, France}
\affiliation{%
Max Planck Institute for Gravitational Physics (Albert Einstein Institute), Am M\"{u}hlenberg 1, Potsdam-Golm 14476, Germany}
\affiliation{%
  School of Mathematics, University of Edinburgh, James Clerk Maxwell Building, Peter Guthrie Tait Road, Edinburgh EH9 3FD, UK}
\author{Sylvain Marsat}
\email[]{sylvain.marsat@l2it.in2p3.fr}
\affiliation{Laboratoire des 2 Infinis - Toulouse (L2IT-IN2P3), Université de Toulouse, CNRS, UPS, F-31062 Toulouse Cedex 9, France}
\author{Jonathan R. Gair}
\affiliation{%
Max Planck Institute for Gravitational Physics (Albert Einstein Institute), Am M\"{u}hlenberg 1, Potsdam-Golm 14476, Germany}
\affiliation{%
  School of Mathematics, University of Edinburgh, James Clerk Maxwell Building, Peter Guthrie Tait Road, Edinburgh EH9 3FD, UK}
\author{Michael L. Katz}
\affiliation{%
Max Planck Institute for Gravitational Physics (Albert Einstein Institute), Am M\"{u}hlenberg 1, Potsdam-Golm 14476, Germany}
\affiliation{NASA Marshall Space Flight Center, Huntsville, Alabama 35811, USA}

\begin{abstract}

Due to the complexities of the Laser Interferometer Space Antenna (LISA), data gaps arising from instrumental irregularities and/or scheduled maintenance are unavoidable. Focusing on merger-dominated massive black hole binary signals, we test the appropriateness of the Whittle-likelihood on gapped data in a variety of cases. From first principles, we derive the likelihood valid for gapped data in both the time and frequency domains. Cheap-to-evaluate proxies to p-p plots are derived based on a Fisher-based formalism, and verified through Bayesian techniques. Our tools allow to predict the altered variance in the parameter estimates that arises from noise mismodeling, as well as the information loss represented by the broadening of the posteriors. The result of noise mismodeling with gaps is sensitive to the characteristics of the noise model, with strong low-frequency (red) noise and strong high-frequency (blue) noise giving statistically significant fluctuations in recovered parameters. We demonstrate that the introduction of a tapering window reduces statistical inconsistency errors, at the cost of less precise parameter estimates. We also show that the assumption of independence between inter-gap segments appears to be a fair approximation even if the data set is inherently coherent. However, if one instead assumes fictitious correlations in the data stream, when the data segments are actually independent, then the resultant parameter recoveries could be inconsistent with the true parameters. The theoretical and numerical practices that are presented in this work could readily be incorporated into global-fit pipelines operating on gapped data.
\end{abstract}

\pacs{Valid PACS appear here}% PACS, the Physics and Astronomy
                             % Classification Scheme.
%\keywords{Suggested keywords}%Use showkeys class option if keyword
                              %display desired
\maketitle

\tableofcontents

\section{Introduction}

On the 20th of June, 2017, the space-based gravitational wave (GW) observatory, the Laser Interferometer Space Antennae (LISA)~\cite{LISA:2017pwj}, was selected as the third large-class mission under the European Space Agency's (ESA) Cosmic Vision program~\cite{LISA:2017pwj}. The LISA space-mission was then adopted by ESA on the 25th of January, 2024, with launch expected to be in $\sim 2034$~\cite{LISA:2024hlh}. The goal of LISA is to observe GWs in the rich $f \in [0.1\,\text{mHz}, 0.1\,\text{Hz}]$ frequency range. GWs at these frequencies are generated by massive black hole binaries (MBHBs)~\cite{colpi2019astro2020sciencewhitepaper}, extreme mass-ratio inspirals (EMRIs)~\cite{Babak:2017tow, berry2019uniquepotentialextrememassratio}, ultra compact binaries (UCBs)~\cite{Littenberg:2019grv} and stochastic backgrounds~\cite{Chen:2018rzo, Bonetti:2020jku, Pozzoli:2023kxy, Ruiter:2007xx, Nelemans:2009hy, Timpano:2005gm, Digman:2022jmp}. In contrast to current ground-based detectors, the LISA instrument is expected to be signal dominated, resulting in a strong cocktail of millions of GW sources of multiple source types all overlapping in both time and frequency. In order to develop and test pipelines for successful source extraction from LISA data, the community have developed ``LISA Data Challenges" of increasing complexity~\cite{LDC_Sangria, Arnaud:2007vr, Arnaud:2007jy}. Recently, there have been major advancements in both computing and LISA data analysis, that have given rise to proposed solutions to the so-called ``global-fit" problem, defined to be the search and characterization of all resolvable GWs buried within the LISA data stream~\cite{Littenberg:2023xpl, Strub:2024kbe, Katz:2024oqg, deng2025modular}. Although impressive -- marking a major milestone in LISA data analysis -- so far none of these group's proposed global-fit solutions were conducted with \emph{realistic} LISA data. Indeed, it was assumed that the underlying noise statistics were Gaussian and (weakly-)stationary, facilitating use of rapid-to-evaluate likelihood models in parameter-inference. In reality, however, it is well known by the community that realistic LISA data will contain noise-transients such as instrumental artefacts (glitches), and b interrupted via data gaps~\cite{Armano:2020zyx,Baghi:2021tfd}. The LISA Pathfinder mission, which demonstrated some of the LISA technology, exhibited such non-stationary features~\cite{Armano:2016bkm}. The \emph{Spritz} data challenge~\cite{LDC_Spritz} was built by the community to test source extraction pipelines, with an underlying Gaussian noise model with unknown spectral density with data corrupted via instrumental artefacts in the form of gaps and glitches. To date and to our knowledge, there has only been a single attempt at the \emph{Spritz} data challenge given by~\cite{Castelli:2024sdb,Castelli:2024sdb}. In their analysis, they use parametric models to detect and subtract glitches while treating gaps in the data using windowing functions. As it stands, the windowing procedure will induce non-stationary features in the underlying noise properties. This renders the usual statistical models utilized for inference of data statistically inconsistent with the data stream itself, and it is therefore important to assess the impact of this mismodeling.

Probabilistic models that describe the data stream must be consistent with the data generating process itself. Since the noise is the only probabilistic quantity that defines the data set, the likelihood used to analyse the data is determined by the model assumed for the underlying noise process. It is often assumed within GW astronomy that the noise process is both Gaussian and stationary, the latter feature giving rise to a \emph{Toeplitz} structure in the TD covariance matrix. An additional simplifying assumption is further placed on the noise process where not only is it Toeplitz, but also \emph{circulant} -- infinite duration or periodic if finite. As demonstrated by Whittle~\cite{whittle:1957}, in the stationary and circulant assumptions, the TD noise-covariance matrix can be diagonalized in a Fourier basis, resulting in a likelihood that can be calculated in $\mathcal{O}(N\log_{2}N)$ operations. The Whittle likelihood is ubiquitous in GW astronomy. Since parameter estimation (PE) schemes usually require many evaluations of a likelihood function, it is clear why one would want to conduct analysis in the FD assuming these (restrictive) conditions are met.

In practice, the assumption that the noise process is circulant is clearly unrealistic for true LISA data: no real data stream is periodic and the LISA dataset will be signal-rich on a finite duration. Attempts to perform TD GW data analyses for stationary non-circulant noise are scarce. One example is given by recent analyses of black hole ring down signals~\cite{siegel2024analyzingblackholeringdownsii, Capano:2021etf, Pitte:2024zbi, Isi:2021iql, Carullo:2019flw}. In the latter case, the short-data segments containing the ring down signal would suffer from edge effects when analyzed in the frequency-domain (FD), which could contaminate the final results through spectral leakage~\cite{harris1978use}. Another related problem is given by premerger analyses for MBHB signals, for which a generalization of whitening and matched filtering has been proposed in~\cite{CabournDavies:2024hea}. In general, time-domain analyses conducted on large scale data sets are usually computationally infeasible, simply due to the computational cost of computing the likelihood, which at best is an $\mathcal{O}(N^{2})$ operation, since the covariance matrix entering the likelihood is not diagonal. 

In the construction of LISA, instrumentalists will try to build an instrument such that its noise properties remain as close to stationary as possible. However, the instrument will not be perfect and we must be ready to account for non-stationary features of the noise when they are present. Examples could be environmental: such as impacts from micro-meteorites that collide with the space-crafts, that will form non-GW induced perturbations to the test masses; or instrumental: failure to discharge the on-board test masses due to charged particles hitting the test-masses. Such artifacts were observed in the LISA pathfinder mission~\cite{Armano:2018kix, LISAPathfinder:2022awx, Baghi:2021tfd}. These noise-transients are called glitches, which, if not accounted for, could feature in PE schemes as biases to recovered parameters~\cite{Spadaro:2023muy}. Such loud noise-transients could be subtracted from the data stream~\cite{Chowdhury:2024jdx, Zackay:2019kkv, LIGOScientific:2017vwq}, or simply \emph{masked}, rendering the glitched data segment unusable.

In this paper, we will focus on instrumental artefacts that result in total losses of data, known as \emph{gaps} in the data stream.
Alongside gaps introduced by glitch-masking, LISA will experience instrumental gaps falling into two categories -- Scheduled and Unscheduled gaps. Scheduled gaps are breaks in the data stream where routine maintenance is performed in order to achieve maximal sensitivity of the instrument. Examples being antennae re-pointing (gap duration $\sim 3$ hours every $\sim14$ days), tilt-to-length coupling constant estimation (gap duration $\sim 2$ days  with frequency $\sim$ four times per year) and point-ahead angle mechanism (PAAM) adjustments (three times per day, lasting $\sim 100$ seconds each). Unscheduled gaps could be more dangerous for LISA data analysis due to their unknown duration and frequency. Examples of unscheduled gaps include instrumental malfunctions, such as collisions between micro-meteorites and the craft that may result in the total loss of the gravitational reference sensor (approximately $\sim 30/\text{yr}$ resulting in $\sim 1$ days worth of lost information for each collision), to minor outages on the instrument, potentially lasting on the order of $\sim 10$ seconds. The lost data types with expected gap frequency/duration are listed in ~\cite{Data_Analysis_Robustness_Report}. It is clear from this discussion that both the frequency and cumulative duration of gaps is large, and so must be accounted for in PE pipelines.

To date there have been a number of studies that have focused on gaps and non-stationary noise in general. On the topic of gaps, there are two families of studies -- the first being data augmentation (gap filling)~\cite{Baghi:2019eqo,Blelly:2021oim, Wang:2024ovi, Mao:2024jad} and the other the windowing procedure~\cite{Carre:2010ra, Talbot:2021igi, Zackay:2019kkv, Edy:2021par}. Data augmentation uses Bayesian (and recently auto-encoder~\cite{Mao:2024jad}) techniques to \emph{fill in} the data during the gap segment, usually learning the behavior of the noise through prior observations. The parameters of the signal are then recovered from the filled-in data set, facilitating use of the rapid-to-evaluate Whittle-likelihood assuming that the original noise process was Gaussian, stationary and circulant. However, the procedure used to fill in the gaps is a computationally expensive procedure and may be inaccurate if the instrument suffers significant changes of state as a result of the gap.  

Another, albeit simpler approach is the windowing procedure. The windowing procedure does not attempt to impute the data in the gap --  a binary mask or smooth window function is applied to the data stream over the gaps. This allows for a coherent data set to be analyzed, at the cost of losing out on the stationary nature of the noise as discussed in ~\cite{Talbot:2021igi, Zackay:2019kkv, Edy:2021par}. Applying a window function to a stationary noise process violates the Toeplitz condition, rendering the overall noise process \emph{non-stationary} making the Whittle-likelihood inconsistent with data generation scheme. A number of studies have applied the windowing procedure, but have not accounted for the change in the underlying statistics of the noise process. Our work presented here corrects for this, providing methodology on how to build probabilistic models that describe the data stream under a variety of gap scenarios in both the TD and FD.

One of the first studies using the windowing procedure to investigate the impact of gaps on LISA-based parameter precision studies was performed by Carré and Porter in ~\cite{Carre:2010ra}. Analysing UCBs, they simulated the effect of gaps, analyzed the impact of leakage and performed a monte-carlo parameter precision study on a large number of UCBs. Their results concluded that gaps in the data stream can lead to loss in precision on parameter estimates. An analysis on MBHBs was performed in~\cite{Dey:2021dem}, which studied the effect of gaps on the inspiral/merger phase. They concluded that the impact of gaps grew in severity the closer the gap was to the merger with parameter precision estimates degrading dramatically. In ~\cite{Wang:2024ovi,Mao:2024jad}, the data augmentation method was applied on MBHBs and reached similar conclusions. In each study~\cite{Carre:2010ra,Dey:2021dem, Wang:2024ovi} gaps were simulated by windowing the data set to zero before and after the gap segment but did not account for the induced correlations between the noise components in the FD. A result of this mismodeling, as shown in~\cite{Edy:2021par}, is that not only is the resultant posterior (co)variance incorrect but the (co)variance of the posterior scattering can be under/over-estimated.

We believe that the work here forms the most general treatment of gaps present in the literature to date. Our methodology can be applied to \emph{any} family of gaps with any duration/frequency and with \emph{any} underlying assumptions about the behavior of the noise across the gap, from coherence to independence. We carefully analyse the windowing procedure and demonstrate how inference can be performed on gapped time-series in both the TD and FD, highlighting advantages of each. Starting from first principles in the TD, we will derive expressions for the signal-to-noise ratio, Fisher matrix and likelihood in the presence of missing data. All of our results are translated into the FD via linear algebra. We will then extend the results from~\cite{Edy:2021par} to derive a Fisher-based approximation to calculate: (1) the expectation of the posterior scattering caused by noise mismodeling and (2) the posterior thinning/widening as a consequence of mismodeling. 

The paper is organized as follows: In Sec.~\ref{sec:conventions}, we will outline conventions and set up notation for our linear algebra formulation of the gaps problem. In Sec.~\ref{subsec:stationary_noise}, we will outline the likelihoods for Gaussian noise in both the TD and FD. In Sec.~\ref{sec:stationary_noise}, we outline what it means for a stochastic process to be Toeplitz and circulant, deriving the familiar form of the Whittle-likelihood. Sec.~\ref{sec:methods} outlines methodology on how to approach gaps: with the marginalization procedure outlined in Sec.~\ref{subsec:analysis_set_up} and the windowing procedure (the two proved to be equivalent) in Sec.~\ref{subsec:windowing_procedure}. We outline efficient numerical schemes to compute the TD and FD covariance matrices in the presence of gaps in Sec.~\ref{subsec:windowed_covariance}. We then derive expressions for the signal-to-noise ratio and Fisher-matrix in Sec.~\ref{subsec:SNR} and Sec.~\ref{subsec:Fisher} respectively. Sec.~\ref{sec:mismodeling} outlines the metrics we will use to assess mismodeling for the gapped noise procedure with any model-covariance matrix in either the time or frequency domain. Specifically, in Sec.~\ref{subsec:mismodeling_gaps} we derive Fisher-based statistics that are used to assess mismodeling data gaps, with Sec.~\ref{subsec:gap_whittle} and Sec.~\ref{subsec:gap_segmented_whittle} focusing on \emph{assuming coherence} and \emph{incoherence} between gapped segments. Sec.~\ref{subsec:mismodeling_measures} derives expressions that can compare the model posterior variance to the true posterior variance Eq.~\eqref{eq:def_Xi} and to the expectation of the scatter of the noise fluctuations due to noise mismodeling given by Eq.~\eqref{eq:def_Upsilon}. In Sec.~\ref{subsec:parameter_bias_cov}, we discuss how the metrics defined by Eqs.~(\ref{eq:def_Xi}--\ref{eq:def_Upsilon}) can be computed in practice. We outline our signal generation, gap placement and noise models in Sec.~\ref{subsec:mbhb_signals_analysis}. We perform an analysis on circulant, Toeplitz and gated noise covariance matrices in Sec.~\ref{subsec:circulant_vs_toeplitz} and in the Sec.~\ref{subsec:pseudo_inverses_practice} and Sec~\ref{subsec:pseudo_inverses_practice_w_taper} we discuss numerical routines on how to handle the degenerate matrices via computing the Moore-Penrose pseudo-inverse. Our results section ~\ref{sec:results} attempts to answer the following questions:
\begin{itemize}
    \item \textbf{[Sec.~\ref{subsec:gap_mismodeling_ratios}]:} For various realistic noise models with different behavior at low and high frequencies, is it reasonable to assume the Whittle-likelihood in the presence of gaps?
    \item \textbf{[Sec.~\ref{subsec:mismodeling_tapering_length}]:} What is the impact of the tapering on both time and frequency domains? Especially during critical observations such as during the merger phase of a MBHB? 
    \item \textbf{[Sec.~\ref{subsec:mismodeling_cadence}:]} Is parameter inference sensitive to the sampling rate of gapped data? 
    \item \textbf{[Sec.~\ref{subsec:mismodeling_independence}]:} What are the consequences of assuming dependence between gap segments when they are truly independent? And vice versa?
    \item \textbf{[Sec.~\ref{subsec:mismodeling_psd_lowf}]:} Since low frequency noise is challenging to estimate -- what is the impact of mismodeling the low-frequency content of the noise process?
    \item \textbf{[Sec.~\ref{subsec:mismodeling_toeplitz_circulant}]:} What are the consequences of assuming that the underlying noise process is circulant (Whittle-like) when the time-series is of finite duration (Toeplitz)? 
\end{itemize}
Since our results are based on a Fisher-matrix approximation, we verify a number of our results using Bayesian techniques in section Sec.~\ref{subsec:Parameter_Estimation}. Our conclusions and scope for future work are presented in Sec.~\ref{subsec:conclusions} and Sec.~\ref{subsec:outlook} respectively. 

We understand that this paper is long and rather technical. For a busy reader, we recommend that they skip to the conclusions section for the answers to the above bullet points in brief. The main results are tabulated in Tabs.~\ref{tab:mismodeling_upsilon_gap},\ref{tab:mismodeling_upsilon_independence},\ref{tab:mismodeling_upsilon_toeplitz} respectively. We use a color scheme of green as acceptable, and any other color as unacceptable for PE. A brief summary of our results can be found in the conclusion Sec.~\ref{sec:conclusions_outlook}.

\subsection{Conventions}\label{sec:conventions}
Time-domain signals $x(t) \approx \{x(t_{i})\}_{i = 0}^{N-1}$ are sampled at discrete times $t_{i} \in [0, \Delta t,\ldots, (N-1)\Delta t]$ with a uniform sampling interval $\Delta t = t_{i+1} - t_{i}$ and with $N$ the size of the TD data segment, of duration $T = N \Delta t$. To facilitate usage of the fast Fourier transform, it is customary to use a segment length that is an integer power of two in length, so that $N = 2^{J}$ for $J\in \mathbb{N}$. We denote the continuous Fourier transform (CFT) through
\begin{equation}\label{eq:CFT}
    \tilde{x}(f) = \mathcal{F}[x(t)] = \int_{-\infty}^{\infty}x(t)e^{-2\pi i f t} \ \text{d}t \,,
\end{equation}
with corresponding inverse
\begin{equation}\label{eq:ICFT}
    x(t) = \mathcal{F}^{-1}[\tilde{x}(f)] = \int_{-\infty}^{\infty}\tilde{x}(f)e^{2\pi i f t} \ \text{d}f.
\end{equation}
For real signals, $\tilde{x}(-f) = \tilde{x}(f)^*$, which implies that the negative frequencies are related to the positive frequencies by conjugation. 

In the discrete domain, $\bm{x}$ will represent the data vector with time interval $\Delta t$, so that $\bm{x}_i = x(t_i)$, while $\tilde{\bm{x}}$ will represent the FD data with frequency interval $\Delta f = 1/T = 1/(N \Delta t)$. Note that we index the FD data from $0$ to $N-1$, with the second half of the vector corresponding to the negative frequencies: for $i=0,\dots,N/2-1$, $\tilde{\bm{x}}_{i} = \tilde{x}(i \Delta f)$, while for  $i=N/2,\dots,N-1$, $\tilde{\bm{x}}_{i} = \tilde{x}((i - N) \Delta f)$. 

We will use the discrete time Fourier transform (DFT) for $N\in 2\mathbb{Z}^{+}$
\begin{align}\label{eq:DFT}
    \tilde{\bm{x}}_j = \Delta t \sum_{i = 0}^{N-1} \bm{x}_i e^{-2\pi i f_{j} t_{i}} \,, 
\end{align}
with the corresponding inverse discrete Fourier transform given by
\begin{equation}\label{eq:inverseDFT}
    \bm{x}_i = \Delta f \sum_{j = 0}^{N-1} \tilde{\bm{x}}_j e^{2\pi i f_{j} t_{i}} \,.
\end{equation}
Here, $t_i = i \Delta t$, $f_j = j \Delta f$, and $t_i f_j = ij/N$ since $\Delta t \Delta f = 1/N$.

Another useful point of view on the DFT is to represent it as a linear algebra transformation:
\begin{subequations}
\begin{align}
	\tilde{\bm{x}} &= \Delta t \sqrt{N} \bm{P} \bm{x} \,, \label{eq:defDFT}\\
	\bm{x} &= \frac{1}{\Delta t \sqrt{N}} \bm{P}^{\dagger} \tilde{\bm{x}} \, \label{eq:defIDFT},
\end{align}
\end{subequations}
where we introduced the DFT transition matrix $\bm{P}$, built from powers of the $N$-th root of unity $\omega \equiv e^{2 i \pi / N}$:
\begin{equation}\label{eq:def_P_jk_matrix}
	\bm{P}_{jk} = \frac{1}{\sqrt{N}} \omega^{-jk} \,.
\end{equation}
The transition matrix $\bm{P}$ is unitary ($\bm{P}^{\dagger} = \bm{P}^{-1}$, where ${}^\dagger$ stands for the hermitian conjugate) and symmetric ($\bm{P}^{T} = \bm{P}$). In our notations, $\bm{P}$ and $\bm{P}^{\dagger}$ are built to be dimensionless with a symmetric normalization, while $\bm{x}$, $\tilde{\bm{x}}$ have the appropriate normalization for data in the TD or FD. This linear algebra point of view is useful for derivations; it is never used for practical computations as matrix-vector multiplications cost $\mathcal{O}(N^2)$ operations while the DFT can be computed in $\mathcal{O}(N \log_{2}N)$ thanks to the Fast Fourier Transform (FFT) algorithm. 

The ensemble average
of a continuous (ergodic) process $X(t)$ is denoted by $\Sigma(t,t') = \langle X(t)X^{\star}(t')\rangle$. In a discretized format, we have 
$\Sigma = \mathbb{E}_{\boldsymbol{x}}[\boldsymbol{x}\boldsymbol{x}^{\dagger}] \approx \langle \boldsymbol{x}\boldsymbol{x}^{\dagger}\rangle$. Here $\mathbb{E}_{\boldsymbol{x}}[\, \cdot \, ]$ represents an expectation with respect to the data generating process that determines $x$. Parameter sets are denoted $\bm{\theta}$, with elements $\theta^{a}\in\bm{\theta}$ and parameter derivatives given by $\partial_{a} := \partial/\partial\theta^{a}$ for $a=1,\dots,d$, with $d$ the dimension of the parameter space.

\section{The noise process}\label{sec:noise_process}
\subsection{Gaussian noise}\label{subsec:stationary_noise}

The output data stream of a GW detector is assumed to be a superposition of noise $n(t)$ and a collection of GW signals belonging to different source types. For simplicity, we will assume that there is a single GW signal $h(t;\bm{\theta}_{0})$ with source parameters $\theta$ buried in pure instrumental noise
\begin{equation}\label{eq:DA_single_data}
    d(t) = h(t;\bm{\theta}_0) + n(t)\,.
\end{equation}  

The data stream \eqref{eq:DA_single_data} contains two key quantities that impact successful extraction and parameter estimation of sources in GW astronomy. The first is a deterministic GW signal, for which accurate (and efficient) model templates can be used to cross-correlate with the data $d(t)$, in order to pick up observational features of the signal. The second is probabilistic detector noise: a culmination of instrumental and environmental fluctuations that perturb the GW detector. Since the noise is the probabilistic feature, it is the noise distribution that determines the likelihood required to perform inference on the signal $h(t;\boldsymbol{\theta})$. 

We will always assume that the noise $n(t)$ is a Gaussian time-series, with zero mean and unspecified covariance. Note that this assumption excludes transient disturbances such as instrumental glitches, known to be common in GW detectors, which depart from Gaussianity.

In discretized notation, we will write for TD data vectors of length $N$
\begin{equation}
    \bm{d} = \bm{h}(\bm{\theta}_0) + \bm{n}\, , 
\end{equation}
and we will represent the noise as following a multivariate normal distribution $\bm{n} \sim \mathcal{N} (\bm{0}, \bm{\Sigma})$, where $\bm{\Sigma}$ is the TD noise covariance matrix of shape $N\times N$ with
\begin{equation} \label{eq:defTDcovariance}
    \bm{\Sigma} = \langle \bm{n} \bm{n}^T\rangle \,.
\end{equation}
The likelihood for the data stream given signal parameters $\bm{\theta}$ is given by $\mathcal{L}(\bm{d} | \bm{\theta}) = p(\bm{n} = \bm{d} - \bm{h}(\bm{\theta}))
$, where $p(\bm{n})$ is the Gaussian probability density for the noise, which gives:
\begin{equation}
    \ln \mathcal{L} (\bm{d} | \bm{\theta}) = -\frac{1}{2} \bm{x}^T \bm{\Sigma}^{-1} \bm{x} - \frac{1}{2} \ln \det \bm{\Sigma} - \frac{N}{2} \ln (2\pi) \,,
    \label{eq:defTDlikelihood}
\end{equation}
where we use the notation $\bm{x} =  \bm{d} - \bm{h}(\theta) = \bm{n}$ for the residual. We will ignore the terms after the first in Eq.~\eqref{eq:defTDlikelihood}, as they appear as normalization constants when considering the likelihood as a function of $\bm{\theta}$; they would need to be considered e.g. for noise estimation.

Since the DFT is a unitary transform of the TD data, the FD likelihood preserves the same Gaussian structure. 
Defining the FD covariance as 
\begin{equation}
    \tilde{\bm{\Sigma}} = \langle \tilde{\bm{n}} \tilde{\bm{n}}^\dagger \rangle \,,
    \label{eq:FDcov}
\end{equation}
from Eq.~\eqref{eq:defDFT}, one can show that the FD covariance is related to the TD covariance via the DFT matrix $\boldsymbol{P}$
\begin{equation}
    \tilde{\bm{\Sigma}} = N \Delta t^2 \bm{P}\bm{\Sigma}\bm{P}^\dagger\,.
    \label{eq:defFDcov}
\end{equation}
The TD covariance matrix is given by
\begin{equation}
\bm{\Sigma} = \frac{1}{N\Delta t^2}\bm{P}^{\dagger}\tilde{\bm{\Sigma}}\bm{P}\,.\label{eq:defIDFTcov}
\end{equation}
Using Eqs.~\eqref{eq:defIDFT} and \eqref{eq:defIDFTcov}, the FD likelihood reads 
\begin{equation}
    \ln \mathcal{L} (\bm{d} | \bm{\theta}) = -\frac{1}{2} \tilde{\bm{x}}^\dagger \tilde{\bm{\Sigma}}^{-1} \tilde{\bm{x}} - \frac{1}{2} \ln \det \left(\frac{\Delta f}{\Delta t} \tilde{\bm{\Sigma}} \right) - \frac{N}{2} \ln \left( 2\pi\right) \,,
    \label{eq:defFDlike}
\end{equation}
where we used the unitary character of $\boldsymbol{P}$ and $N=1/(\Delta f\Delta t)$ to rewrite the determinant term. 

In this section we have restricted ourselves to one data stream but we will generalize our results to the full set of $A$, $E$ and $T$ time-delay-interferometry (TDI) streams that LISA will produce in Sec.~\ref{sec:application_MBHB}.

\subsection{Stationary noise}\label{sec:stationary_noise}
For real ergodic stationary processes, averaging a stochastic process over time through expectation $\mathbb{E}_{n}$ is equivalent to an ensemble average $\langle \cdot \rangle$. For now, we make the assumption that $n(t)$ is a Gaussian and stationary process implying that neither the mean or variance change over time and the auto-covariance function $C_{n}$ only depends on the lag $\tau$~\cite{Cutler:1994ys} 
\begin{align}\label{eq:auto_covariance}
C_{n}(\tau) = \langle n(t)n(t + \tau)\rangle\, .
\end{align}
From this auto-covariance, the noise Power Spectral Density (PSD) $S_n$ can be defined as
\begin{equation}
	\frac{1}{2} S_{n}(f) = \int d\tau \, e^{-2 i \pi f \tau} C_n(\tau) \,,
	\label{eq:defPSD}
\end{equation}
with the factor 2 being a matter of convention, for $S_{n}$ seen as a 1-sided PSD over positive frequencies. The PSD is real and positive: $S_n(f) \geq 0$. 

Using the continuous Fourier transform~\eqref{eq:CFT}, it can be shown that the ensemble average between the noise at two different frequencies $f,f'$ of either sign is related to the one-sided Power Spectral Density (PSD) through~\cite{wiener1930generalized,khintchine1934korrelationstheorie}
\begin{subequations}\label{eq:wiener_khintchine}
\begin{align}
    \Sigma_{n} (f, f') &= \langle \tilde{n}(f)\tilde{n}^{\star}(f')\rangle = \frac{1}{2}\delta(f - f')S_{n}(f) \,, \label{DA_eq:uncorrelated_frequencies} \\ 
    R_{n} (f, f') &= \langle\tilde{n}(f)\tilde{n}(f')\rangle = \frac{1}{2}\delta(f + f')S_{n}(f)\, . \label{DA_eq:zero_realisations_stationary}
\end{align}
\end{subequations}
Here Eqs.~\eqref{DA_eq:uncorrelated_frequencies} and \eqref{DA_eq:zero_realisations_stationary} are a statement that, for positive frequencies $f>0$ and $f'>0$, the frequency components among stationary noise components are uncorrelated. Further, $R_{n}(f,f') = 0$ in that case and the variance of the real and imaginary components of the noise process $\tilde{n}(f)$ are non-zero and equal. 

We now turn to the discrete domain in order to derive the usual Whittle-likelihood used commonly throughout GW astronomy. The direct translation of the stationarity of the noise is that there exists a vector $\bm{\sigma}$ representing the auto-covariance $C_n(\tau)$, such that
\begin{equation}\label{eq:deftoeplitz}
    \bm{\Sigma}_{ij} = \bm{\sigma}_{|i - j|} \,, \quad |i-j|=0,\dots,N-1 \,.
\end{equation}
The covariance matrix above has a \emph{Toeplitz} structure: entries of the covariance are constant along each diagonal of the matrix. The matrix must satisfy additional constraints arising from the fact that $C_n(\tau)$ must be a valid covariance function that can be written as the inverse Fourier transform of a real
and positive spectral density.

The Whittle-likelihood is obtained by imposing a further assumption on the covariance, that of a \emph{circulant} structure:
\begin{equation}\label{eq:defcirculant}
    \bm{\sigma}_{N-i} = \bm{\sigma}_{i} \,, \quad i=1,\dots,N-1 \,.
\end{equation}
This means that the covariance entries are not only constants along diagonals, but now these diagonals wrap around modulo $N$. In terms of physical interpretation, the Toeplitz structure is appropriate for an observation of a part of a much longer time series, while the circulant model assumes that the observed data represents the entirety of the stochastic process, by enforcing periodicity (which is never physical). In practice, one often considers a circulant process defined on a time interval $T$ much longer than the observed data, which itself is then a snapshot of that circulant process with a Toeplitz covariance. The choice of the circulant length $T$ will affect the lowest frequency $1/T$ accessible to the process, and will in principle change the low-frequency content of the autocorrelation function -- although we do not explore this in the present paper. 

In App.~\ref{app:diagonal_cov_matrix_derivation_circulant}, we demonstrate by direct calculation that  (with no summation over repeated indices)
\begin{equation}\label{eq:PSigmaPdag_stat_cov}
(\bm{P}\bm{\Sigma} \bm{P}^{\dagger})_{ij} = \sqrt{N} \delta_{ij} (\bm{P}\bm{\sigma})_{i}\,,
\end{equation} 
where we have imposed the circulant condition in Eq.~\eqref{eq:defcirculant}. One can now relate the PSD $S_{n}(f)$ to the Fourier transform of the auto-covariance function
\begin{equation}\label{eq:PSD_terms_of_sig_tilde}
\frac{1}{2}S_{n}(f_{j}) = \sqrt{N}\Delta t (\bm{P} \bm{\sigma})_{j} = \tilde{\bm{\sigma}}_{j}\,.
\end{equation}
Allowing one to read off the elements of $\tilde{\bm{\sigma}}_{j}$.
\begin{equation}
  \tilde{\bm{\sigma}}_{j} = \begin{cases} \frac{1}{2} S_{n} (j \Delta f) \; \text{for} \; j=0,\dots,N/2-1, \\
                               \frac{1}{2} S_{n} ((j-N) \Delta f) \; \text{for} \; j=N/2,\dots,N-1\,,
                   \end{cases}
   \label{eq:sigmatildePSD}
\end{equation}
with the components of $\tilde{\bm{\sigma}}_j$ being real and positive, and showing the same symmetry as Eq.~\eqref{eq:defcirculant}. The property $\tilde{\bm{\sigma}}_j \in \mathbb{R}$ is automatic for a circulant structure, while the condition $\tilde{\bm{\sigma}}_j \geq 0$ can be seen as a condition for $\bm{\sigma}$ to represent an admissible autocorrelation function, since the covariance $\bm{\Sigma}$ must be positive (as does its FD counterpart $\tilde{\bm{\Sigma}}$). In practice, we typically build the time domain $\bm{\sigma}$ from the IFT of a positive PSD, not the other way around.

Finally, the use of Eqs.~\eqref{eq:PSigmaPdag_stat_cov}, \eqref{eq:PSD_terms_of_sig_tilde} and \eqref{eq:sigmatildePSD}, one obtains the familiar FD diagonal matrix valid for both stationary and circulant noise 
\begin{equation}\label{eq:FDcovarianceWhittle}
    \tilde{\bm{\Sigma}} = N \Delta t^{2} (\bm{P} \bm{\Sigma} \bm{P}^{\dagger}) = N \Delta t \, \mathrm{Diag} (\tilde{\bm{\sigma}}) = \mathrm{Diag} \left( \frac{S_n}{2 \Delta f} \right) \,,
\end{equation}
where the values of the PSD along the diagonal are organized according to Eq.~\eqref{eq:sigmatildePSD}. This is the discrete equivalent to the continuous (and infinite duration) expression Eq.~\eqref{DA_eq:uncorrelated_frequencies}.

Note however that we have to impose the circulant condition to arrive at this diagonal structure, which is never realistic in practice: our data segment is an excerpt from a continuous physical process that, even if stationary, has no reason to be periodic. As discussed in~\cite{gray2006toeplitz}, Toeplitz and circulant matrices are asymptotically equivalent, implying that the Toeplitz matrices are asymptotically diagonalizable in a Fourier basis. This means that the Whittle-based likelihood better approximates the exact TD likelihood in the limit of an infinite observing time. We must therefore see the Whittle-likelihood as an approximation in that sense. This \emph{Toeplitz} versus \emph{circulant} problem is usually alleviated by considering data segments that are long enough to encompass the relevant signal with some margin, and by applying a smooth taper at each end of the data, enforcing periodicity of the time-segment. We will discuss circulant and Toeplitz processes in more details in Sec.~\ref{sec:analysis_noise_cov}, and investigate mis-modelling of such processes in the results Sec.~\ref{subsec:mismodeling_toeplitz_circulant}.  

From the diagonal structure given by Eq.~\eqref{eq:FDcovarianceWhittle}, we can now derive the usual Whittle log-likelihood used commonly throughout GW astronomy. The matrix $\tilde{\bm{\Sigma}}^{-1}$ is diagonal and readily inverted , and we can gather positive ($j=1,\dots,N/2-1$) and negative ($j=N/2+1,\dots,N-1$) frequency contributions. The FD likelihood~\eqref{eq:defFDlike} becomes, with $S_n^j \equiv S_n(j \Delta f)$,
\begin{align}
    \ln \mathcal{L} (\bm{d} | \theta) &= -2 \Delta f \sum_{j=1}^{N/2-1} \frac{\left| \tilde{\bm{h}}_j(\theta) - \tilde{\bm{d}}_j \right|^2}{S_n^j} - \sum_{j=1}^{N/2-1} \ln \frac{S_n^j}{\Delta t} \nonumber\\
    & - \frac{\Delta f}{S_n^0} \left| \tilde{\bm{h}}_0(\theta) - \tilde{\bm{d}}_0 \right|^2 - \frac{\Delta f}{S_n^{\frac{N}{2}}} \left| \tilde{\bm{h}}_{\frac{N}{2}}(
    \theta) - \tilde{\bm{d}}_{\frac{N}{2}} \right|^2 \nonumber\\
    & - \frac{1}{2} \ln \frac{S_n^0}{\Delta t} - \frac{1}{2} \ln \frac{S_n^{\frac{N}{2}}}{\Delta t} - \frac{N}{2}\ln 2\pi \,.
    \label{eq:DiscreteWhittle}
\end{align}
This is the Whittle-likelihood in discretized form, where we kept all the additive constants in $\ln\mathcal{L}$. When performing a Bayesian PE of the source parameters with a known PSD, the terms depending only on $S_n$ are normalization constants that can be ignored. Note also that the terms at $j=0$ and $j=N/2$ are not relevant for detectors having a finite frequency range of sensitivity. This expression~\eqref{eq:DiscreteWhittle} is efficient to compute, with $\mathcal{O}(N)$ operations for the sum and $\mathcal{O}(N\log_{2}  N)$ for the FFT.

In continuous notations, introducing the usual noise-weighted inner product
\begin{equation}\label{eq:def_innerproduct_functions}
    ( a | b ) = 4 \mathrm{Re} \int_{f > 0} df \; \frac{\tilde{a}(f)\tilde{b}^{*}(f)}{S_n(f)} \,,
\end{equation}
the result~\eqref{eq:DiscreteWhittle} above is the discretized version of
\begin{equation}\label{eq:FunctionalWhittle}
    \ln \mathcal{L}_{\rm Whittle} (\bm{d} | \bm{\theta}) = -\frac{1}{2} ( h(\bm{\theta}) - d | h(\bm{\theta}) - d ) \,,
\end{equation}
up to an additive constant incorporating PSD-dependent terms as an overall normalization.

The Whittle-likelihood \eqref{eq:DiscreteWhittle} is suitable for parameter inference provided the noise process is both Gaussian and circulant. As we will see, the impact of gaps destroys the circulant (and Toeplitz) nature of the TD noise covariance matrix, making the effective overall noise process as non-stationary.

\section{Direct modelling of data gaps}\label{sec:methods}

\subsection{Analysis set-up}\label{subsec:analysis_set_up}

We will now present the \emph{direct} approach to data gaps, where we marginalize over the missing data, while retaining statistical consistency. We will assume that we know the noise covariance of the underlying stochastic process representing the instrumental noise, and we treat the missing data entries as random variables that will be marginalized over, according to the joint probability distribution for the full process. We will also relate this \emph{direct} approach to the \emph{windowing} approach. We recall that, throughout this paper, we keep the assumption that the noise process is Gaussian.

Representing gaps can be done most easily in the TD and in the language of linear algebra. We present in App.~\ref{app:likelihood_using_masks_permutations} a derivation for a generic gap configuration, but for simplicity we will consider here a setting with a single gap inside the data set. For illustrative purposes, consider such a data set $\bm{x} = (\bm{x}_{0}, \bm{x}_{1}, \bm{x}_{2})$ of length $N=N_0 + N_1 + N_2$ with corresponding covariance matrix in block matrix form
\begin{equation}
\bm{\Sigma} = \begin{pmatrix}
\bm{\Sigma}_{00} & \bm{\Sigma}_{01} & \bm{\Sigma}_{02} \\
\bm{\Sigma}_{01}^{T} & \bm{\Sigma}_{11} & \bm{\Sigma}_{12} \\
\bm{\Sigma}_{02}^{T} & \bm{\Sigma}_{12}^{T} & \bm{\Sigma}_{22}
\end{pmatrix}\,.
\end{equation}
Now consider a gap in the data set, which we will represent by setting $\bm{x}_{1} = 0$, with observed data given by $\bm{x}_{0}$ and $\bm{x}_{2}$. In the \emph{direct} approach to data gaps, we marginalize over the missing data. The marginal joint probability distribution for the observed noise $\bm{n}_{0}$ and $\bm{n}_{2}$ is 
\begin{equation}\label{eq:marginalLike}
	p(\bm{n}_{0},\bm{n_{2}}) = \int d\bm{n}_{1} \, p(\bm{n}_{0}, \bm{n}_{1},\bm{n}_{2}) \,,
\end{equation}
where the joint probability $p(\bm{n}_{0}, \bm{n}_{1}, \bm{n}_{2})$ is described by the full covariance $\bm{\Sigma}$. Marginalizing a multidimensional Gaussian distribution is done by directly truncating its covariance, resulting in a $N_0 + N_2$ dimensional matrix 
\begin{equation}
\bm{\Sigma}^{\text{marg}} = \begin{pmatrix}
\bm{\Sigma}_{00} & \bm{\Sigma}_{02} \\
\bm{\Sigma}_{02}^{T}  & \bm{\Sigma}_{22}
\end{pmatrix}\,,
\end{equation}
and the likelihood therefore features the inverse of the $(N_{0} + N_{2}) \times (N_{0} + N_{2})$ matrix above. 

Assuming that $\bm{\Sigma}_{00}$ and the Shur complement of $\bm{\Sigma}_{00}$, i.e., $\bm{S} = (\bm{\Sigma}_{22} - \bm{\Sigma}_{02}^{T}\bm{\Sigma}_{00}^{-1}\bm{\Sigma}_{02})$ are invertible, the usual $2\times2$ block-inverse for a symmetric matrix reads
\begin{subequations}
\label{eq:2x2_inverse_identity}
\begin{align}
(\bm{\Sigma}^{\text{marg}})^{-1} &= \begin{pmatrix} \bm{T}_{00} & \bm{T}_{02} \\ \bm{T}_{02}^T & \bm{T}_{22} \end{pmatrix} \,,\\
\bm{T}_{00} &= \bm{\Sigma}_{00}^{-1} + \bm{\Sigma}_{00}^{-1}\bm{\Sigma}_{02}\bm{S}^{-1}\bm{\Sigma}_{02}^{T}\bm{\Sigma}_{00}^{-1}\,, \\
\bm{T}_{02} &= -\bm{\Sigma}_{00}^{-1}\bm{\Sigma}_{02}\bm{S}^{-1}\,,\\
\bm{T}_{22} & = \bm{S}^{-1}\,,
\end{align}
\end{subequations}
which can be used to obtain
\begin{align}
    \ln \mathcal{L}_{\rm gap} (\bm{x} | \bm{\theta}) =& -\frac{1}{2} \bm{x}_{0}^{T} \bm{T}_{00} \bm{x}_{0}  - \bm{x}_{0}^T \bm{T}_{02} \bm{x}_2\\
    &-\frac{1}{2} \bm{x}_{2}^T \bm{T}_{22} \bm{x}_{2} + \mathrm{const.}\,,
	\label{eq:marginalised_likelihood_td}
\end{align}
where as usual we use $\bm{x}_j = \bm{h}_j(\bm{\theta}) - \bm{d}_j$. We see that this direct approach is straightforward in the TD for Gaussian noise: we simply have to replace the full covariance with a sub-covariance. However, we still need to draw a connection to the FD which requires $N$-sized data segment, and to the windowing approach.

\subsection{The windowing procedure}\label{subsec:windowing_procedure}

Introducing a window (or gating) function $w(t)$ can be represented by an $N\times N$ diagonal matrix $\bm{W} = \mathrm{Diag}(w(i\Delta t), i=0,\dots,N-1)$. In the case where $w(t)$ is everywhere non-zero, $\bm{W}$ is invertible. This is the case where $w$ is a simple deterministic modulation, with no loss of information. Assuming the data analyst is given the modulated output $\bm{W} \bm{x}$, they can still rewrite the likelihood using
\begin{equation*}
	\bm{x}^{T} \bm{\Sigma}^{-1} \bm{x} = \left(\bm{W} \bm{x}\right)^{T} \left( \bm{W} \bm{\Sigma} \bm{W}\right)^{-1} \left(\bm{W} \bm{x}\right) \quad (\text{if} \; w \neq 0) \,.
\end{equation*}
Note that the covariance for the modulated data $\bm{W} \bm{\Sigma} \bm{W}$ will lose its Toeplitz structure in general; even though the underlying $\bm{x}$ is drawn from a stationary process, such modulated data is manifestly not stationary anymore -- but a modulation can be undone through inference schemes. For studies on non-stationary noise given modulated noise processes, we refer the reader to~\cite{Edy:2021par} for further discussion.

Data gaps cannot be undone as they erase information, and they can be represented by setting to zero the corresponding entries in $\bm{W}$ such that $t\in \mathrm{gap}$. While we could have $w$ implemented via a smooth taper around the gaps with $w(t) \in [0,1]$, here we will simply consider a gating\footnote{Whenever we refer to a window as a gating window we are referring to a rectangular window where $w(t) = 1$ outside the gap and $w(t) = 0$ inside the gap.} window of the form
\begin{equation}
	\bm{W} = \begin{pmatrix}
					\mathds{1} & \bm{0} & \bm{0} \\
					\bm{0} & \bm{0} & \bm{0} \\
	               \bm{0} & \bm{0} & \mathds{1}\end{pmatrix}
	               \,,
\end{equation}
and we have
\begin{align}\label{eq:Block_matrix_structure_gaps_TD}
\bm{W}\bm{\Sigma}\bm{W} &= \begin{pmatrix}
\bm{\Sigma}_{00} & \bm{0} & \bm{\Sigma}_{02} \\
\bm{0}  & \bm{0}  & \bm{0}  \\
\bm{\Sigma}_{02}^{T} & \bm{0} & \bm{\Sigma}_{22}
\end{pmatrix} \,,
\end{align}
which is non-invertible, representing the erasure of information by the introduction of gaps.

However, the notion of Moore-Penrose pseudo-inverse~\cite{moore1920reciprocal, penrose1955generalized} will allow us to find a direct connection to the result in Eq.~\eqref{eq:marginalLike} for the likelihood marginalized over missing data values. For any matrix $\bm{A}$ (with no prerequisites, $\bm{A}$ can even be non-square), there exists a unique pseudo-inverse $\bm{A}^{+}$ satisfying the four properties
\begin{subequations}
\begin{align}
	 \bm{A}^{+} \bm{A} \bm{A}^{+} &= \bm{A}^{+} \,, \\
	 \bm{A} \bm{A}^{+} \bm{A} &= \bm{A} \,, \\
	 \left(\bm{A} \bm{A}^{+} \right)^{\dagger} &= \bm{A} \bm{A}^{+} \,, \\
	 \left(\bm{A}^{+} \bm{A} \right)^{\dagger} &= \bm{A}^{+} \bm{A} \,.
\end{align}
	\label{eq:defpseudoinverse}
\end{subequations}
The pseudo-inverse coincides with the normal inverse when the matrix is invertible. The following properties of the pseudo-inverse, that can be checked from the definition, will be useful for us: if $\bm{B}$ is unitary ($\bm{B}^{\dagger} \bm{B} = \mathds{1}$), $(\bm{A}\bm{B})^{+} = \bm{B}^{+} \bm{A}^{+} = \bm{B}^{\dagger} \bm{A}^{+}$ and similarly $(\bm{B}\bm{A})^{+} = \bm{A}^{+} \bm{B}^{+} = \bm{A}^{+}\bm{B}^{\dagger}$. Scalar multiplication works as for a normal inverse: $(\alpha \bm{A})^{+} = \bm{A}^{+} / \alpha$ for $\alpha \in \mathbb{C}$, $\alpha \neq 0$. Finally, for a block-diagonal matrix, the pseudo-inverse respects the block-diagonal structure. In practice, one general algorithm to compute the pseudo-inverse~\cite{press2007numerical} consists in first computing the singular value decomposition (SVD) of the matrix, identifying the zero singular values (SVs) and setting their reciprocal to zero when approximating the matrix inverse. This will be discussed in more detail in Sec.~\ref{subsec:pseudo_inverses_practice}.

For the $3\times 3$ block structure that we have, the pseudo-inverse can be calculated directly through repeated application of the $2\times 2$ block inverse formula, or checked a posteriori, resulting in 
\begin{equation}
(\bm{W}\bm{\Sigma}\bm{W}) ^{+} = \begin{pmatrix}  \bm{T}_{00} & \bm{0} & \bm{T}_{02} \\
\bm{0}  & \bm{0}  & \bm{0}  \\
\bm{T}_{02}^T & \bm{0} & \bm{T}_{22}
\end{pmatrix}\,,
\label{eq:pseudo_inverse_window}
\end{equation}
with block matrices given as blocks of the $2\times 2$ block-inverse in Eq.~\eqref{eq:2x2_inverse_identity}. Notice that the block diagonal structure has been preserved. We can verify that 
\begin{equation}\label{eq:pseudo_inv_mult_non_inv_identity}
(\bm{W}\bm{\Sigma}\bm{W})(\bm{W}\bm{\Sigma}\bm{W}) ^{+} = \begin{pmatrix}
\mathds{1} & \bm{0} & \bm{0} \\
\bm{0}  & \bm{0}  & \bm{0}  \\
\bm{0} & \bm{0} & \mathds{1}
\end{pmatrix}\,,
\end{equation}
showing that $(\bm{W}\bm{\Sigma}\bm{W}) ^{+}$ is indeed the unique pseudo-inverse of $(\bm{W}\bm{\Sigma}\bm{W})$, satisfying all four properties in Eq.~\eqref{eq:defpseudoinverse}. The likelihood for gated data in the TD reads
\begin{align}
    \ln \mathcal{L}_{\rm gap} =& (\bm{W}\bm{x})^T(\bm{W}\bm{\Sigma}\bm{W})^{+}(\bm{W}\bm{x}) \nonumber \\
    =& -\frac{1}{2} \bm{x}_{0}^{T} \bm{T}_{00} \bm{x}_{0}  - \bm{x}_{0}^T \bm{T}_{02} \bm{x}_2 \label{eq:gated_likelihood_td_W} -\frac{1}{2} \bm{x}_{2}^T \bm{T}_{22} \bm{x}_{2} \,,
\end{align}
which is precisely the same form as the TD likelihood Eq.~\eqref{eq:marginalised_likelihood_td} where we marginalized out the data corresponding to the gap segment $\bm{x}_{1}$. This is an important point: it shows that marginalizing the data (via truncating the noise covariance matrix) is equivalent to treating the data coherently while filling the gap segment with zeros. The operation on the space of vectors via $(\bm{W}\bm{\Sigma}\bm{W})^{+}$ or $(\bm{\Sigma}^\text{marg})^{-1}$ is equivalent in all cases.

We arrive at the gated likelihood
\begin{tcolorbox}[colframe=black, colback=white, arc=4mm, boxsep=0mm, left=2mm, right=1mm, top=0cm, bottom=4mm]
\begin{subequations}\label{eq:windowed_likelihood_td}
\begin{align}
	\ln\mathcal{L}_{\text{gate}} (\bm{d} | \bm{\theta}) &=-\frac{1}{2}\bm{x}^{T} \left(\bm{W}\bm{\Sigma} \bm{W}\right)^{+} \bm{x}\,,\\
    &= -\frac{1}{2}(\bm{W}\bm{x})^{T} \left(\bm{W}\bm{\Sigma} \bm{W}\right)^{+} (\bm{W}\bm{x})\,, \label{eq:windowed_likelihood_td_withW}
\end{align}
\end{subequations}
\end{tcolorbox}
with $\bm{x} = \bm{d} - \bm{h}$. In the above, the data set is $\bm{d}=(\bm{d}_{0}, \bm{d}_{1}, \bm{d}_{2})$ and the model template is $\bm{h} = (\bm{h}_{0}, \bm{h}_{1}, \bm{h}_{2})$; note here that $\bm{d}_{1}$, $\bm{h}_{1}$ do not contribute the the likelihood. Assuming that the correct pseudo-inverse has been computed, then its presence will \emph{annihilate} any data within the gap segment. 

This is a point that warrants further discussion. The data stream of LISA will be a collection of data products produced via first-stage L0/L1 data pipelines, which will result in time-ordered data sets~\cite{LISA:2024hlh}. The missing data segments are segments of data that are not present in these data sets. Glitches (or instrumental/environmental noise transients) may contaminate the data stream so violently that they may need to be masked and thus removed entirely. Throughout this section we have made the convenient choice to fill in those missing data segments with zeros in order to ``connect" data products $\bm{d}_{0}$ and $\bm{d}_{2}$ allowing for a \emph{coherent} data set $\bm{d} = (\bm{d}_{0},\bm{0},\bm{d}_{2})$ to be analyzed. It is important to understand that this is already placing an assumption that the segments are correlated, which may not be the case in practice. This will be investigated later in Sec.~\ref{subsec:mismodeling_independence}. We have shown that provided the covariance matrix of the noise accounts for missing-data as zeros, then the content of the data stream is irrelevant during the gated segment. For instance, in terms of the input gated data $\bm{W}\bm{d} = (\bm{d}_{0}, \bm{0}, \bm{d}_{2})$, then Eq.~\eqref{eq:windowed_likelihood_td_withW} reads equivalently: 
\begin{tcolorbox}[colframe=black, colback=white, arc=4mm, boxsep=0mm, left=2mm, right=1mm, top=0mm, bottom=2mm]
\begin{equation}\label{eq:windowed_likelihood_td_gate}
    \ln \mathcal{L}_{\rm gate} (\bm{d} | \bm{\theta}) = -\frac{1}{2} (\bm{W}\bm{d} - \bm{h})^{T} (\bm{W}\bm{\Sigma} \bm{W})^{+}(\bm{W}\bm{d} - \bm{h})\,,
\end{equation}
\end{tcolorbox}
ignoring additive constants. Thus, the content of the data and model template during the gap segment are erased entirely. In reality, the data stream during the gated segment will not exist and could be represented via \texttt{NaNs} or $\bm{d}_{1} = 0$, the point here is that the choice would not matter. For inference purposes, we could use the full model template, fill the gap segment with zeros or even arbitrary values and always get the same result for the likelihood.

Note that we were assuming a gating window above, with $w(t)=0$ or $w(t)=1$. For a more general tapering window with $w(t) \in [0,1]$ outside the gap, we can repeat the same derivation with
\begin{align}
	\bm{W} &= \begin{pmatrix}
					\bm{W}_{0} & \bm{0} & \bm{0}\\
					\bm{0} & \bm{0} & \bm{0} \\
	               \bm{0} & \bm{0} & \bm{W}_{2}\end{pmatrix}
	               \,, \label{eq:time_domain_tapering_window_matrix}\\
	\bm{W}\bm{\Sigma} \bm{W} &= \begin{pmatrix}
					\bm{W}_{0}\bm{\Sigma}_{00}\bm{W}_{0} & \bm{0} &  \bm{W}_{0}\bm{\Sigma}_{02}\bm{W}_{2}\\
					\bm{0} & \bm{0} & \bm{0} \\
\bm{W}_{2}\bm{\Sigma}_{02}^T\bm{W}_{0} & \bm{0} & \bm{W}_{2}\bm{\Sigma}_{22}\bm{W}_{2}  
	               \end{pmatrix} \label{eq:time_domain_noise_cov_windowed}\,,
\end{align}
with $\bm{W}_{0}$ and $\bm{W}_{2}$ invertible, we would obtain the same result as in Eq.~\eqref{eq:windowed_likelihood_td_withW}. Note that when the window is not a gate, we need to modify Eq.~\eqref{eq:windowed_likelihood_td_gate} to also apply the tapering outside of the gaps to the templates $\bm{h}$.

We conclude our time-domain prescription of data gaps with a particularly counter-intuitive point about gaps with smooth tapers applied pre and post-gap segment. For the smooth tapering matrix defined in Eq.~\eqref{eq:time_domain_tapering_window_matrix}, it's easy to show that 
\begin{align}
(\bm{W}\bm{W}^{+}) = (\bm{W}^{+}\bm{W}) = \begin{pmatrix}
\mathds{1} & \bm{0} & \bm{0} \\
\bm{0}  & \bm{0}  & \bm{0}  \\
\bm{0} & \bm{0} & \mathds{1}
\end{pmatrix}\,, 
\end{align}
leading to the observation that 
\begin{align}
\ln \mathcal{L}_{\text{gap}} &\propto -\frac{1}{2}(\bm{W}\bm{x})^T (\bm{W}\bm{\Sigma}\bm{W})^{+}(\bm{W}\bm{x}) \\
& = \begin{pmatrix} \bm{x}_{1} & \bm{x}_{2} & \bm{x}_{3} \end{pmatrix} \begin{pmatrix}
\bm{\Sigma}_{00} & \bm{0} & \bm{\Sigma}_{02} \\
\bm{0}  & \bm{0}  & \bm{0}  \\
\bm{\Sigma}_{02}^{T} & \bm{0} & \bm{\Sigma}_{22}
\end{pmatrix}^{+} \begin{pmatrix} \bm{x}_{1} \\ \bm{x}_{2} \\ \bm{x}_{3} \end{pmatrix} \\
&\equiv \ln \mathcal{L}_{\text{gate}} \label{eq:stat_consistency_gate_taper}\,.
\end{align}
The above argument proves that there is no loss of statistical consistency between tapered and gated processes. Assuming that the pseudo-inverse has been correctly computed, the tapering scheme will result in \emph{no loss of information} of the signal. From the time-domain perspective, there is little reason to taper your data since the gated and tapered process will result in the same likelihood calculation. However, tapering may prove advantageous in the FD in order to mitigate edge-effects as a result of working with finite-duration time-series. 

We will now translate this gap treatment to the FD, using the version of the formalism that uses windowed $N$-size data. Starting from Eq.~\eqref{eq:windowed_likelihood_td}, using the DFT definitions in Eqs.~\eqref{eq:defDFT}, we have
\begin{equation}
	\bm{W} \bm{x} = \frac{1}{\sqrt{N} \Delta t} \bm{P}^\dagger \widetilde{\bm{W} \bm{x}} \,,
\end{equation}
for $\widetilde{\bm{W} \bm{x}} = \tilde{\bm{W}}\tilde{\bm{x}}$. Since $\bm{P}$ is unitary, we can absorb it inside the pseudo-inverse as 
\begin{equation}
	\bm{P} \left(\bm{W}\bm{\Sigma} \bm{W}\right)^{+}  \bm{P}^{\dagger} = \left(\bm{P} \bm{W} \bm{\Sigma} \bm{W} \bm{P}^{\dagger} \right)^{+} \,.
\end{equation}
Using Eq.~\eqref{eq:defFDcov} for the FD covariance matrix, and introducing the notation (note that this is not the diagonal matrix built from the vector $\tilde{\bm{w}}_{j}$) 
\begin{equation}\label{eq:defWtilde}
	\tilde{\bm{W}} = \bm{P} \bm{W} \bm{P}^{\dagger} \,,
\end{equation}
we can further rewrite the above as
\begin{equation}
	\frac{1}{N \Delta t^2} \left(\bm{P} \bm{W} \bm{\Sigma} \bm{W} \bm{P}^{\dagger} \right)^{+} =  \left(\tilde{\bm{W}} \tilde{\bm{\Sigma}} \tilde{\bm{W}}\right)^{+} \,.
\end{equation}
We finally obtain the FD likelihood in the gap case, marginalized over missing data, 
\begin{equation}\label{eq:windowed_likelihood_fd}
	\ln \mathcal{L}_{\rm gap} (\bm{d} | \bm{\theta}) = -\frac{1}{2}\left(\widetilde{\bm{W}\bm{x}}\right)^{\dagger} \left( \tilde{\bm{W}} \tilde{\bm{\Sigma}} \tilde{\bm{W}}\right)^{+} \left(\widetilde{\bm{W} \bm{x}}\right) \,, 
\end{equation}
with $\bm{x} = \bm{d} - \bm{h}(\bm{\theta})$ and ignoring additive constants.

Coming back to the implementation choice, we have a counter-intuitive freedom in how we choose to fill the missing data values for the template. It is natural to fill in the model template $\bm{h}$ with zeros, as the most convenient choice. For particular GW model templates, however, we can have situations where the windowing would strongly affect the shape of the Fourier transform of the signal. This is already the case for a MBHB, but it would be much worse for a quasi-monochromatic GB signal. If a smooth taper were applied to the TD GB signal, the GBs FD spectrum would no longer be compact. According to our argument, however, we can choose to not apply the window to the template, allowing us to use existing codes producing FD templates with no modification. This implementation would read:
\begin{tcolorbox}[colframe=black, colback=white, arc=4mm, boxsep=0mm, left=2mm, right=1mm, top=0mm, bottom=2mm]
\begin{equation}\label{eq:windowed_likelihood_fd_alt}
	\ln \mathcal{L}_{\rm gate} (\bm{d} | \bm{\theta}) = -\frac{1}{2}(\widetilde{\bm{W} \bm{d}} - \tilde{\bm{h}})^{\dagger}( \tilde{\bm{W}} \tilde{\bm{\Sigma}} \tilde{\bm{W}})^{+} (\widetilde{\bm{W}\bm{d}} - \tilde{\bm{h}}) \,,
\end{equation}
\end{tcolorbox}
which may be a more favorable likelihood model to compute for models built directly in the FD~\cite{speri2024fast, Khan:2015jqa, Husa:2015iqa, Cotesta:2020qhw}. For smooth tapers, it is necessary to include the window in the model templates $\widetilde{\bm{W}\bm{h}}$ in Eq.\eqref{eq:windowed_likelihood_fd_alt}.

\subsection{Computing the windowed covariance}
\label{subsec:windowed_covariance}
If one were to compute the FD covariance with gated data and transform to the TD covariance, then one is making the hidden assumption that the data is just missing observations in the middle of some continuous process. In other words, the two data segments are \emph{dependent} and coherent with one another. In this section, we will provide algorithms that can be used to compute the noise covariance matrix assuming coherence between observed data segments, which translates both in the noise level being assumed to be the same before and after the gap, and in the presence of off-diagonal blocks in the TD covariance.

In the above, we found that the likelihood can be written down in terms of the covariance matrices for the windowed noise process in the time or FD,
\begin{align}
    \left\langle (\bm{W}\bm{n}) (\bm{W} \bm{n})^T \right\rangle &= \bm{W}\bm{\Sigma}\bm{W} \,,\nonumber\\
    \left\langle (\widetilde{\bm{W}\bm{n}}) (\widetilde{\bm{W} \bm{n}})^\dagger \right\rangle &= \tilde{\bm{W}} \tilde{\bm{\Sigma}} \tilde{\bm{W}} \,.
\end{align}
We will now present how these matrices can be computed in practice.

In the TD, the computation of the windowed covariance is straightforward. In functional notations, we have
\begin{equation}
    \langle w(t)n(t) w(t')n(t')\rangle = w(t)w(t') C_n(t,t') \,,
\end{equation}
and the translation of the above direct product in discrete notations is, for $i,j = 0,\dots,N-1$,
\begin{align}
    (\bm{W}\bm{\Sigma}\bm{W})_{ij} &= \bm{w}_i \bm{w}_j \bm{\Sigma}_{ij}\,,
\end{align}
where $\bm{w}$ is the vector of values for the window. 

To build the circulant TD noise covariance matrix, an inverse discrete-Fourier transform of the PSD $S_{n}(f)$ is computed, which then forms the first row of the matrix. As the matrix is circulant, one can easily construct the matrix via Eq.~\eqref{eq:defcirculant}. Constructing the Toeplitz matrix is slightly more involved, since Eq.~\eqref{eq:defPSD} is a definition that assumes infinite observation times. In practice, we construct the inverse discrete-Fourier transform of the PSD $S_{n}(f)$ assuming twice the observation time. 
Half of the time-array is chosen to be the first row of the matrix and, applying Eq.~\eqref{eq:deftoeplitz}, one can construct the resultant Toeplitz matrix. When gaps are present, one simply truncates the rows and columns corresponding to the time-indices where the gap is present (see Eq.~\eqref{eq:Block_matrix_structure_gaps_TD}). Note that this process is \emph{not invariant} to the choice of the length of the extended circulant process, of which our Toeplitz process is a snapshot. Choosing more than twice the target length would change the low-frequency content of the autocorrelation function. We did not explore this aspect in the present study.

In the FD, the product with the window becomes a convolution, according to 
\begin{equation}
    \widetilde{wn}(f) = \int \text{d}u \; \tilde{w}(u) \tilde{n}(f - u) \,.
\end{equation}
Using Eq.~\eqref{DA_eq:zero_realisations_stationary} for stationary noise, the functional FD covariance becomes~\cite{Zackay:2019kkv, Talbot:2021igi}
\begin{align}
    \Sigma_{wn}(f,f') &= \langle \widetilde{wn}(f) \widetilde{wn}(f')^{*} \rangle \nonumber\\
    &= \frac{1}{2} \int \text{d}u \; S_n(u) \tilde{w}(f - u) \tilde{w}(f' - u)^{*} \,.
\end{align}
In linear algebra notations, this result will translate into a discrete convolution. First, from our definition for $\tilde{\bm{W}}$,
\begin{equation}
	\tilde{\bm{W}}_{ij} = \left( \bm{P} \bm{W} \bm{P}^{\dagger}\right)_{ij} = \frac{1}{N} \sum_{k} \omega^{k(i-j)} \tilde{w}_{k} \,.
\end{equation}
For discrete convolutions, it is useful to introduce periodic indices, to be interpreted modulo $N$. We will denote them with an overline, $\tilde{\bm{w}}_{\overline{k}} = \tilde{\bm{w}}_{k \mod N }$. Using these notation allows to simplify intermediate sums while ignoring edge effects on indices. Using the diagonal structure for the FD covariance given by Eq.~\eqref{eq:FDcovarianceWhittle} in the circulant case, we obtain the convolution
\begin{equation}
	\left( \tilde{\bm{W}} \tilde{\bm{\Sigma}} \tilde{\bm{W}}\right)_{ij} = \frac{\Delta f}{2} \sum_{k=0}^{N-1} S_{n}^{k} \tilde{\bm{w}}_{\overline{i-k}} \tilde{\bm{w}}_{\overline{j-k}}^{*}\,.
	\label{eq:WSigmaW}
\end{equation}
The above matrix can be computed more efficiently than with the apparent $\mathcal{O}(N^3)$ of this formula, by using the FFT for convolutions. Namely, for each periodic diagonal $j - i = p \mod N$, if we define the vector $\bm{v}^{(p)}_i = \tilde{\bm{w}}_{\overline{i}} \tilde{\bm{w}}_{\overline{i+p}}^{*}$, then the computation for this diagonal can be seen as the computation of the convolution $\sum S_n^k v^{(p)}_{i - k}$. The latter convolution can be computed by taking the IFFT of the product of the two vector's FFT, for an overall cost of $\mathcal{O}(N\log_{2}N)$. This needs to be done for each diagonal, which gives the overall scaling $\mathcal{O}(N^2 \log_{2} N)$or computing the full matrix\footnote{Similarly, given knowledge of the TD noise covariance matrix the act of $\tilde{\bm{\Sigma}} \propto \bm{P}\bm{\Sigma}\bm{P}^{\dagger}$ is equivalent to taking an FFT on the \emph{columns} of the matrix $\bm{\Sigma}$ (pre-multiplying by $\bm{P}$) and then an IFFT on the post-matrix rows (equivalent to post-multiplying by $\bm{P}^{\dagger} = \bm{P}^{-1}$). The cost of this operation scales like $\mathcal{O}(N^2 \log_{2}N)$ for data lengths a power of two. The algorithm in the text computes each diagonal separately, which could be advantageous for band-dominated matrices, should we truncate the matrix to a preset number of diagonals.}.

Computing the pseudo-inverse of that matrix, however, requires computing its SVD in $\mathcal{O}(N^3)$, as we will describe in Sec.~\ref{subsec:pseudo_inverses_practice}. Below in Alg.~\ref{algorithm:FD_covariance} we provide pseudo-code that can be used to compute the FD noise covariance matrix assuming a Circulant and Gaussian noise process. When building the frequency-domain gated covariance matrix for Toeplitz processes, we found it simpler to construct the time-domain covariance $\bm{\Sigma}^{\text{toe}}$ and compute $\tilde{\bm{\Sigma}} = (N\Delta t)^{-2}\bm{P}\bm{\Sigma}^{\text{toe}}\bm{P}^{\dagger}$. Analytical formulae for the FD version of the Toeplitz matrix are given in App.\ref{app:The_Toeplitz_Case_Derivation}, Eq.~\eqref{eq:Toeplitz_Matrix}. The TD and FD covariances for both Circulant and Toeplitz processes will be illustrated and discussed in Sec.\ref{sec:analysis_noise_cov}. 

\begin{algorithm}
\begin{algorithmic}
\label{algorithm:FD_covariance}
\caption{Fourier Domain Noise Covariance}
\Require $S_n$ PSD values of length $N/2 + 1$, $\bm{w}$ window array in TD of length $N$ and $\Delta t$ the sampling interval.
\Ensure $\bm{\tilde{\Sigma}}$ (Fourier-domain covariance matrix, complex array)
\State $N \gets \text{length}(\bm{w})$
\State $\Delta f \gets 1 / (N\Delta t)$
\State Compute Fourier transform: $\bm{\tilde{w}} \gets \Delta t \cdot\text{FFT}(\bm{w})$
\State Initialize $\bm{0} \gets \bm{u} \in \mathbb{C}^{N\times 1}$ 
\State Complex $\bm{\tilde{\Sigma}} \gets \text{Zero Matrix}$ of size $N \times N$.
\State $\bm{u}[0:N/2+1] \gets \bm{S_n}$,
\State $\bm{u}[N/2+1:N] \gets \bm{S_n}[1:-1]$ with $\bm{S_n}$ reversed
\For{diagonal $\in \{0, \ldots , N/2\}$}
    \State $\bm{w}_{\text{shift}} \gets \text{permute}(\bm{w}, \ \text{-diagonal})^{*}$ 
    \State $\bm{v} \gets \bm{w} \cdot \bm{w}_{\text{shift}}$
    \State  $\bm{\hat{u}} \gets \text{FFT}(\bm{u})$
    \State $\bm{\hat{v}} \gets \text{FFT}(\bm{v})$
    \State $\text{D} \gets (\Delta f / 2) \cdot \text{IFFT}(\bm{\hat{u}} \cdot \bm{\hat{v}})$
    \State Set diagonal of $\bm{\tilde{\Sigma}} \gets D$ 
    \If{diagonal $\neq$ (N/2)}
        \State Assign $D^{*}$ to mirrored diagonal entries
    \EndIf
\EndFor \\
\Return $\bm{\tilde{\Sigma}}$
\end{algorithmic}
\end{algorithm}

\subsection{Signal-to-noise ratio}
\label{subsec:SNR}

In this section, we rewrite the usual definitions for the Wiener filtering and Signal-to-Noise Ratio (SNR) in the TD, using the same linear algebra notations as in previous sections, and adapt the derivation to the presence of gaps in the data.

For a given filter $k(t)$, we consider the $L^2$ inner product of this filter with the data stream $\int dt \; k(t) d(t)$. We introduce the ``signal'' and ``noise'' as the expectation value of the inner product when a signal $h(t;\bm{\theta})$ is present in the data, and its standard deviation when the signal is absent (the expectation value of the noise being zero), respectively. The signal-to-noise ratio (SNR) is then
\begin{equation}
\label{eq:def_SNR_functions}
    \left(\frac{S}{N}\right) = \frac{ \int_{-\infty}^{\infty} dt \; k(t) h(t;\bm{\theta}) }{\sqrt{\left\langle \left( \int_{-\infty}^{\infty} dt \;  k(t)n(t) \right)^{2} \right\rangle}},
\end{equation}
and we wish to optimize this SNR by best matching the chosen filter $k$ to the target signal $h(t;\bm{\theta})$, assumed to be known.

In linear algebra notations, the filter and waveform templates become vectors $\bm{k}, \bm{h} \in \mathbb{R}^N$ (we omit the $\bm{\theta}$-dependency in $\bm{h}$) and the SNR is
\begin{equation}
\label{eq:def_SNR_linalg}
    \left(\frac{S}{N}\right) = \frac{\bm{k}^T \bm{h}}{\sqrt{\left\langle \left(\bm{k}^T \bm{n} \right)^2 \right\rangle}} = \frac{\bm{k}^T \bm{h}}{\sqrt{ \bm{k}^T \bm{\Sigma} \bm{k} }} \,,
\end{equation}
where we used the TD noise covariance matrix in Eq.~\eqref{eq:defTDcovariance} to rewrite the denominator.

For $\bm{\Sigma}$ symmetric and positive definite, there exists precisely one positive definite and symmetric square-root matrix\footnote{This would still hold with positive-semi-definite matrices.} $\bm{\Sigma}^{1/2}$ such that $\bm{\Sigma} = \bm{\Sigma}^{1/2} \bm{\Sigma}^{1/2}$~\cite{horn2012matrix}. Denoting the inverse square root as $\bm{\Sigma}^{-1/2}$ and using the Cauchy-Schwarz inequality,
\begin{equation}
    \left| \bm{k}^T \bm{h} \right| = \left| \left( \bm{\Sigma}^{1/2} \bm{k} \right)^T \left(\bm{\Sigma}^{-1/2} \bm{h}\right) \right| \leq \sqrt{\bm{k}^T \bm{\Sigma} \bm{k}} \sqrt{\bm{h}^T \bm{\Sigma}^{-1} \bm{h}} \,,
\end{equation}
with the bound being saturated for the two vectors being collinear, that is to say for $\bm{\Sigma}^{1/2} \bm{k} \propto \bm{\Sigma}^{-1/2} \bm{h}$, or equivalently for
\begin{equation}
    \bm{k} \propto \bm{\Sigma}^{-1} \bm{h} \,.
\end{equation}
Coming back to the expression of the SNR~\eqref{eq:def_SNR_linalg}, this gives the optimal square SNR as 
\begin{equation}\label{eq:SNRopt}
    \rho_{\rm opt}^2 = \left(\frac{S}{N}\right)^2_{\rm opt} = \bm{h}^T \bm{\Sigma}^{-1} \bm{h} \,.
\end{equation}

Working in the FD rather than in the TD, The expressions Eqs~\eqref{eq:defIDFT} and~\eqref{eq:defIDFTcov} allow us to write Eq.~\eqref{eq:SNRopt} in the form
\begin{equation}\label{eq:SNRopt_FD}
    \rho_{\rm opt}^2 = \tilde{\bm{h}}^\dagger \tilde{\bm{\Sigma}}^{-1} \tilde{\bm{h}} \,.
\end{equation}
For stationary noise (and in the circulant case), $\tilde{\bm{\Sigma}}$ is diagonal and this inner product becomes a simple sum, as in Eq.~\eqref{eq:DiscreteWhittle}, which is the discrete equivalent of the usual expression $\rho_{\rm opt}^2 = (h|h)$ using the functional inner product defined by Eq.~\eqref{eq:def_innerproduct_functions}.

Now, considering the case of data gaps, we can define the observed data stream via $\bm{W}\bm{h}$ and marginalised TD covariance matrix $\bm{W}\bm\Sigma\bm{W}$. The SNR is then 
\begin{equation}
    \left(\frac{S}{N}\right) = \frac{\bm{k}^T \bm{W}\bm{h}}{\sqrt{\left\langle \left(\bm{k}^T \bm{W}\bm{n} \right)^2 \right\rangle}} = \frac{\bm{k}^T \bm{W}\bm{h}}{\sqrt{ \bm{k}^T \bm{W}\bm{\Sigma}\bm{W}\bm{k} }}\,.
\end{equation}
As in the case of the likelihood, we can either work with an explicit expression of the pseudo-inverse in a $3\times 3$ block structure as in Sec.~\ref{sec:methods}, or use the reordering notations of App.~\ref{app:likelihood_using_masks_permutations}, and obtain for the optimal SNR
\begin{equation}\label{eq:SNRoptgap}
    (\rho_{\rm opt}^{\rm gap})^2 = (\bm{W}\bm{h})^T (\bm{W}\bm{\Sigma}\bm{W})^{+} (\bm{W}\bm{h}) \,.
\end{equation}
In the above, the factor $\bm{W}$ in $\bm{W}\bm{h}$ is only mandatory if the window is not a pure gating window but includes a smooth tapering on the segments outside the gaps. We remind the reader that the presence of $(\bm{W}\Sigma\bm{W})^{+}$ kills all imputed information within the gap segment, implying that we are free to use the non-gated templates for SNR computations. Similarly, the string of equalities leading to the statement in Eq.~\eqref{eq:stat_consistency_gate_taper} applies here, showing that the SNR of the gated signal is identical to the SNR of the tapered process. Hence the only portion of the signal that results in information loss is only the segment of zero data -- during the gap. This phenomena will be verified in Fig.~\ref{fig:cov_vs_pseudo_inv_eps} found in Sec.~\ref{subsec:pseudo_inverses_practice_w_taper}.

We will see in later sections that tapered waveforms with assumed Whittle-based diagonal covariances could significantly impact the SNR of the source, especially when information-rich content of the signal is being erased.

The translation to the FD is the same as for the likelihood, and reads
\begin{equation}\label{eq:SNRoptgapFD}
    (\rho_{\rm opt}^{\rm gap})^2 = (\widetilde{\bm{W}\bm{h}})^\dagger (\tilde{\bm{W}}\tilde{\bm{\Sigma}}\tilde{\bm{W}})^{+} (\widetilde{\bm{W}\bm{h}}) \,.
\end{equation}

\subsection{Fisher matrices}
\label{subsec:Fisher}
The Fisher matrix (FM) formalism is a useful tool for approximate PE, and we will use it extensively in the rest of the paper. We repeat here the derivation of the consequences of the linearized signal approximation, adapted to the case of data with gaps. 

We start by writing the linearized signal approximation for the Whittle-likelihood~\eqref{eq:FunctionalWhittle} and with the functional overlap~\eqref{eq:def_innerproduct_functions}, for simplicity of notation before adapting to the gap case. We expand the signal at linear order in deviations from the true parameters $\bm{\theta}_0$ as
\begin{equation}
    h(\bm{\theta}) = h(\bm{\theta}_0) + \Delta \bm{\theta}^a \partial_a h(\bm{\theta}_0) + \mathcal{O}(\Delta \bm{\theta}^2) \,.
\end{equation}
Given a noise realization so that $\bm{d} = \bm{h} + \bm{n}$ and in the absence of waveform uncertainties the likelihood becomes a quadratic form~\cite{Finn:1992wt} in $\Delta \bm{\theta}$:
\begin{align}
    \ln \mathcal{L} &= -\frac{1}{2} (\Delta \bm{\theta}^a - \Delta \hat{\bm{\theta}}^{a}_{\text{bf}})\bm{\Gamma}_{ab}(\Delta\bm{\theta}^b - \Delta \hat{\bm{\theta}}^{b}_{\text{bf}}) \nonumber  \label{eq:LSA_like}\\
    & \qquad -\frac{1}{2}(n|n)  - (\partial_{a}h|n)(\bm{\Gamma}^{-1})_{ab}(\partial_{b}h|n) \\
    & \qquad \qquad + \mathcal{O}(\Delta \bm{\theta}^3) \nonumber
\end{align}
with quantities 
\begin{align}\label{eq:FM}
    \bm{\Gamma}_{ab} &= (\partial_a h | \partial_b h)\,, \\
    \Delta \hat{\bm{\theta}}_{\rm bf}^a &= \left( \bm{\Gamma}^{-1} \right)_{ab} (\partial_b h | n) \,. \label{eq:noise_fluctuation_n}
\end{align}
Notice that the first term in Eq.~\eqref{eq:LSA_like} is of the same form of a multivariate-Gaussian distribution with parameter covariance matrix given by $\bm{\Gamma}^{-1}$ centered on ``best-fit" parameters $\Delta \hat{\bm{\theta}}^{a}_{\text{bf}}$. 
Direct calculation\footnote{Under specific regularity conditions (see page 64 in~\cite{Burke:2021xrg}), it can be shown that $\mathbb{E}_{\bm{n}}\left[-\partial_{a}\partial_{b} \log p(\boldsymbol{\theta}|d)\right] = \mathbb{E}_{\bm{n}}\left[\partial_{a} \log p(\boldsymbol{\theta}|d)\partial_{b} \log p(\boldsymbol{\theta}|d)\right]$ with latter expression the formal definition of the FM. A short calculation shows that $\bm{\Gamma}_{ab} = \mathbb{E}_{\bm{n}}\left[\partial_{a} \log p(\boldsymbol{\theta}|d)\partial_{b} \log p(\boldsymbol{\theta}|d)\right] = (\partial_{a} h | \partial_{b} h)$.} shows that 
\begin{align*}
\mathbb{E}_{\bm{n}}\left[-\partial_{a}\partial_{b} \log p(\boldsymbol{\theta}|d)\right] &= \big\langle \left(\partial_{a}\partial_{b}h|n\right)\big\rangle + \big\langle\left(\partial_{a}h|\partial_{b}h\right)\big\rangle \\
&= (\partial_{a}h|\partial_{b}h) \\
& = \bm{\Gamma}_{ab}
\end{align*}
allowing us to identify the matrix $\bm{\Gamma}$ in Eq.~\eqref{eq:FM} as the expectation of the observed information matrix, usually called the FM.
The FM $\bm{\Gamma} \sim \rho^{2}$ and $\Delta\boldsymbol{\theta}^{a}_{\text{bf}} \sim \rho^{-1}$, meaning that the first term in \eqref{eq:LSA_like} is an $\mathcal{O}(\rho)$ quantity whereas the last two terms are $\mathcal{O}(1)$ quantities. This reasoning implies that in the limit of high signal-to-noise ratio, those latter terms can be neglected and the likelihood can be approximated as a multi-variate Gaussian distribution with parameter covariance matrix given by $\bm{\Gamma}^{-1}$ and maximum likelihood estimate given by $\Delta\hat{\bm{\theta}}^{a}_{\text{bf}}$~~\cite{Flanagan:1997kp, Miller:2005qu, Cutler:2007mi}. These best-fit parameters are unbiased $\langle \Delta \hat{\bm{\theta}}_{\rm bf} \rangle = 0$, and with variance given by the Fisher covariance itself,
\begin{align}
    \left\langle \Delta \hat{\bm{\theta}}_{\rm bf}^a \Delta \hat{\bm{\theta}}_{\rm bf}^b \right\rangle &= \left( \bm{\Gamma}^{-1} \right)_{ac} \left( \bm{\Gamma}^{-1} \right)_{bd} \left\langle ( \partial_c h |n ) ( n| \partial_d h) \right\rangle \nonumber\\
    &= \left( \bm{\Gamma}^{-1} \right)_{ac} \left( \bm{\Gamma}^{-1} \right)_{bd} \bm{\Gamma}_{cd} \nonumber\\
    &= \left( \bm{\Gamma}^{-1} \right)_{ab} \,,
\end{align}
where we used $\langle (a|n)(n|b)\rangle = (a|b)$, easily derived using Eq.~\eqref{DA_eq:uncorrelated_frequencies}. We will refer to these ``best-fit" quantities $\Delta \hat{\bm{\theta}}^{a}_{\text{bf}}$ as ``noise biases"
\footnote{The quantities $\Delta\hat{\bm{\theta}}^{a}_{\text{bf}}$ are not biases in the statistical sense, since $\langle \Delta\hat{\bm{\theta}}^{a}_{\text{bf}}\rangle = 0$. However, as LISA will observe a single noise realization that in principle will ``shift" recovered parameters away from the true parameters, these noise fluctuations represent the ``bias" in individual observations.}, in contrast to the \emph{Cutler-Vallisneri} biases that arise in the context of waveform systematics~\cite{Cutler:2007mi}.

Coming back to the gap case, in linear algebra notations, we start from the likelihood in Eq.~\eqref{eq:windowed_likelihood_td} and apply the same linearization of the signal around the true parameters. We obtain again a quadratic form, with the new FM
\begin{equation}
    \bm{\Gamma}^{\rm gap}_{ab} = \left(\bm{W} \partial_a \bm{h}\right)^T (\bm{W} \bm{\Sigma} \bm{W})^+ \left(\bm{W} \partial_b \bm{h}\right) \,,
\end{equation}
and best-fit parameters
\begin{equation}\label{eq:cv_correct_gap}
    \Delta \hat{\bm{\theta}}_{\rm bf,gap}^a = \left(\bm{\Gamma}_{\rm gap}^{-1} \right)_{ab} \left(\bm{W} \partial_b \bm{h}\right)^T (\bm{W} \bm{\Sigma} \bm{W})^+ \left(\bm{W} \bm{n}\right) \,,
\end{equation}
which provide an unbiased estimator for the parameters with variance 
\begin{widetext}
\begin{align}\label{eq:correct_gap_bias_variance_derivation}
    \left\langle \Delta \hat{\bm{\theta}}_{\rm bf,gap}^a \Delta \hat{\bm{\theta}}_{\rm bf,gap}^b \right\rangle &= \left( \bm{\Gamma}^{-1}_{\rm gap} \right)_{ac} \left( \bm{\Gamma}^{-1}_{\rm gap} \right)_{bd} \left\langle \left(\bm{W} \partial_c \bm{h}\right)^T (\bm{W} \bm{\Sigma} \bm{W})^+ \left(\bm{W} \bm{n}\right) \left(\bm{W} \bm{n}\right)^T (\bm{W} \bm{\Sigma} \bm{W})^+ \left(\bm{W} \partial_d \bm{h}\right) \right\rangle \nonumber\\
    &= \left( \bm{\Gamma}^{-1}_{\rm gap} \right)_{ac} \left( \bm{\Gamma}^{-1}_{\rm gap} \right)_{bd} \left(\bm{W} \partial_c \bm{h}\right)^T (\bm{W} \bm{\Sigma} \bm{W})^+ \bm{W} \bm{\Sigma} \bm{W} (\bm{W} \bm{\Sigma} \bm{W})^+ \left(\bm{W} \partial_d \bm{h}\right) \nonumber\\
    &= \left( \bm{\Gamma}^{-1}_{\rm gap} \right)_{ac} \left( \bm{\Gamma}^{-1}_{\rm gap} \right)_{bd} \left(\bm{W} \partial_c \bm{h}\right)^T (\bm{W} \bm{\Sigma} \bm{W})^+ \left(\bm{W} \partial_d \bm{h}\right) \nonumber\\
    &= \left( \bm{\Gamma}^{-1}_{\rm gap} \right)_{ac} \left( \bm{\Gamma}^{-1}_{\rm gap} \right)_{bd} \bm{\Gamma}^{\rm gap}_{cd} \nonumber\\
    &= \left( \bm{\Gamma}^{-1}_{\rm gap} \right)_{ab} \,,
\end{align}
\end{widetext}
where we have used $\langle \left(\bm{W} \bm{n}\right) \left(\bm{W} \bm{n}\right)^T \rangle = \bm{W} \bm{\Sigma} \bm{W}$ and one of the defining properties of the pseudo-inverse, $\bm{A}^+ \bm{A} \bm{A}^+ = \bm{A}^+$.
The equivalent expressions in the FD are obtained with the same rewriting that we used for the likelihood, using the unitarity of $\bm{P}$ to absorb it under the pseudo-inverse. This results in
\begin{equation}
    \tilde{\bm{\Gamma}}^{\rm gap}_{ab} = \left(\widetilde{\bm{W} \partial_a \bm{h}}\right)^\dagger (\tilde{\bm{W}} \tilde{\bm{\Sigma}} \tilde{\bm{W}})^+ \left(\widetilde{\bm{W} \partial_b \bm{h}}\right) \,,
\end{equation}
and
\begin{equation}
    \Delta \hat{\bm{\theta}}_{\rm bf,gap}^a = \left(\tilde{\bm{\Gamma}}_{\rm gap}^{-1} \right)_{ab} \left(\widetilde{\bm{W} \partial_b \bm{h}} \right)^\dagger (\tilde{\bm{W}} \tilde{\bm{\Sigma}} \tilde{\bm{W}})^+ \left(\widetilde{\bm{W} \bm{n}} \right) \,.
\end{equation}

\section{Quantifying noise mismodeling errors}\label{sec:mismodeling}

Having explored how to write down the correct likelihood, SNR and FM in the presence of data gaps (and possibly of a tapering window), we will now present tools to assess the impact that mismodeling the statistics of the noise could have on PE.

\subsection{Effects of mismodeling the noise process}
\label{subsec:mismodeling_gaps}

In all mismodeling cases that we will consider, the approximate model for the likelihood $\mathcal{L}'$ will remain Gaussian-like, in the sense that it will possess the structure
\be\label{eq:model_likelihood_generic}
    \ln\mathcal{L}^{\prime} (\bm{d}|\bm{\theta}) = -\frac{1}{2} \left( \bm{W} \bm{x} \right)^T \left(\bm{\Sigma}'\right)^{+} \left( \bm{W} \bm{x} \right) \,.
\ee
We have written the model's inverse covariance $\left(\bm{\Sigma}'\right)^{+}$ generically as a pseudo-inverse and in the TD, but in applications we might be using an invertible covariance (e.g. when using the Whittle-likelihood), and we might be working in the FD. To connect to the correct modelling case, in this notation, this inverse covariance absorbs the effect of windows: $\left(\bm{\Sigma}'\right)^{+}_{\rm correct} = \left(\bm{W} \bm{\Sigma} \bm{W}\right)^{+}$.

Note that we do not consider the case where we apply the window to the data but not the template, as we were able to do in Eq.~\eqref{eq:windowed_likelihood_fd_alt} when using the correct covariance with the gating window. This would complicate notations and also presumably represent an additional source of errors whenever using an approximate inverse covariance; in all applications, we will apply the same window (gating or tapering) to the data stream and to the templates, $\bm{W}\bm{x} = \bm{W} \bm{h}(\bm{\theta}) - \bm{W}\bm{d}$.

The \emph{wrong} likelihood defined in Eq.~\eqref{eq:model_likelihood_generic} does not encode the correct statistics for the windowed noise process. Since it retains the usual quadratic structure, we can repeat the steps leading to the derivation of the FM and noise bias. Using the linearized signal approximation, we obtain the FM
\begin{equation}
    \bm{\Gamma}^{\prime}_{ab} = \left( \bm{W} \partial_a \bm{h} \right)^T \left(\bm{\Sigma}'\right)^{+} \left( \bm{W} \partial_b \bm{h} \right) \,,
\end{equation}
that will differ in general from the correct FM $\bm{\Gamma}_{ab}$. We also obtain the following noise bias, for a given noise realization $\bm{n}$,
\begin{equation}\label{eq:CV_bias_model}
    \left(\Delta \hat{\bm{\theta}}_{\rm bf}^a \right)' = \left(\bm{\Gamma}^{\prime}\right)^{-1}_{ab} \left( \bm{W} \partial_b \bm{h} \right)^T \left(\bm{\Sigma}'\right)^{+} \left( \bm{W} \bm{n} \right) \,.
\end{equation}
These best-fit parameters should not be interpreted as giving only the best-fit in $L^2$ sense of the template $\bm{h}(\bm{\theta})$ to the data $\bm{d}$: the model's pseudo-inverse covariance $\left(\bm{\Sigma}'\right)^{+}$ intervenes in the computation. Inaccuracies in this covariance will impact the best-fit solution.

Over many realisations of the noise, the best-fit parameters are still unbiased by construction, with $\Big\langle \left(\Delta \hat{\bm{\theta}}_{\rm bf}^a \right)' \Big\rangle = 0 $, but they will have a variance that does not match the expected parameter covariance represented by the FM and encoded in the posterior distributions,
\begin{widetext}
\begin{align}\label{eq:derivation_bias_variance}
    \left\langle \left( \Delta \hat{\bm{\theta}}_{\rm bf}^a \right)' \left( \Delta \hat{\bm{\theta}}_{\rm bf}^b \right)' \right\rangle &= \left(\bm{\Gamma}^{\prime}\right)^{-1}_{ac} \left(\bm{\Gamma}^{\prime}\right)^{-1}_{bd} \left\langle \left(\bm{W} \partial_c \bm{h}\right)^T \left(\bm{\Sigma}'\right)^{+} \left(\bm{W} \bm{n}\right) \left(\bm{W} \bm{n}\right)^T \left(\bm{\Sigma}'\right)^{+} \left(\bm{W} \partial_d \bm{h}\right) \right\rangle \nonumber\\
    &= \left(\bm{\Gamma}^{\prime}\right)^{-1}_{ac} \left(\bm{\Gamma}^{\prime}\right)^{-1}_{bd} \left(\bm{W} \partial_c \bm{h}\right)^T \left(\bm{\Sigma}'\right)^{+} \bm{W} \bm{\Sigma} \bm{W} \left(\bm{\Sigma}'\right)^{+} \left(\bm{W} \partial_d \bm{h}\right) \,,
\end{align}

where the inner matrix products do not simplify further, in general $\left(\bm{\Sigma}'\right)^{+} \bm{W} \bm{\Sigma} \bm{W} \left(\bm{\Sigma}'\right)^{+} \neq (\bm{W} \bm{\Sigma} \bm{W})^{+}$, so that this \emph{parameter bias covariance} will not match the model's Fisher covariance (nor the correct Fisher covariance):
\be
    \left\langle \left( \Delta \hat{\bm{\theta}}_{\rm bf}^a \right)' \left( \Delta \hat{\bm{\theta}}_{\rm bf}^b \right)' \right\rangle \neq \left(\bm{\Gamma}'\right)_{ab}^{-1} \,. \label{eq:param_fluctuations_not_equal_FM}
\ee
Only in the case where $\bm{\Sigma}' \equiv \bm{W}\bm{\Sigma}\bm{W}$, i.e., where the model noise covariance matrix matches the noise covariance matrix of the underlying noise process, would we achieve a simplification and thus equality in Eq.~\eqref{eq:param_fluctuations_not_equal_FM}. In the FD, Eq.~\eqref{eq:derivation_bias_variance} reads
\begin{equation}\label{eq:freq_domain_derivation_bias_variance}
\left\langle \left( \Delta \hat{\bm{\theta}}_{\rm bf}^a \right)' \left( \Delta \hat{\bm{\theta}}_{\rm bf}^b \right)' \right\rangle 
    = \left(\tilde{\bm{\Gamma}}^{\prime}\right)^{-1}_{ac} \left(\tilde{\bm{\Gamma}}^{\prime}\right)^{-1}_{bd} \left(\widetilde{\bm{W} \partial_c \bm{h}}\right)^\dagger \left(\tilde{\bm{\Sigma}}'\right)^{+} \tilde{\bm{W}} \tilde{\bm{\Sigma}} \tilde{\bm{W}} \left(\tilde{\bm{\Sigma}}'\right)^{+} \left(\widetilde{\bm{W} \partial_d \bm{h}}\right) \,,
\end{equation}
with $\tilde{\bm{\Gamma}}$ given by the Whittle-based FM computed in the FD. We remark that Eq.~\eqref{eq:freq_domain_derivation_bias_variance} was derived in~\cite{Edy:2021par} that explored mismodeling non-stationary processes in ground-based detectors as stationary in the context of modulations and bursts, but not gaps.
\end{widetext}

\subsection{Coherent Whittle-likelihood}\label{subsec:gap_whittle}

In practice, as discussed in Sec.~\ref{subsec:windowing_procedure}, for data with gaps one could make the choice to treat the data segment as a whole, using the Whittle-likelihood for that full segment, while simply filling the missing data in the gap with zeroes. We can either use a gating window $w_g$, or apply a tapering window $w_t$ smoothing each side of the gap. This approach has been used in a number of works~\cite{Porter:2009wi,Talbot:2021igi, Edy:2021par, Dey:2021dem, Castelli:2024sdb, Zackay:2019kkv, Mao:2024jad, Wang:2024ovi}. Thus the approximate likelihood, denoted with a prime, would read
\begin{equation}\label{eq:approx_likelihood_coherent}
    \ln\mathcal{L}^{\prime} (\bm{d}|\bm{\theta}) = -\frac{1}{2} \left( \bm{W} \bm{x} \right)^T \bm{\Sigma}^{-1} \left( \bm{W} \bm{x} \right) \,,
\end{equation}
with $\bm{x} = \bm{h}(\bm{\theta}) - \bm{d}$. In practice, this allows for an efficient  computation in the FD,
\begin{equation}
    \ln\mathcal{L}^{\prime}(\bm{d}|\bm{\theta}) = -\frac{1}{2} \left( \widetilde{\bm{W} \bm{x}} \right)^\dagger \tilde{\bm{\Sigma}}^{-1} \left( \widetilde{\bm{W} \bm{x}} \right) \,,
\end{equation}
where $\tilde{\bm{\Sigma}}$ is diagonal, as for the Whittle-likelihood, with components given by \eqref{eq:FDcovarianceWhittle}, and where the window would be applied in the TD to the residual $\bm{x}$ before taking an FFT. We could instead use the Toeplitz form of Eq.~\eqref{eq:approx_likelihood_coherent} and transform the likelihood into the FD. A major disadvantage would be that the favored diagonal structure of the FD likelihood would be lost (a result of assuming Toeplitz, but not circulant, covariances) and this is rarely, if ever, done in practice. For this reason, we will use the Circulant condition to facilitate the use of the rapid-to-evaluate Whittle-likelihood. Later in the results Sec.~\ref{sec:results} we will test mismodeling errors when assuming the underlying process is circulant when the true noise-process is Toeplitz.

We can apply the results of Sec.~\ref{subsec:mismodeling_gaps} with the replacement $(\bm{\Sigma}')^+ = \bm{\Sigma}^{-1}$, and distinguish two cases:
\begin{itemize}
    \item $\bm{W} = \bm{W}_g$, a gating window, meaning we directly fill the gaps with zeros and leave the rest of the data untouched;
    \item $\bm{W} = \bm{W}_t$, a smooth tapering window: we taper each gap edge using the Planck window~\eqref{eq:def_planck_window} with a lobe length that we will allow to vary from 1 minute to 2 hours.
\end{itemize}
The naive gating approach will give us a worst-case estimate of the impact of the gap. The tapering approach will reduce the impact of the mismodeling, but will also erase some of the information conveyed by the data inside the tapering lobes, depending on the lobe length\footnote{The loss of information during the tapering scheme is a feature of mismodeling the data using the Whittle-like circulant covariance matrix. If the model covariance were consistent with the tapered noise, no information would be lost. See Eq.~\eqref{eq:stat_consistency_gate_taper} and surrounding text for a reminder.}. 

The results~of Sec.~\ref{subsec:mismodeling_gaps} give in this context
\begin{widetext}
\begin{subequations}\label{eq:fisher_cv_bias_whittle_list_of_equations}
\begin{align}
    \bm{\Gamma}^{\prime}_{ab} &= \left( \bm{W} \partial_a \bm{h} \right)^T \bm{\Sigma}^{-1} \left( \bm{W} \partial_b \bm{h} \right) \,,\label{eq:fisher_whittle} \\
    \left(\Delta \hat{\bm{\theta}}_{\rm bf}^a \right)' &= \left(\bm{\Gamma}^{\prime}\right)^{-1}_{ab} \left( \bm{W} \partial_b \bm{h} \right)^T \bm{\Sigma}^{-1} \left( \bm{W} \bm{n} \right) \,, \label{eq:fisher_cv_bias_whittle} \\
    \left\langle \left( \Delta \hat{\bm{\theta}}_{\rm bf}^a \right)' \left( \Delta \hat{\bm{\theta}}_{\rm bf}^b \right)' \right\rangle
    &= \left(\bm{\Gamma}^{\prime}\right)^{-1}_{ac} \left(\bm{\Gamma}^{\prime}\right)^{-1}_{bd} \left(\bm{W} \partial_c \bm{h}\right)^T \bm{\Sigma}^{-1} \bm{W} \bm{\Sigma} \bm{W} \bm{\Sigma}^{-1} \left(\bm{W} \partial_d \bm{h}\right) \,. \label{eq:fisher_cv_bias_scatter_whittle}
\end{align}
\end{subequations}
\end{widetext}
Eqs.~\eqref{eq:fisher_cv_bias_scatter_whittle} and \eqref{eq:fisher_whittle} will be used extensively in the results Sec.~\ref{subsec:gap_mismodeling_ratios}, namely to test the appropriateness of the Whittle-likelihood on gated and tapered data for PE purposes.

\subsection{Segmented Whittle-likelihood}\label{subsec:gap_segmented_whittle}

The previous Whittle model for the likelihood in Sec.~\ref{subsec:gap_whittle} treats the data segment as a whole (and thus coherent). Another approach would be to segment the data, treating each data segment between gaps as independent from the others, and using the Whittle-likelihood for stationary (and circulant) noise on each of these data segments. We will also use the same noise model before and after the gap segments. Assuming incoherence between data segments gives an advantage of working with data chunks of limited size where the data gaps simply amount to ignoring certain segments. The caveats are that:
\begin{itemize}
    \item One neglects all noise correlations between separate segments. As we will see in details in Sec.~\ref{subsec:mismodeling_independence}, this loss of information is in fact very minor.
    \item One uses the Whittle-likelihood on short segments, which comes with errors in itself at the segment edges. As in the previous section, this can be alleviated by tapering, but tapering itself erases some information contained in the data.
    \item One will observe a loss of resolvable frequencies, impacting sensitivity for low-frequency gravitational-wave sources. Assuming LISA is sensitive to sources for $f\in(10^{-4},10^{-1})$, this becomes problematic for data segments of length $T \lesssim 3\,\text{hours}$.
\end{itemize}    
Note that this approach is perhaps the closest to short-time Fourier transform or time-frequency methods. 

In our gap-in-the-middle notations, this approach is equivalent to taking
\begin{align}\label{eq:block_circulant_assume_indep}
	\left(\bm{\Sigma}'\right)^+ &= \begin{pmatrix}
					(\bm{\Sigma}_{00}^{\rm circ})^{-1} & \bm{0} & \bm{0}\\
					\bm{0} & \bm{0} & \bm{0} \\
                    \bm{0} & \bm{0} & (\bm{\Sigma}_{22}^{\rm circ})^{-1}  
	               \end{pmatrix}\,,
\end{align}
where $\bm{\Sigma}_{11}^{\rm circ}$, $\bm{\Sigma}_{33}^{\rm circ}$ are both circulant matrices built from the same PSD (assumed to be known), for the segment lengths $N_1$ and $N_3$. Note that these block matrices are not sub-matrices of $\bm{\Sigma}$, i.e. $\bm{\Sigma}_{11}^{\rm circ} \neq \bm{\Sigma}_{11}$, even if $\bm{\Sigma}$ is itself Toeplitz or circulant. To reduce edge effects, windows $\bm{W}_1$, $\bm{W}_3$ are used to taper both ends of each segment.

In a more general notation for $K$ segments indexed by $I=1,\dots,K$, the model likelihood is then a sum over the segments that are not part of the set of gaps,
\begin{subequations}
\begin{align}
    \ln\mathcal{L}^{\prime} (\bm{d}|\bm{\theta}) &= -\frac{1}{2} \sum_{I \notin {\rm gaps}} \left( \bm{W}_I \bm{x}_I \right)^T (\bm{\Sigma}_{II}^{\rm circ})^{-1} \left( \bm{W}_I \bm{x}_I \right) \,, \\
    &= -\frac{1}{2} \sum_{I \notin {\rm gaps}} \left( \widetilde{\bm{W}_I \bm{x}_I} \right)^\dagger (\tilde{\bm{\Sigma}}_{II}^{\rm circ})^{-1} \left( \widetilde{\bm{W}_I \bm{x}_I} \right) \,,
\end{align}
\end{subequations}
with $\bm{x}_I = \bm{h}_I(\bm{\theta}) - \bm{d}_I$ the residual on the segment $I$. In the FD, each $\tilde{\bm{\Sigma}}_{II}^{\rm circ}$ is diagonal and the FFT can be used; the length of each segment $N_I$ is decided by the gap locations and will not be a power of 2 in general. This will have an impact on speed of the discrete Fourier transform, making it at worst an $\mathcal{O}(N^{2})$ operation rather than a $\mathcal{O}(N\log_{2}N)$ operation.

The results~of Sec.~\ref{subsec:mismodeling_gaps} for the model's FM and noise bias also become sums over $I$, in the same fashion: 
\begin{subequations}\label{eq:fisher_cv_bias_segmented}
\begin{align}
    \bm{\Gamma}^{\prime}_{ab} &= \sum_{I \notin {\rm gaps}} \left( \bm{W}_I \partial_a \bm{h}_I \right)^T \left(\bm{\Sigma}_{II}^{\rm circ}\right)^{-1} \left( \bm{W}_I \partial_b \bm{h}_I \right) \,, \\
    \left(\Delta \hat{\bm{\theta}}_{\rm bf}^a \right)' &= \left(\bm{\Gamma}^{\prime}\right)^{-1}_{ab} \sum_{I \notin {\rm gaps}} \left( \bm{W}_I \partial_b \bm{h}_I \right)^T \left(\bm{\Sigma}_{II}^{\rm circ}\right)^{-1} \left( \bm{W}_I \bm{n}_I \right) \,,
\end{align}
\end{subequations}

The bias covariance, however, is a double sum over all segments that are observed: 
\begin{widetext}
\begin{align}
    \left\langle \left( \Delta \hat{\bm{\theta}}_{\rm bf}^a \right)' \left( \Delta \hat{\bm{\theta}}_{\rm bf}^b \right)' \right\rangle &= \left(\bm{\Gamma}^{\prime}\right)^{-1}_{ac} \left(\bm{\Gamma}^{\prime}\right)^{-1}_{bd} \sum_{I \notin {\rm gaps}} \sum_{J \notin {\rm gaps}} \left\langle \left(\bm{W}_I \partial_c \bm{h}_I\right)^T \left(\bm{\Sigma}_{II}^{\rm circ}\right)^{-1} \left(\bm{W}_I \bm{n}_I\right) \left(\bm{W}_J \bm{n}_J\right)^T \left(\bm{\Sigma}_{JJ}^{\rm circ}\right)^{-1} \left(\bm{W}_J \partial_d \bm{h}_J\right) \right\rangle \nonumber\\
    &= \left(\bm{\Gamma}^{\prime}\right)^{-1}_{ac} \left(\bm{\Gamma}^{\prime}\right)^{-1}_{bd} \sum_{I \notin {\rm gaps}} \sum_{J \notin {\rm gaps}} \left(\bm{W}_I \partial_c \bm{h}_I\right)^T \left(\bm{\Sigma}_{II}^{\rm circ}\right)^{-1} \bm{W}_I \bm{\Sigma}_{IJ} \bm{W}_J \left(\bm{\Sigma}_{JJ}^{\rm circ}\right)^{-1} \left(\bm{W}_J \partial_d \bm{h}_J\right) \,.
\end{align}
\end{widetext}
Indeed, even if our model assumes segment independence, cross-terms between segments appear because the true underlying noise process $\bm{n}$ will have in general correlations between different segments, represented by $\bm{\Sigma}_{IJ}$ with $I\neq J$.

\subsection{Mismodeling measures}
\label{subsec:mismodeling_measures}

Recall that our framework is different from mismodeling induced by systematic errors in the waveform models~\cite{Vallisneri:2007ev}. In the latter case, the part of the shift between best-fit and true parameters due to waveform errors is deterministic, although it will vary from source to source. Here, our biases in Eq.~\eqref{eq:CV_bias_model} are zero-mean, but have the wrong statistics~\eqref{eq:derivation_bias_variance}.

In our case, we have two types of mismodeling errors introduced by our use of the wrong statistical model for the likelihood:
\begin{itemize}
    \item A mismatch between the width of the posteriors obtained with the wrong model likelihood and correct model likelihood.
    \item A mismatch between the scatter of best-fit parameters when varying the noise realization and the width of the modelled posteriors.
\end{itemize}

The first type of error motivates the introduction of \emph{noise mismodeling posterior width ratio} measuring the ratio of 1$\sigma$ errors between the incorrect and correct analyses, as approximated in the Fisher approach:
\begin{tcolorbox}[colframe=black, colback=white, arc=4mm, boxsep=0mm, left=2mm, right=1mm, top=0mm, bottom=2mm]
\begin{equation}\label{eq:def_Xi}
    \Xi_a = \frac{\sqrt{\left(\bm{\Gamma}'\right)_{aa}^{-1}}}{\sqrt{\left(\bm{\Gamma}\right)_{aa}^{-1}}} \,,
\end{equation}
\end{tcolorbox}
where $\left(\bm{\Gamma}\right)_{aa}^{-1}$ is the correct Fisher covariance. The measured posterior width ratio is qualitatively similar (although not identical, as correlations could change) to a loss of SNR. For instance, we might want to alleviate the effect of gaps by introducing an aggressive tapering window around them, but this would come at the expense of a loss of information reflected by values $\Xi > 1$.

The second type of errors lead us to introduce the \emph{noise-mismodelling posterior scatter-to-width ratio}
\begin{tcolorbox}[colframe=black, colback=white, arc=4mm, boxsep=0mm, left=2mm, right=1mm, top=0mm, bottom=2mm]
\begin{equation}\label{eq:def_Upsilon}
    \Upsilon_a = \frac{\sqrt{\left\langle \left( \Delta \hat{\bm{\theta}}_{\rm bf}^a \right)' \left( \Delta \hat{\bm{\theta}}_{\rm bf}^a \right)' \right\rangle}}{\sqrt{\left(\bm{\Gamma}'\right)_{aa}^{-1}}}\,.
\end{equation}
\end{tcolorbox}
We will sometimes refer to Eq.~\eqref{eq:def_Upsilon} in short by the \emph{scatter-to-width ratio}. When forming the product $\Xi_a \Upsilon_a$, the approximate FM $\bm{\Gamma}'$ cancels out and we end up with a ratio. The numerator is the variance of an unbiased estimator of the posterior co-variance, while the denominator is the FM for the true noise process, $\bm{\Gamma}$; this means that the Cramer-Rao inequality~\cite{rao1992information, cramer1999mathematical,Finn:1992wt} applies and ensures that
\be\label{eq:xi_upsilon_bound}
    \Xi_a \Upsilon_a \geq 1 \,.
\ee

The scatter-to-width ratio $\Upsilon_a$ will be the most important quantity we will compute. It is a measure of the statistical consistency of the use of an incorrect statistical model for the noise, and provides a quick-to-compute proxy for a probability-probability (PP) plot. The numerator measures the scatter of best-fit parameters, while the denominator measures the width of the posterior, both obtained with the same approximate likelihood. A value $\Upsilon_a > 1$ indicates that the scatter in the posterior mode is larger than the posterior width; therefore, for most noise realizations, the true parameters will be excluded by the posteriors, leading to statistically significant random biases. For a concrete example of a case where $\Upsilon_a > 1$, see Fig.~\ref{fig:corner_correct_vs_whittle} and Fig.~\ref{fig:biascorner_incorrect_gap_pe_vs_cv} below in the results section Sec.~\ref{sec:results}. A value $\Upsilon_a \simeq 1$ is the target, it would indicate that the bias scatter and posterior width are comparable, see Fig.~\ref{fig:corner_correct_vs_whittle_taper10min} for an example. When using the correct likelihood, we have exactly $\Upsilon_a = 1$, as shown by Eq.~\eqref{eq:correct_gap_bias_variance_derivation}.

We will use these $\Upsilon_a$ for diagonal elements of the covariance that are scaled by the approximate Fisher covariance so they can be understood to represent ``number of sigmas'' with respect to the approximate posterior. We could also compare off-diagonal components (without the square roots), representing correlations between parameters. In practice we found that the values of $\Upsilon_a$ and $\Xi_a$ for different physical parameters $\theta^a$ showed the same overall trends, and in tables we will show simply their average over parameters
\be\label{eq:def_upsilonbar_xibar}
    \overline{\Upsilon} \equiv \frac{1}{d} \sum_{a=1}^{d} \Upsilon_a\,, \quad \overline{\Xi} \equiv \frac{1}{d} \sum_{a=1}^{d} \Xi_a \,.
\ee

An important remark in contrasting the roles of the posterior width ratio $\Xi_a$ and scatter-to-width ratio $\Upsilon_a$ is the following: what happens if one uses a wrong noise model that under- or over-estimates the overall noise level, while having the correct correlations? This would translate as $\bm{\Sigma}' = \lambda^2 \bm{\Sigma}$ (or for the PSD, $S_n^{\prime} = \lambda^2 S_n$) for some factor $\lambda > 0$ (quadratic for the covariance, linear for the noise). Fisher covariances would scale as $(\bm{\Gamma}')^{-1} = \lambda^2 \bm{\Gamma}^{-1}$, marking a widening in the approximate posterior. However, noise biases~\eqref{eq:CV_bias_model} are insensitive to $\lambda$; these max-likelihood biases are sensitive to the noise spectral shape, not to the noise level (similarly, in the context of waveform modeling errors, Cutler-Vallisneri biases are SNR-independent). We would therefore have $\Xi_a = \lambda$, $\Upsilon_a = 1/\lambda$, and $\Xi_a \Upsilon_a = 1$, saturating the bound in Eq.~\eqref{eq:xi_upsilon_bound}. These scalings are in line with intuition: for $\lambda>1$, we overestimate the noise, the posteriors would be too wide, encompassing the true parameters too often; for $\lambda<1$, we underestimate the noise, the posteriors are too narrow and they often exclude the true parameters. As useful as this example is, we stress that in the rest of the paper, we focus on the mismodeling of noise correlations, not of the overall noise level. We will see examples where $\Xi_a \simeq 1$ but $\Upsilon_a > 1$ and conversely. We present examples in Tables \ref{tab:mismodeling_upsilon_gap} and \ref{tab:mismodeling_xi_gap}, computed via the Whittle-likelihood to analyse gated data for a noise process with a large red-noise (low frequency) component.

\subsection{Computing the parameter bias covariance}\label{subsec:parameter_bias_cov}

In general, we will be in a situation where computations with the correct likelihood are expensive, while using the incorrect model likelihood is much cheaper. While the ratio of Fisher covariances $\Xi_a$ can perhaps be approximated by the ratio of SNRs, the scatter-to-width ratio $\Upsilon_a$ is the most important and informative quantity and requires computing the parameter bias covariance defined by Eq.~\eqref{eq:derivation_bias_variance}. Different approaches can be used to compute it in practice:
\begin{itemize}
    \item A \emph{direct computation} of Eq.~\eqref{eq:derivation_bias_variance}, which requires in general to compute the $N\times N$ matrix $\left(\bm{\Sigma}'\right)^{+} \bm{W} \bm{\Sigma} \bm{W} \left(\bm{\Sigma}'\right)^{+}$, which can be challenging for large $N$.
    \item A \emph{Monte-Carlo computation} of the variance, by simulating many realizations of the noise $\bm{n}$ and using the noise biases in Eq.~\eqref{eq:CV_bias_model} for estimating the biases, using only the model likelihood.
    \item Full Bayesian PE using a set of noise realizations, producing directly max-likelihood parameters as best-fit parameters (without the linearized signal approximation).
\end{itemize}
In the second approach, we bypass the limitations imposed by working with $N\times N$ matrices where all that is required is the ability simulate noise realizations $\bm{n}$ according to the correct covariance $\Sigma$. The computation of each noise bias itself then only involves the \emph{model} likelihood, not the correct one. If we can also compute the associated Fisher covariance, this means that we can compute $\Upsilon_a$, with only the cheaper model likelihood; by contrast, $\Xi_a$ requires the Fisher covariance for the correct likelihood.

Full Bayesian PE sampling is expensive in itself, potentially also for the model likelihood, and can be used as a check of the linear signal approximation estimate for biases. In our analysis, we will verify through multiple PE simulations that the three approaches are consistent with each other. This is demonstrated in Fig.~\ref{fig:biascorner_incorrect_gap_pe_vs_cv}. If doing a series of Bayesian PE runs, we would be close to producing a PP-plot -- with the difference that a PP-plot requires in principle to randomize also the input parameters of the signal, according to the priors used. 

It is important to emphasize that the computation of the parameter bias covariance~\eqref{eq:derivation_bias_variance} does not rely on any assumption about the underlying noise process: the noise $\bm{n}$ can also be \emph{non-Gaussian}~\cite{Antonelli:2021vwg}.
The assumption that enters the derivation is that the model likelihood itself is Gaussian. In the direct computation approach, we only need access to the variance of the noise process (before applying windows) $\langle \bm{n}\bm{n}^T \rangle$. In the Monte-Carlo approach, we only need to be able to simulate multiple draws of $\bm{n}$. We do remark however that, when $\bm{n}$ is non-Gaussian, the distribution of the estimator itself may no longer be Gaussian so it's statistical properties may not be entirely represented by the covariance matrix.

Finally, although we focus here on non-stationarity in the data stream in the form of data gaps, we stress that the tools described above could apply to other forms of mismodeling: mismodeling of the PSD itself, non-stationarities in the form of a modulated galactic binary foreground or evolving instrumental PSDs, and non-Gaussian features such as non-Gaussianity in the foreground or instrumental glitches.

\section{Waveform and noise models}\label{sec:application_MBHB}

\subsection{MBHB signals and analysis setup}\label{subsec:mbhb_signals_analysis}

\begin{figure*}
    \centering
    \includegraphics[width = \textwidth]{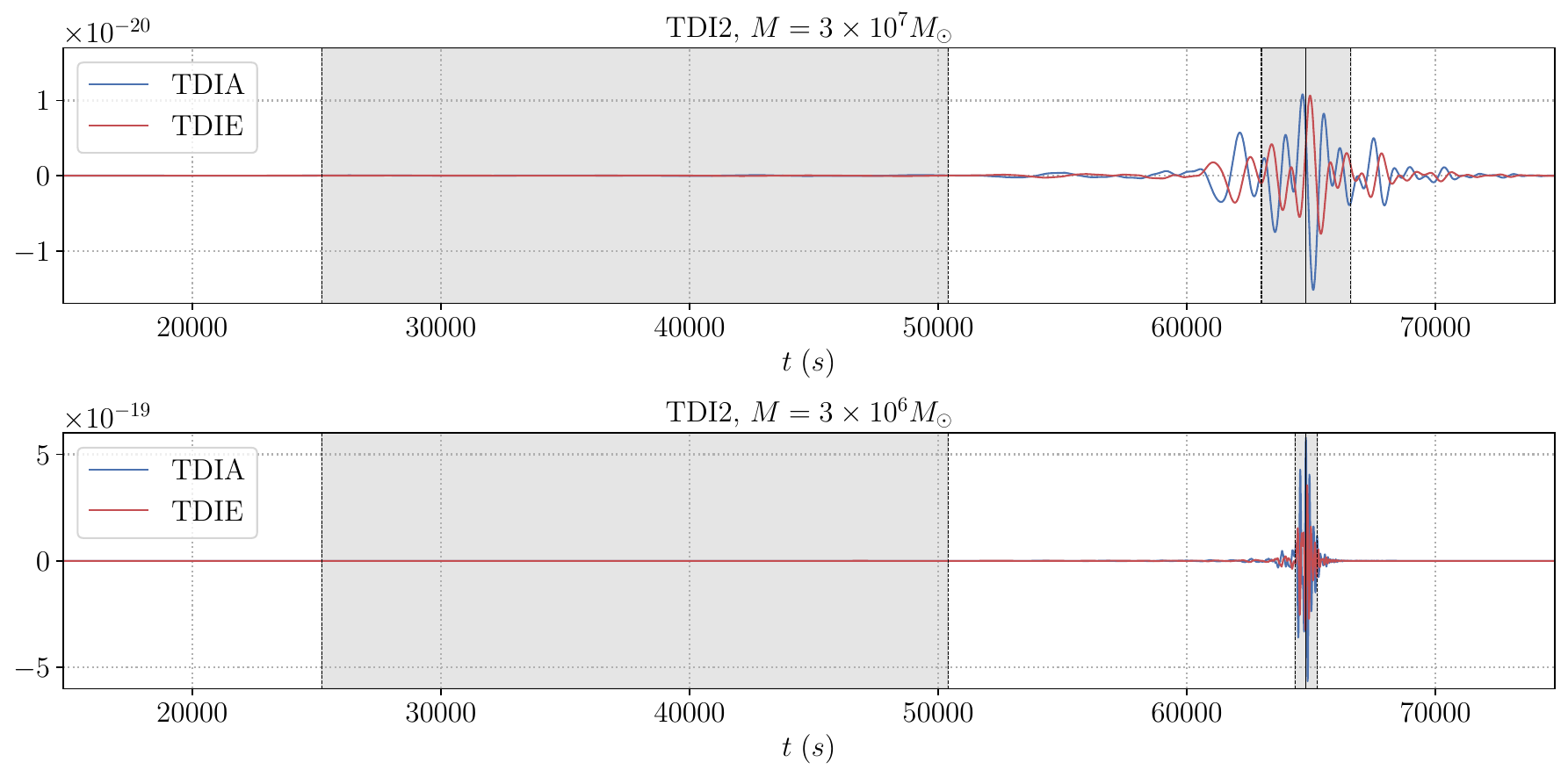}
    \caption{\textbf{(Top Panel):} We present the M3e7 case with parameters given in the top row of Tab.~\ref{tab:mbhb_params}. The full extent of the data is $[0,81920]\mathrm{s}$ and the greyed out regions indicate the gap placements (Given in Tab.\ref{tab:mbhb_gap_params}) we will consider throughout this work. The blue/red is the second generation TDI variable A/E given by the response of the instrument to the incoming MBH radiation. \textbf{(Bottom Panel):)} The same as the top panel, but with MBH/gap parameters given in the bottom row of Tab.~\ref{tab:mbhb_params} and Tab.~\ref{tab:mbhb_gap_params} respectively.}
    \label{fig:signal_with_gaps_tdi2_M3e7_M3e6}
\end{figure*}

\begin{table}
\begin{tabular}{l | c|c|c|c|c|c|c|c|c|c|c}
\toprule
System & $M_z / M_\odot$ & $q$ & $\chi_1$ & $\chi_2$ & $z$ & $t_c$ & $\iota$ & $\varphi$ & $\lambda_L$ & $\beta_L$ & $\psi_L$ \\
\midrule
\textbf{M3e7} & $3 \times 10^7$ & \multirow{2}{*}{2} & \multirow{2}{*}{0.5} & \multirow{2}{*}{0.5} & \multirow{2}{*}{1} & \multirow{2}{*}{0} & \multirow{2}{*}{$\pi$/3} & \multirow{2}{*}{1.1} & \multirow{2}{*}{0.8} & \multirow{2}{*}{0.3} & \multirow{2}{*}{1.7} \\
\textbf{M3e6} & $3\times 10^6$  &  &  &  &  &  &  &  &  &  & \\
\bottomrule
\end{tabular}
\caption{Physical parameters of our two MBHB systems: redshifted mass $M_z$, mass ratio $q$, aligned spin components $\chi_1$,$\chi_2$, redshift $z$, coalescence time $t_c$, inclination $\iota$, azimuthal phase $\varphi$, LISA-frame sky longitude and latitude $\lambda_L$,$\beta_L$, LISA-frame polarization $\psi_L$.}
\label{tab:mbhb_params}
\end{table}

\begin{table*}
\begin{tabular}{l | c c c c c c c c}
\toprule
System & $T$ & $\Delta t$ & $N$ & $t_0$ & $\Delta T_{\rm merger}^{\rm gap}$ & $\Delta T_{\rm inspiral}^{\rm gap}$ & Merger gap & Inspiral gap  \\
\midrule
\textbf{M3e7} & \multirow{2}{*}{$81920\mathrm{s}$} & \multirow{2}{*}{$10\mathrm{s}$} & \multirow{2}{*}{8192} & \multirow{2}{*}{$64800\mathrm{s}$}  & $1\mathrm{hr}$ & \multirow{2}{*}{$7\mathrm{hr}$} & $[63000\mathrm{s}, 66600\mathrm{s}]$ & \multirow{2}{*}{$[25200\mathrm{s}, 50400\mathrm{s}]$} \\
\textbf{M3e6} &  &  &  &  & $15\mathrm{min}$ &  & $[64350\mathrm{s}, 65250\mathrm{s}]$  \\
\bottomrule
\end{tabular}
\caption{Segment length and gap configurations for our two MBHB systems. Note that the merger gap has a different length for the two masses: 1hr for M3e7 and 15min for M3e6.}
\label{tab:mbhb_gap_params}
\end{table*}

\begin{table}
\begin{tabular}{l | c c c}
\toprule
SNR & No gap & Merger gap & Inspiral gap  \\
\midrule
\textbf{M3e7} & 2238 & 1372 & 2234 \\
\textbf{M3e6} & 4811 & 1361 & 4807 \\
\bottomrule
\end{tabular}
\caption{SNR of our MBHB systems in different gap scenarios.}
\label{tab:mbhb_gap_snr}
\end{table}

We apply our methods to the case of MBHB observations by LISA. At high masses, these signals are short and dominated by the merger. We will consider two MBHB systems at redshift $z=1$, with redshifted masses $M=3\times 10^7 M_\odot$ and $M=3\times 10^6 M_\odot$; their parameters, otherwise identical, are given in Tab.~\ref{tab:mbhb_params}. The lighter signal will extend to higher frequencies and allows us to explore the importance of the frequency content of the signal itself. Focusing on high-mass systems allows us to capture most of the SNR (SNR values are given in Tab.~\ref{tab:mbhb_gap_snr}) in a short data segment where we can apply the full treatment of gaps by marginalization described in Sec.~\ref{sec:methods}, which requires to work with $N\times N$ matrices. This full treatment will give us a crucial point of reference to assess modelling errors with the tools of Sec.~\ref{sec:mismodeling}.

We consider different gap scenarios, described in full in Tab.~\ref{tab:mbhb_gap_params}. In order to gauge the importance of gap location, we introduce either a 7h gap during the inspiral, which represents perhaps the most realistic instrumental setting (particularly if the gap is planned), or a gap centered right on the merger, which represents the worst-case scenario of an unlucky unplanned gap; note however that the most massive signals are short and vulnerable even to planned gaps in the absence of an advance detection. For the gap at merger, we choose a \emph{different} duration of 1h for $M=3\times 10^7 M_\odot$ and 15min for $M=3\times 10^6 M_\odot$; the merger signal is longer at high masses, and this choice is done in order to arrive at roughly the same SNR after introducing the gap. The SNR values obtained with and without the gaps are listed in~\ref{tab:mbhb_gap_snr}. For tapering, we employ a Planck window as described in App.~\ref{app:window}.

The waveforms themselves are produced with the FD waveform model \texttt{IMRPhenomHM}~\cite{London:2017bcn} with higher harmonics \{(2,2),(2,1),(3,3),(3,2),(4,4),(4,3)\}. We restrict the parameter space to spins that are aligned with the angular momentum. Although more recent and more accurate models are available (e.g.~\cite{Garcia-Quiros:2020qpx}), we do not expect the waveform choice to have an impact on the noise mismodeling errors (beyond qualitative differences like the inclusion of higher modes, as they change the morphology of the posteriors, see ~\cite{Marsat:2020rtl} for more details). We do not consider waveform systematics in the present study, and the waveform model is identical for simulated data and analysis templates.

In our analysis, we generate the MBHB GW signals and compute the LISA response in the FD using \texttt{lisabeta}~\cite{Marsat:2020rtl}. We approximate LISA orbits as having constant and equal armlengths. Again, we do not expect that more realistic LISA orbits and response would change our results about the effects of gaps significantly, as the same approximations enter the simulations and the templates.

Bayesian parameter estimation is performed using the parallel-tempered code \texttt{ptemcee}~\cite{Vousden:2016eeu}, that is based on the ensemble sampling technique~\cite{Goodman:2010dyf, Foreman-Mackey:2012any}.

\subsection{LISA response and TDI}\label{subsec:tdi}

We will give a minimal account of the Time-Delay Interferometry (TDI) observables that form the LISA data stream (see~\cite{LISA:2024hlh} for an introduction to the LISA mission), as we wish to highlight later on the importance of the noise spectral content, from red noise to blue noise. We use the notations of~\cite{Marsat:2020rtl}. In the equal-armlength approximation, the one-arm laser frequency shift observables, $y_{slr} = (\nu_s - \nu_r) / \nu$, measured between emitting spacecraft $s$ and receiving spacecraft $r$ along the link $l$ take the form~\cite{Estabrook:1975jtn} (setting $c=1$):
\be\label{eq:defyslr}
	y_{slr} = \frac{1}{2} \frac{n_{l} \otimes n_{l}}{1 - k\cdot n_{l}} : \left[ H(t - L - k\cdot p_{s}) - H(t - k\cdot p_{r}) \right] \,,
\ee
where $k$ is the GW propagation unit vector, $n_l$ is the unit vector from $s$ to $r$ along the LISA arm, $p_s$, $p_r$ are the spacecraft positions at some time $t$, and $H$ is the GW in $3\times 3$ matrix form (see the notations of~\cite{Marsat:2020rtl}). The notation $\otimes \ldots :$ indicates that $H$ is contracted twice with $n_l$.

TDI is a construction that reproduces in post-processing interferometric configurations which reduce the otherwise dominant laser noise, and is crucial to enable LISA observations (\cite{armstrong1999time, Tinto:1999yr, Dhurandhar:2001tct, Cornish:2003tz, Shaddock:2003dj, Tinto:2003vj, Vallisneri:2004bn}, see~\cite{tinto2014time} for a review). Multiple variables can be constructed, see, e.g.,~\cite{Vallisneri:2004bn}. First-generation TDI cancels laser-noise for static, unequal-arms. Second-generation TDI is necessary to provide the needed cancellations for realistic orbits, with armlengths changing over time~\cite{Cornish:2003tz, Shaddock:2003dj, Tinto:2003vj, Vallisneri:2004bn}. For recent work on TDI including higher-generation TDI, we refer the reader to Refs.~\cite{Bayle:2018hnm, Tinto:2022zmf, Tinto:2023ouy, Houba:2024ysm}.

The full TDI expressions simplify drastically in the approximation of equal-armlengths, treating all delays as equal in both directions along each arm. There is then a single delay in the construction, $L/c \simeq 8\,\mathrm{s}$ for $L=2.5\,\mathrm{Gm}$. With the notation $y_{sr,nL} = y_{sr}(t - nL)$ (dropping the middle index), the first-generation and second-generation TDI Michelson observables $X_1$, $X_2$ read:
\begin{align}
    X_0 &= y_{31} + y_{13,L} - \left( y_{21} + y_{12,L} \right) \,, \nonumber\\
	X_1 &= X_0 - X_{0,2L}  \,, \nonumber\\
    X_2 &= X_1 - X_{1,4L} \label{eq:TDI_construction_X}\,.
\end{align}
Other Michelson variables $Y$, $Z$ are obtained by cyclic permutations of the spacecraft. Here, $X_0$ can be regarded as a 0-th generation Michelson TDI. Geometrically (see Fig.~3 of~\cite{Vallisneri:2004bn}), it would correspond to a single pass along each arm; such a combination does not cancel laser noise for unequal arms.

Uncorrelated combinations $A$, $E$ and $T$~\cite{Prince:2002hp} can be built from $X$, $Y$, $Z$ (of any generation) as
\begin{align}\label{eq:defAET}
	A &= \frac{1}{\sqrt{2}} \left( Z - X \right) \,, \nonumber\\
	E &= \frac{1}{\sqrt{6}} \left( X - 2Y + Z \right) \,, \nonumber\\
	T &= \frac{1}{\sqrt{3}} \left( X + Y + Z \right) \,.
\end{align}
These combinations have the advantage of featuring uncorrelated instrumental noise, under assumptions about the equality of noise levels in the different spacecraft. This means that these three channels can be treated as three independent detectors. This is only true for equal armlengths, and we would have to consider correlations across channels in any realistic setting for the instrument; we will retain this approximation for simplicity here. The channel $T$ is strongly suppressed at low frequencies, while the two channels $A$, $E$ are not.

As illustrated by~\eqref{eq:TDI_construction_X}, successive TDI generations correspond in first approximation to discrete derivatives of the data stream. In the FD, delays are simple phase factors and these finite differences take simple forms. In~\cite{Marsat:2020rtl}, one of us introduced a notation for rescaled TDI variables, factoring out these frequency-dependent terms as
\begin{align}\label{eq:def_aet}
	\tilde{a}, \tilde{e} &= \frac{e^{-2i\pi fL}}{i \sqrt{2} \sin (2\pi f L)}\times \tilde{A}, \tilde{E} \,,
\end{align}
and a different rescaling for $\tilde{T}$, which is unnecessary for this work.

In our study, we will use these different TDI generations to explore the importance of red and blue noise; a derivative changes the spectrum of a variable by one power of $f$, and second-generation TDI is therefore a process that has a blue tilt compared to first-generation. Our nomenclature for tables and figures is: 
\begin{itemize}
    \item TDI0: channels $A_0$, $E_0$ given by the combinations~\eqref{eq:defAET} from $X_0$, $Y_0$, $Z_0$ given in~\eqref{eq:TDI_construction_X}; they differ by a constant factor from the variables given in~\eqref{eq:def_aet} as $\tilde{a},\tilde{e} = -\sqrt{2} \tilde{A}_0,\tilde{E}_0$ (while $\tilde{t}$ of~\cite{Marsat:2020rtl} is not the same as $\tilde{T}_0$);
    \item TDI1: channels $A_1$, $E_1$ given by the combinations~\eqref{eq:defAET} from $X_1$, $Y_1$, $Z_1$ given in~\eqref{eq:TDI_construction_X};
    \item TDI2: channels $A_2$, $E_2$ given by the combinations~\eqref{eq:defAET} from $X_2$, $Y_2$, $Z_2$ given in~\eqref{eq:TDI_construction_X}.
\end{itemize}

We will also ignore the $T$ channel entirely for our low-frequency signals. Assuming independence across the $A$, $E$ TDI channels (of any generation), the log-likelihood becomes a sum over the two channels. For instance, \eqref{eq:windowed_likelihood_td} becomes
\be
    \ln\mathcal{L}_{\text{gap}} = -\frac{1}{2} \sum_{C=A,E} (\bm{W}\bm{x}_C)^{T} \left(\bm{W}\bm{\Sigma}_C \bm{W}\right)^{+} (\bm{W}\bm{x}_C) \,,
\ee
with in fact an identical noise covariance for the two channels, $\bm{\Sigma}_A = \bm{\Sigma}_E$. Similarly, all expressions of Secs.~\ref{sec:methods} and~\ref{sec:mismodeling} for the SNR, Fisher matrices, noise biases, with and without mismodeling, are similarly given with sums $\sum_{A,E} (\dots)$.

Strictly speaking, gaps would differently affect the different TDI generations. As shown in Eq.~\eqref{eq:TDI_construction_X}, the TDI construction involves delayed combinations of the base observables $y_{sr}$. This leads to (a priori) a loss of information at the edges of the missing base data, leading to a wider data gap, an effect that is worse for higher-order TDI constructions. For TDI2, the maximum delay that intervenes in~\eqref{eq:TDI_construction_X} is $7 L /c\simeq 58\mathrm{s}$ (1 for TDI0, +2 for TDI1, +4 for TDI2). In the gap case, this leads to the SNR being slightly different for the three TDI versions. We leave these differences aside in the present study, and the gaps are treated as covering the same time interval in all TDI generations. We refer the reader to Ref.~\cite{Houba:2024ysm} for more discussions related to how gaps impact TDI variables for various generations.

\subsection{Noise models}

\begin{figure}
\includegraphics[width = 0.45\textwidth]{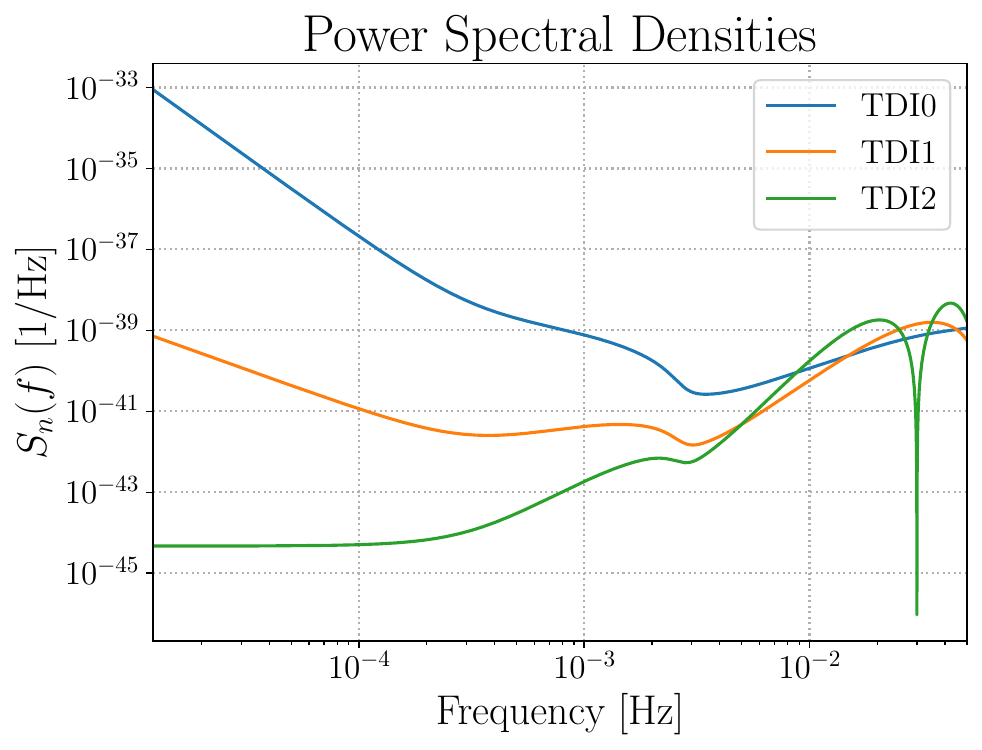}
    \caption{The blue, orange and green curves depict the behavior of the noise curves given by $S^{\text{TDI0}}_{n}(f)$ \eqref{eq:noise_curve_TDI0}, $S^{\text{TDI1}}_{n}(f)$  \eqref{eq:noise_curve_TDI1} and $S^{\text{TDI2}}_{n}(f)$ \eqref{eq:noise_curve_TDI2} respectively. The white dwarf background is present and it has been assumed that LISA has been operational for one year.}
    \label{fig:noise_curves}
\end{figure}

Throughout our analysis we will focus on three noise curves each with different properties. We will use the \texttt{SciRDV1} noise curve given in the LISA mission requirements document~\cite{LISAsr:18aa}. 
We will define the instrument related single-link optical metrology noise $P_{\text{OMS}}$ and test mass acceleration noise $P_{\text{acc}}$ by (restoring factors of $c$)

\begin{widetext}
\begin{align}
P_{\text{OMS}}(f) &= (15 \times 10^{-12}\text{m})^2\left(1 + \left(\frac{2\, \text{mHz}}{f}\right)^4\right) \left(\frac{2 \pi f}{c}\right)^2\,\text{Hz}^{-1} \label{eq:OMS} \\
P_{\text{acc}}(f) &= (3\times 10^{-15}\, \text{m} \,\text{s}^{-2})^2\left(1 + \left(\frac{0.4\,\text{mHz}}{f}\right)^2\right)\left(1 + \left(\frac{f}{8\,\text{mHz}}\right)^4\right) \left(\frac{1}{2 \pi c f}\right)^{2} \,\text{Hz}^{-1} \label{eq:acc}
\end{align}
The noise curves for the $A$ and $E$ channels in each TDI configuration, TDI0, TDI1 and TDI2 respectively, are given by 
\begin{subequations}
\begin{align}
S^{A_0, E_0}_{n}(f) &= 2 \left[2 (3 + 2\cos x + \cos 2x) P_{\text{acc}}(f) + (2 + \cos x)P_{\text{OMS}}(f) \right] \label{eq:noise_curve_TDI0}\,, \\
S^{A_1, E_1}_{n}(f) &= 4 \sin^2 x \, S^{A_0, E_0}_{n}(f) \,, \label{eq:noise_curve_TDI1} \\
S^{A_2, E_2}_{n}(f) &= 4 \sin^2 2x \, S^{A_1, E_1}_{n}(f) \,.\label{eq:noise_curve_TDI2}
\end{align}
\end{subequations}
where $x = 2\pi f L/c$, $L = 2.5\ \text{Gm}$ is the length of the LISA arms and $c$ the speed of light. We also account for the presence of the galactic foreground $S_{c}(f)$, which is folded into $S_{n}(f) \mapsto S_{n}(f) + S_{c}(f)$, according to~\cite{Karnesis:2021tsh,Babak:2021mhe}. In this work, we assume a LISA observation duration $T_{\rm obs}=4\mathrm{yr}$ to set the level of this foreground.
\end{widetext}
With cadence $\Delta t = 10$ seconds over a $\sim 22.475$ hour long interval, plots for the different noise curves  are given in Figs.~(\ref{eq:noise_curve_TDI0} - \ref{eq:noise_curve_TDI2}).

\begin{figure*}
    \centering
    \includegraphics[width = \textwidth]{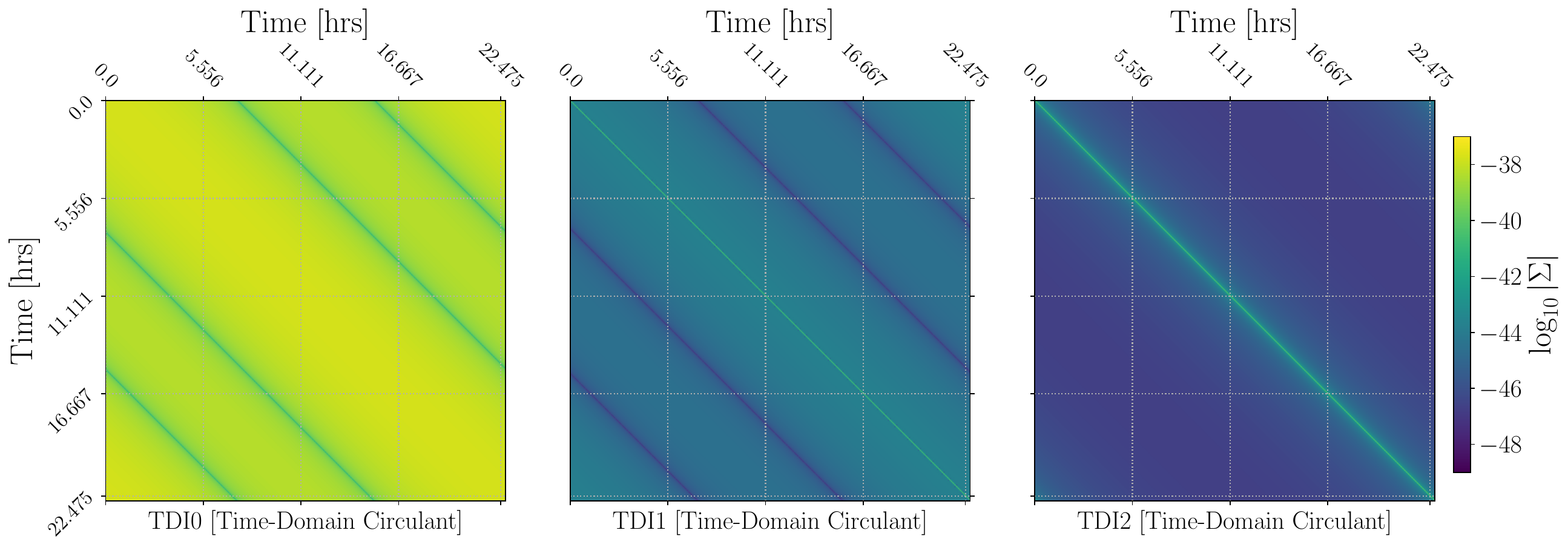}
    \caption{\textbf{(Left to right panels:)} Time-domain and \emph{circulant} noise covariance matrices for TDI0/1/2 respectively.}
\label{fig:circulant_td_all_matrices}
\end{figure*}
\begin{figure*}
    \centering
    \includegraphics[width = \textwidth]{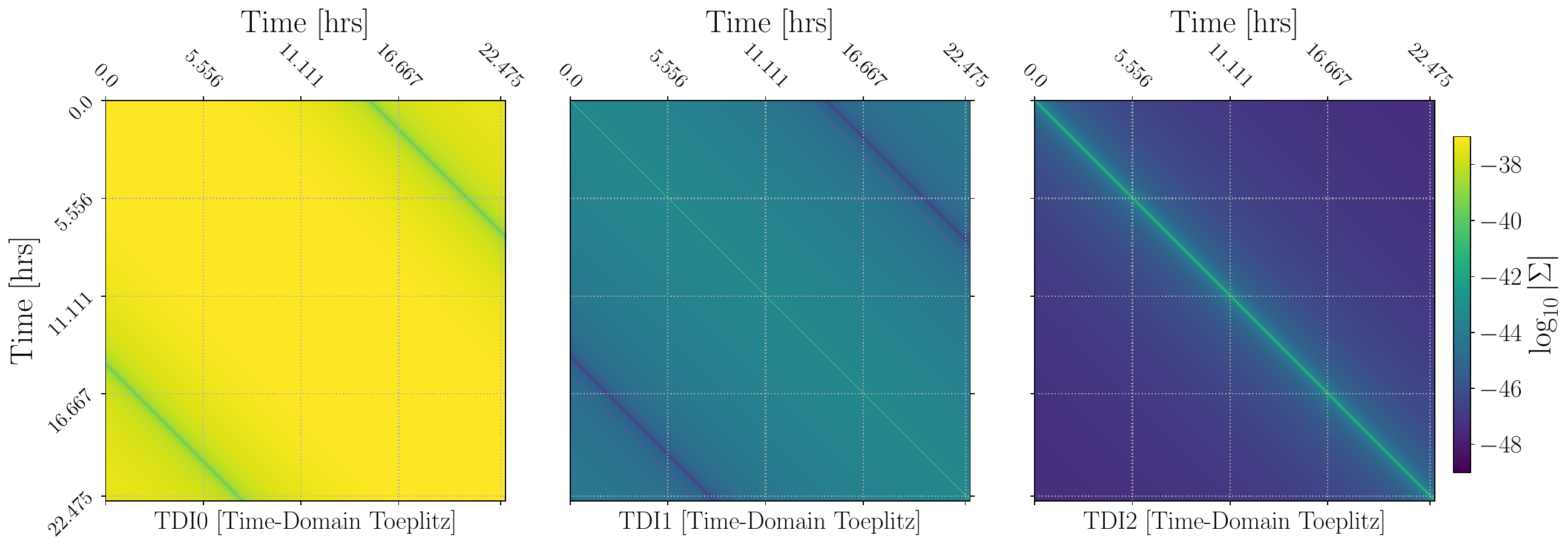}
    \caption{\textbf{(Left to right panels:)} Time-domain and \emph{Toeplitz} noise covariance matrices for TDI0/1/2 respectively.}
\label{fig:toeplitz_td_all_matrices}
\end{figure*}
Noise described by the $S_{n}^{\text{TDI0}}$ PSD, yields a powerful red-noise component at low frequencies, proportional to $\sim 1/f^4$. Time-domain noise generated from $S_{n}^{\text{TDI0}}$ would describe low frequency oscillations of the arms with noise amplitudes many orders of magnitude higher than noise generated from the TDI1 and TDI2 noise curves respectively. Similarly, the TDI1 noise curve $S_{n}^{\text{TDI1}}(f)$ yields a non-trivial red noise component $\sim 1/f^2$. The TDI2 noise curve $S_{n}^{\text{TDI2}}$, in contrast to both TDI0/1 curves, is a white noise process for low frequencies $f\rightarrow 0$. Each process demonstrates the same blue noise behavior in general for high frequencies, $\sim f^{2}$. In contrast to TDI0, both TDI1 and TDI2 have ``zero-crossing" behavior that increase in number for higher frequencies. In~\cite{Marsat:2020rtl}, TDI0 was constructed so that is avoids such zero-crossings and hence numerical instabilities from computing ratios $0/0$. For more discussion, please refer to Eqs.(29--33) in~\cite{Marsat:2020rtl}.

\section{Analysis of the noise covariance matrices}\label{sec:analysis_noise_cov}

With noise models described by Eqs.~(\ref{eq:noise_curve_TDI0}-\ref{eq:noise_curve_TDI2}), gap structure in Tab.\ref{tab:mbhb_gap_params} and tapering function~\eqref{eq:def_planck_window}, and with formalisms developed in Sections \ref{subsec:analysis_set_up} and \ref{subsec:windowed_covariance}, we are now in a position to numerically investigate the impact of gaps on the noise covariance matrix in both the time and FD. Sec.~\ref{subsec:circulant_vs_toeplitz} illustrates the differences between circulant and Toeplitz structures. Sec.~\ref{subsec:noise_cov_matrix_w_gaps} demonstrates the impact of gaps on the covariance matrices, manifesting as a breakdown of the TD Toeplitz structure leading to an overall \emph{non-stationary} process. In Sec.~\ref{subsec:pseudo_inverses_practice}, we outline our numerical schemes detailing how we compute pseudo-inverses in practice. Throughout this section, we will consider an observation time of $T \sim 22.435\,\text{hrs}$ with cadence $\Delta t = 10\,\text{s}$ resulting in a data stream of length $N = 8192$ with lowest resolvable frequency $\Delta f \sim 1/T \approx 10^{-5}\,\text{Hz}$.
\subsection{Circulant and Toeplitz Processes}\label{subsec:circulant_vs_toeplitz}

Plots of the circulant and Toeplitz matrices are given respectively in Figs.~(\ref{fig:circulant_td_all_matrices}-\ref{fig:toeplitz_td_all_matrices}). In Fig.~\ref{fig:circulant_td_all_matrices}, the matrices are circulant~\eqref{eq:defcirculant}, obeying not only the Toeplitz rule of constant entries along each diagonal, with all entries deduced from the first row of the matrix, but also showing a circular symmetry: each row is permuted by one element to the right with respect to the previous row, wrapping around the edge. The first row is also symmetric with respect to its middle point. Notice that the matrices in Fig.~\ref{fig:toeplitz_td_all_matrices} maintain the Toeplitz property in Def.~$\eqref{eq:deftoeplitz}$ but lack circular symmetry: different rows are not related by a cyclic permutation.
\begin{figure}
    \centering
    \includegraphics[width = 0.475\textwidth]{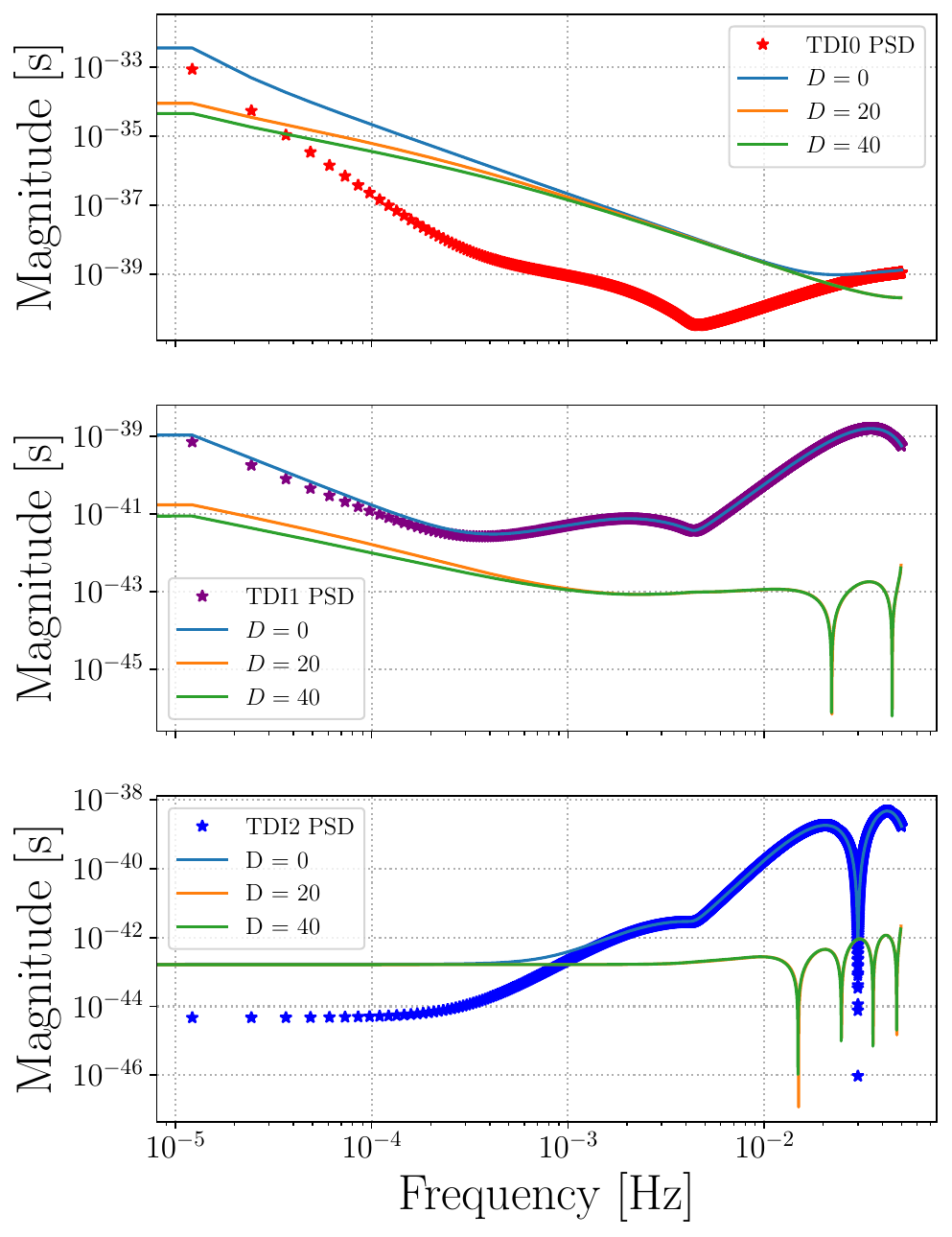}
    \caption{\textbf{(Top plot):} Comparison between (in red) TDI0 PSD, the main diagonal of circulant noise process, with the main diagonal ($D = 0$, blue) and sub-leading diagonals ($D = \{20,40\},$ orange and green respectively) of the same noise curve but given by a Toeplitz process $\bm{\tilde{\Sigma}}^{\text{toe}}$. \textbf{(Middle and bottom plot):} The same as the top plot but for (in purple) TDI1 noise curves and (in blue) TDI2 noise curves respectively. mismodeling the Toeplitz process as a circulant process (most common in practice) appears to be worse for TDI0 and TDI2 than the TDI1 noise curve. The Toeplitz TDI1 process looks to be well approximated by the corresponding circulant process.}
\label{fig:toeplitz_vs_PSD}
\end{figure}
There are clear differences in the TD noise covariance matrices between the two cases. The FD can shed further light on the differences between the two stationary processes. We can compute 
\begin{align}
\tilde{\bm{\Sigma}}^{\text{circ}} &= N\Delta t^2 (\bm{P}\bm{\Sigma}^{\text{circ}}\bm{P}^{\dagger}) = \ \text{Diagonal Matrix}\,, \label{eq:noise_cov_sec_fourier_Sigma_circ} \\
\tilde{\bm{\Sigma}}^{\text{toe}} &= N\Delta t^2 (\bm{P}\bm{\Sigma}^{\text{toe}}\bm{P}^{\dagger}) = \text{Banded Matrix}\,.\label{eq:noise_cov_sec_fourier_Sigma_toe}
\end{align}
The matrix $\tilde{\bm{\Sigma}}^{\text{circ}}$ is a diagonal matrix with elements given by the noise curves Eqs.(\ref{eq:noise_curve_TDI0}--\ref{eq:noise_curve_TDI2}). Converting the Toeplitz TD noise covariance matrices to their FD analogs $\tilde{\bm{\Sigma}}^{\text{toe}}$ results in a \emph{banded matrix} with sub-leading off-diagonal elements. This result was expected given the calculation present in App.~\ref{app:diagonal_cov_matrix_derivation_circulant} leading to Eq.~\eqref{eq:Toeplitz_Matrix}. A plot of the main diagonal of the circulant matrix $\tilde{\bm{\Sigma}}^{\text{circ}}$ compared to the leading and sub-leading diagonals of $\tilde{\bm{\Sigma}}^{\text{toe}}$ is given in Fig.~\ref{fig:toeplitz_vs_PSD}. For TDI1, it appears that the assumption of a circulant process when the underlying process is Toeplitz appears reasonable for our range of observable frequencies. At high frequencies one obtains a near perfect match between the Toeplitz and circulant based FD matrices, with minor deviations at low frequency. The leading diagonals are subdominant, with power decreasing monotonically. For TDI0, the noise is (significantly) under-estimated which would result in tighter posteriors and significantly larger bias-scatters with respect to the true posterior. In the context of Sec.~\ref{subsec:mismodeling_measures}, we would expect the scatter-to-width ratios $\Upsilon_{a} \gg  1$. We find the same result for TDI2, but perhaps to a slightly lesser degree than TDI0. We remark that this is consistent with our later findings in Tab.\ref{tab:mismodeling_upsilon_toeplitz} found in the results Sec.~\ref{sec:results}, where we perform parameter inference assuming the underlying noise process as circulant when in reality it is Toeplitz.

In our work, unless stated otherwise, we will assume that the underlying noise process is both stationary, circulant and gaps modeled as missing data. This is to remain in line with the general community that use the Whittle-likelihood on the usual assumption that the noise is circulant, and it allows us to isolate the effects of gap mismodeling from the effects of Toeplitz/circulant mismodeling.

\begin{figure*}
    \centering
    \includegraphics[width = \textwidth]{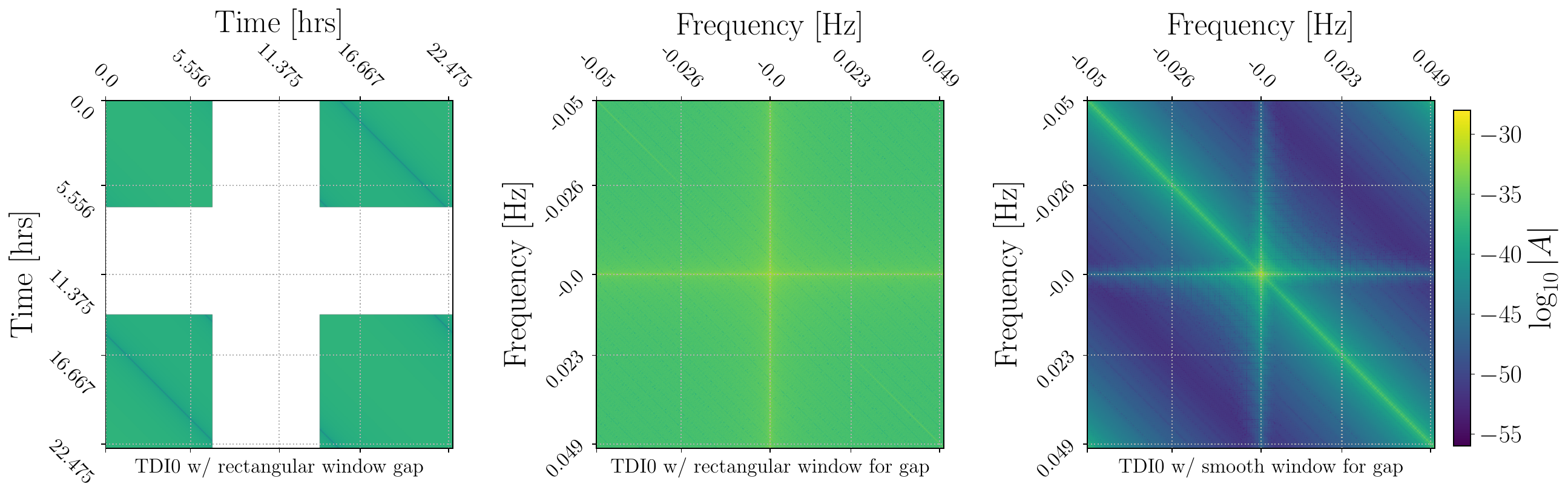}
    \caption{\textbf{(Left panel:)} Time-domain gated noise covariance matrix for a circulant process with a TDI0 noise curve. Here the gap is represented via zeros in the noise covariance matrix. \textbf{(Middle panel:)} The FD representation of the left-most panel. \textbf{(Right panel:)} The FD noise-covariance matrix of the left-most panel but with a smooth 30 minute taper at each endpoint of the gap segment. Notice that the powerful red-noise component of TDI0 yields significant leakage (and thus correlations) amongst the frequency components of the middle panel. The right-most panel has precisely the same power as the middle panel, but more concentrated along the diagonals (and $0-f$ frequency correlations).}
\label{fig:TDI0_td_fd_gap_smooth_harsh}
\end{figure*}
\begin{figure*}
    \centering
    \includegraphics[width = \textwidth]{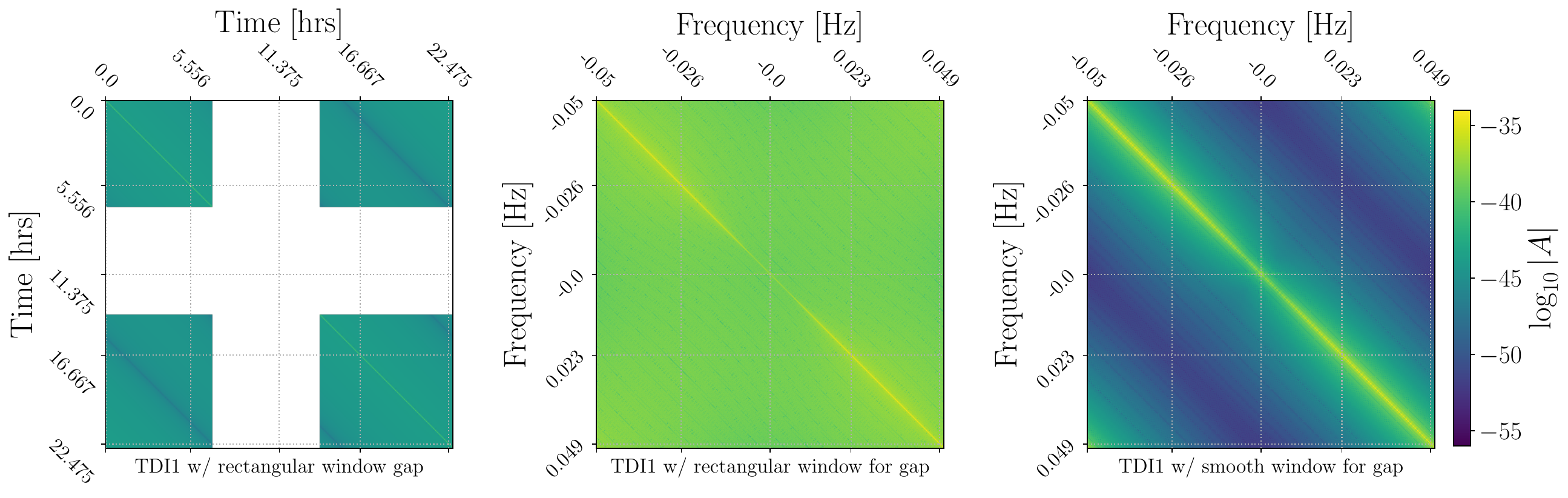}
    \caption{The same set up as Fig.~\ref{fig:TDI0_td_fd_gap_smooth_harsh} but for TDI1. The middle panel, in contrast, retains diagonal dominance but much of the power leaks out into neighboring frequency bins. This effect is reduced significantly if a 30 minute taper is applied as shown by the right-most panel.}
\label{fig:TDI1_td_fd_gap_smooth_harsh}
\end{figure*}
\begin{figure*}
    \centering
    \includegraphics[width = \textwidth]{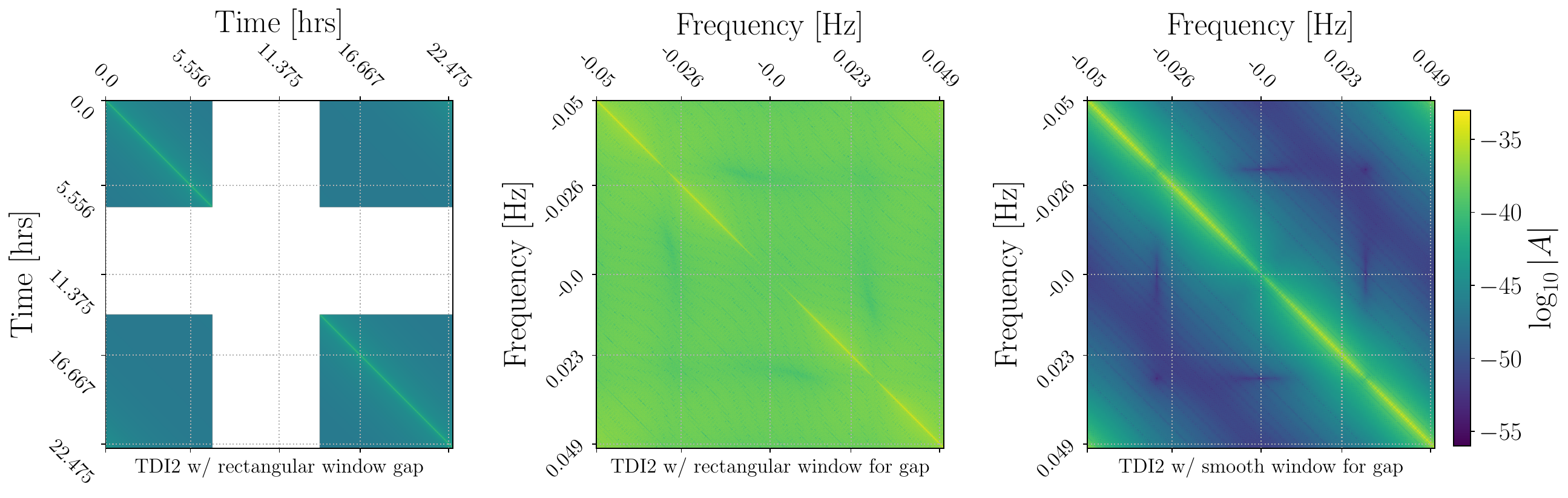}
    \caption{The same set up as Fig.~\ref{fig:TDI0_td_fd_gap_smooth_harsh} but for TDI2. The zero crossings of the PSD (as shown in Fig.~\ref{fig:noise_curves}) are seen along the main diagonal of the middle panel and correlations with that zero crossing are shown by the fainter patches in both the middle and right panels. Like the right-most panels of Fig.~\ref{fig:TDI0_td_fd_gap_smooth_harsh} and Fig.~\ref{fig:TDI1_td_fd_gap_smooth_harsh}, the leakage effects are reduced significantly only if a smooth taper is applied.}
\label{fig:TDI2_td_fd_gap_smooth_harsh}
\end{figure*}
\subsection{Noise Covariance Matrices with Gaps}\label{subsec:noise_cov_matrix_w_gaps}

The left-most panels in Figs.~(\ref{fig:TDI0_td_fd_gap_smooth_harsh} -- \ref{fig:TDI2_td_fd_gap_smooth_harsh}) visually demonstrate the impact of a single gap on the TD noise covariance matrix with noise curves TDI0/1/2 respectively. Notice that both the circulant and Toeplitz structure, as depicted via Fig.~\ref{fig:circulant_td_all_matrices}, is now lost, and instead a rank deficient matrix (see Eq.\eqref{eq:Block_matrix_structure_gaps_TD}) features. %Since the circulant property of the noise process is lost, the Whittle-based likelihood loses validity. 
The individual diagonal sub-matrices each have a Toeplitz structure and are \emph{not} circulant. The off-diagonal sub-matrices represent correlations of the noise process pre and post gap segment. If these off-diagonal matrices were neglected, then the analyst would be making the assumption that the noise pre and post gap is independent. This was mentioned in Sec.~\ref{subsec:gap_segmented_whittle}, where the analyst may decide to make the approximation that the individual diagonal sub-matrices are circulant. Notice, that since the circulant property is lost in the individual main diagonal sub-matrices, their FD covariances cease to be diagonal, and the Whittle-likelihood loses validity. These noise mismodeling investigations will be discussed in more detail in the results Sec.~\ref{subsec:mismodeling_independence}.

The TD noise covariance matrices can also be expressed in the FD. For TDI0, the middle panel of Fig.~\ref{fig:TDI0_td_fd_gap_smooth_harsh} is the FD version of the TD gated noise covariance matrix given by the left panel of Fig.~\ref{fig:TDI0_td_fd_gap_smooth_harsh}, computed via Alg.~\ref{algorithm:FD_covariance}. The excess power at low frequencies $f\approx 10^{-5}\,\text{Hz}$, given by the steep rise of the red-noise dominated noise curve, correlates strongly with higher frequencies as $f \rightarrow 0.05\,\text{Hz}$. Introducing a taper with lobes of length 30 minutes mitigates the leakage effects, restoring the banded nature of the FD noise covariance matrix. The correlations between low and high frequencies are still present and this is due to the enormous red-noise component of the noise-curve for TDI0. For TDI1, the middle and right panels of Fig.~\ref{fig:TDI1_td_fd_gap_smooth_harsh} demonstrate to a slightly lesser degree the effects of leakage, with tapering focusing the power more along the main diagonals. The same for TDI2 is observed in Fig.~\ref{fig:TDI2_td_fd_gap_smooth_harsh}, but with the additional feature of correlated zero-crossings between frequency components. In the limit as $\Delta t \rightarrow 0^{+}$ (higher Nyquist frequency) and/or infinite observation times similar features would be observed in TDI1, but not TDI0.

The tapering scheme, in theory, should not impact observables involving inner products with the noise covariance matrices in the FD assuming the model covariance matrix is consistent with the true noise covariance matrix. The likelihood (and SNR) should be independent of the tapering scheme and information lost should only happen during the gated segment. The band-diagonal nature of the FD noise covariance matrices for TDI1 and TDI2 could be exploited to accelerate likelihood computations. However, we will see in the next Sec.~\ref{subsec:pseudo_inverses_practice} that the introduction of a smooth taper makes the computation of the pseudo-inverse challenging. 

\subsection{Pseudo-inverses with gated functions}\label{subsec:pseudo_inverses_practice}

The gated TD noise-covariance matrix has a block-like structure, which facilitates a relatively simple computation of the pseudo-inverse. The gated FD noise-covariance matrix has no obvious structure to take advantage of. Below we will outline a numerical scheme that allows for an \emph{exact} calculation of the pseudo-inverse for FD noise covariance matrices assuming a simple gating function. The results are then extended to windows with non-trivial lobes in Sec.~\ref{subsec:pseudo_inverses_practice_w_taper}.

\begin{figure}
    \centering
    \includegraphics[width = 0.475\textwidth]{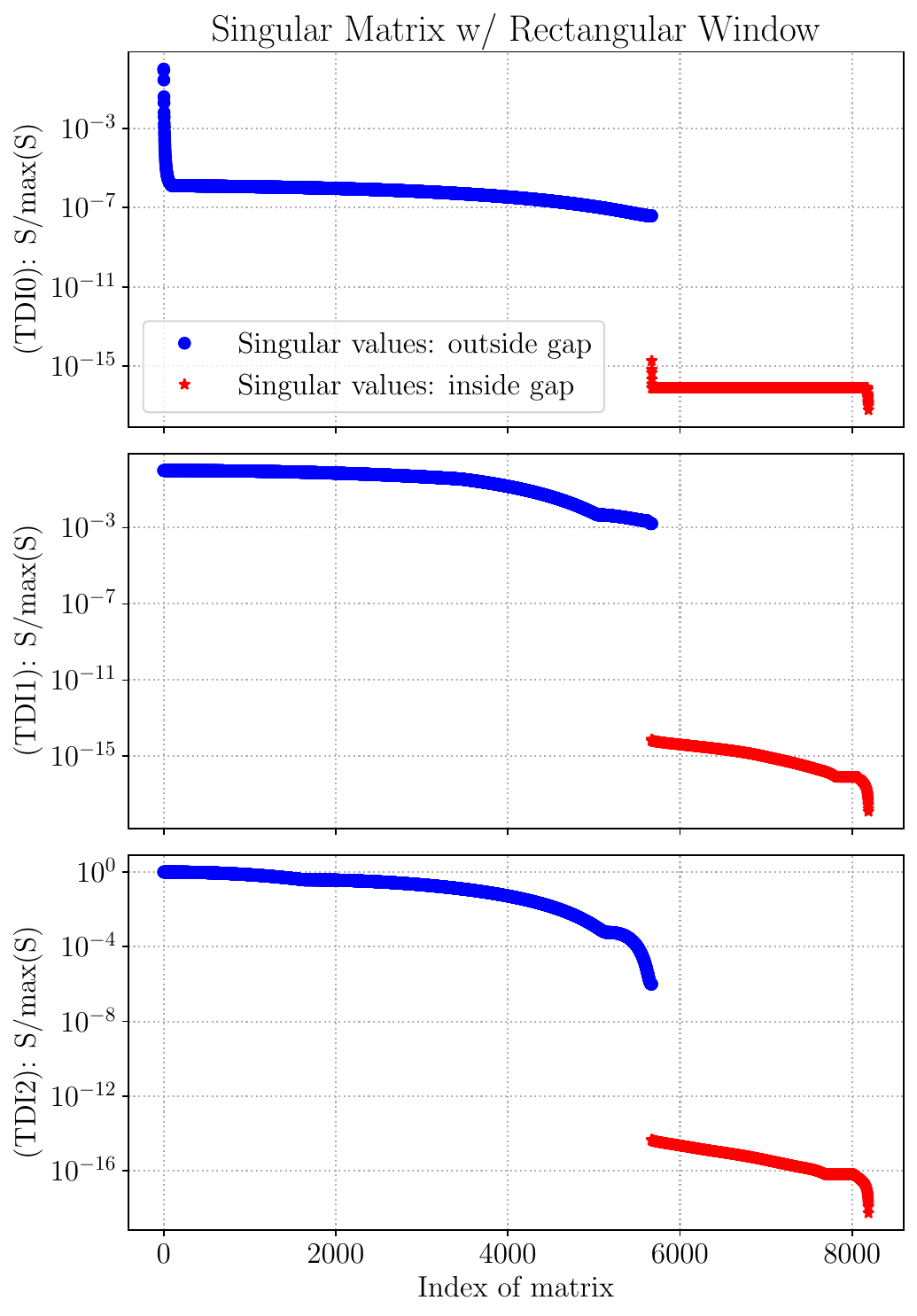}
    \caption{\textbf{(Top plot to bottom plot):} Normalised SVs of the FD noise covariance matrix for TDI0/1/2 respectively. In red/blue are the SVs that are retained/discarded. Since we use a rectangular window, we know precisely that the zero-SVs correspond to the number of zeros in the window function $w_{\text{rect}}$. }
\label{fig:TDI0_TDI1_TDI2_singular_vals_rect}
\end{figure}

In App.~\ref{app:proof_eigenvalues_matrices} we prove that the number of zero-eigenvalues in the FD gated noise-covariance matrix is equivalent to the number of zero rows (or columns) in the TD gated covariance matrix. Following~\cite{press2007numerical}, one can compute the Singular Value Decomposition (SVD) $\tilde{\bm{\Sigma}}^{\text{gate}} = \bm{U}\bm{\Lambda}\bm{V}^{\dagger}$ to obtain the pseudo-inverse as $\tilde{\bm{\Sigma}}^{+} = \bm{V} \bm{\Lambda}^{+} \bm{U}^{\dagger}$. Since the matrix $\tilde{\bm{\Sigma}}^{\text{gate}}$ is hermitian symmetric, the number of zero-eigenvalues are equivalent to the number of zero-singular values in the matrix $\bm{S}$. Fig.~\ref{fig:TDI0_TDI1_TDI2_singular_vals_rect} shows a plot of the normalized singular matrix $\bm{S}/\text{max}(\bm{S})$ computed via the SVD of $\tilde{\bm{\Sigma}}^{\text{gate}}$ for each of the TDI0/1/2 noise curves using a rectangular window. The parameters of the rectangular window are given by the last column of Tab.~\ref{tab:mbhb_gap_params}. On all panels in Fig.~\ref{fig:TDI0_TDI1_TDI2_singular_vals_rect}, the red curves are the SVs that correspond to the gap. The blue curves are the SVs that point towards data points that exist between gap segments. The top plot is for TDI0, where we observe a steep drop in the SVs that represents the strong red-noise component of the noise curve. This feature is not present in TDI1 or TDI2. We understand from previous arguments that the number of zero-SVs are precisely the number of time-bins where the window function is zero. In practice we set these zero-SVs as infinite so they will not contribute in matrix calculations using the inverse $\bm{\Lambda}^{+}$. The routine outlined here will always calculate the \emph{unique} pseudo-inverse for the FD covariance matrix for simple gating functions.   

Computing the pseudo-inverse in the TD is trivial in our set up. The pseudo-inverse of the TD noise covariance matrix is simply the inverse of the individual block Toeplitz matrices as discussed in section \ref{subsec:analysis_set_up}. For a generic gap configuration, the same method applies after a reordering of the matrix, as shown in App.~\ref{app:likelihood_using_masks_permutations}. This becomes more challenging if a smooth taper is applied, since the action of the taper increases the condition number of the matrix. One could apply a similar logic of the discussion surrounding \eqref{eq:regualrised_singular_matrix} for the TD matrix to compute the pseudo-inverse. Whenever calculations are conducted in the TD, we will never taper the data to ensure that our results are as numerically stable as possible.
a
\subsection{Pseudo-inverses with tapering functions}\label{subsec:pseudo_inverses_practice_w_taper}
\begin{figure}
    \centering
    \includegraphics[width = 0.48\textwidth]{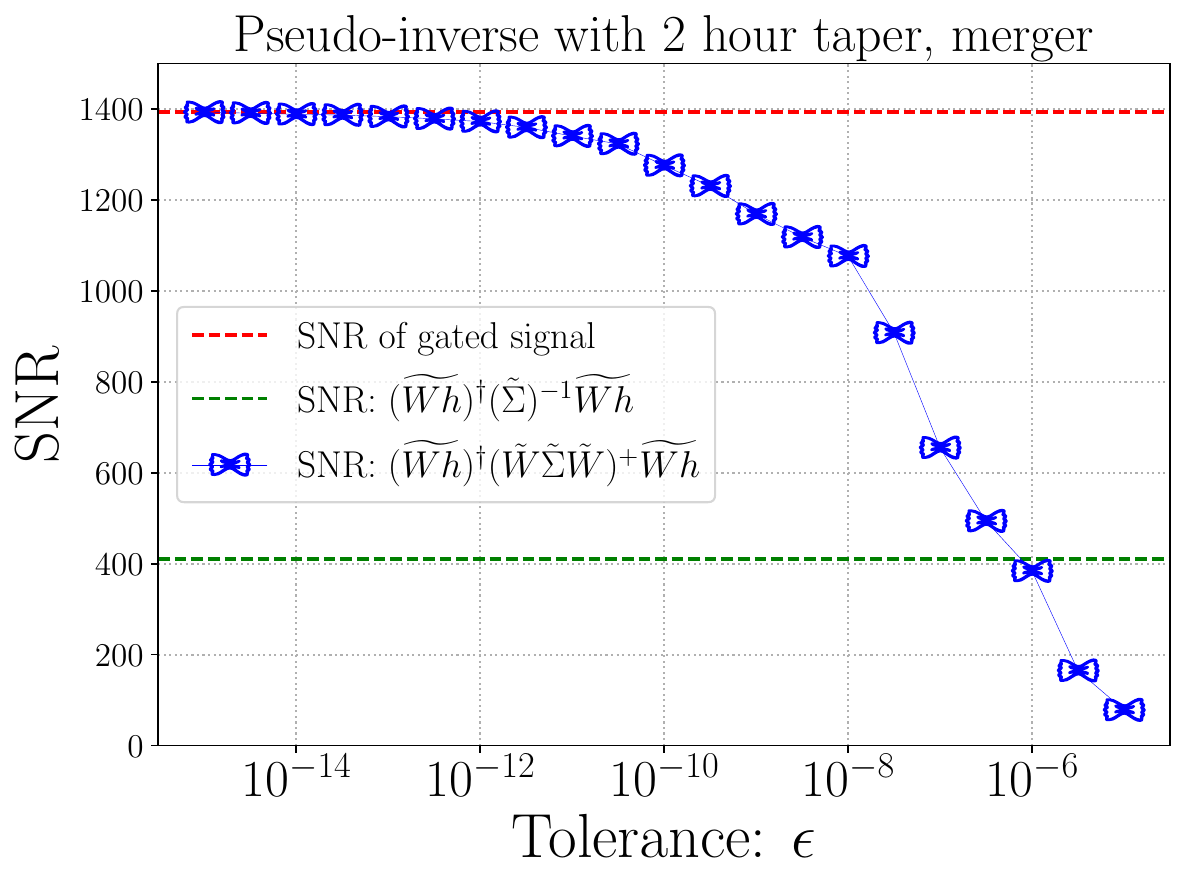}
    \caption{\textbf{(Red horizontal dashed line):} The SNR of the M3e7 MBH with one hour gap at merger using a gating function, with pseudo-inverse computed using only the (blue) SVs retained in the bottom panel of Fig.~\ref{fig:TDI0_TDI1_TDI2_singular_vals_rect}. \textbf{(Green horizontal dashed line):} The SNR of the MBH using a smooth taper with 2 hour length lobes with Whittle-based noise covariance. \textbf{(Blue farfalle):} The SNR computations now using a smooth tapering function (see the yellow tapered MBH signal in Fig.~\ref{fig:zoomed_in_taper_plot}) with pseudo-inverse computed with increasing accuracy as $\epsilon \rightarrow 0$.}
\label{fig:pseudo_inv_vs_SNR}
\end{figure}
It was seen in the right-most panels of Figs.~(\ref{fig:TDI0_td_fd_gap_smooth_harsh}--\ref{fig:TDI2_td_fd_gap_smooth_harsh}), that there was clear impact of tapering on the structure of the FD noise-covariance matrix. The presence of a taper marginally complicates the computation of the pseudo-inverse, since there is now a smooth drop towards the SVs that cause the bad behavior of the noise covariance matrix. As such, there is no clear criterion on how many SVs to remove when smooth tapers are introduced (see~\cite{Talbot:2021igi} for more discussion). 
To get by this, we normalize the SVs by $\bm{S}/\text{max}(\bm{S}) \in[1,0)$ and define a tolerance $0 \leq \epsilon \ll 1$, such that
\begin{equation}\label{eq:regualrised_singular_matrix}
    \bm{\Lambda}_{i}^{+} \mapsto \begin{cases}
        \bm{\Lambda}_{i}^{-1} & \bm{S}_{i}/\text{max}(\bm{S}) > \epsilon  \\
        0 &\text{otherwise.}
    \end{cases}
\end{equation}
The ideal choice of $\epsilon$ would be small enough such that maximal information during the tapering scheme is retained \emph{and} the matrix is well behaved. 
\begin{figure*}
    \centering
    \includegraphics[width = \textwidth]{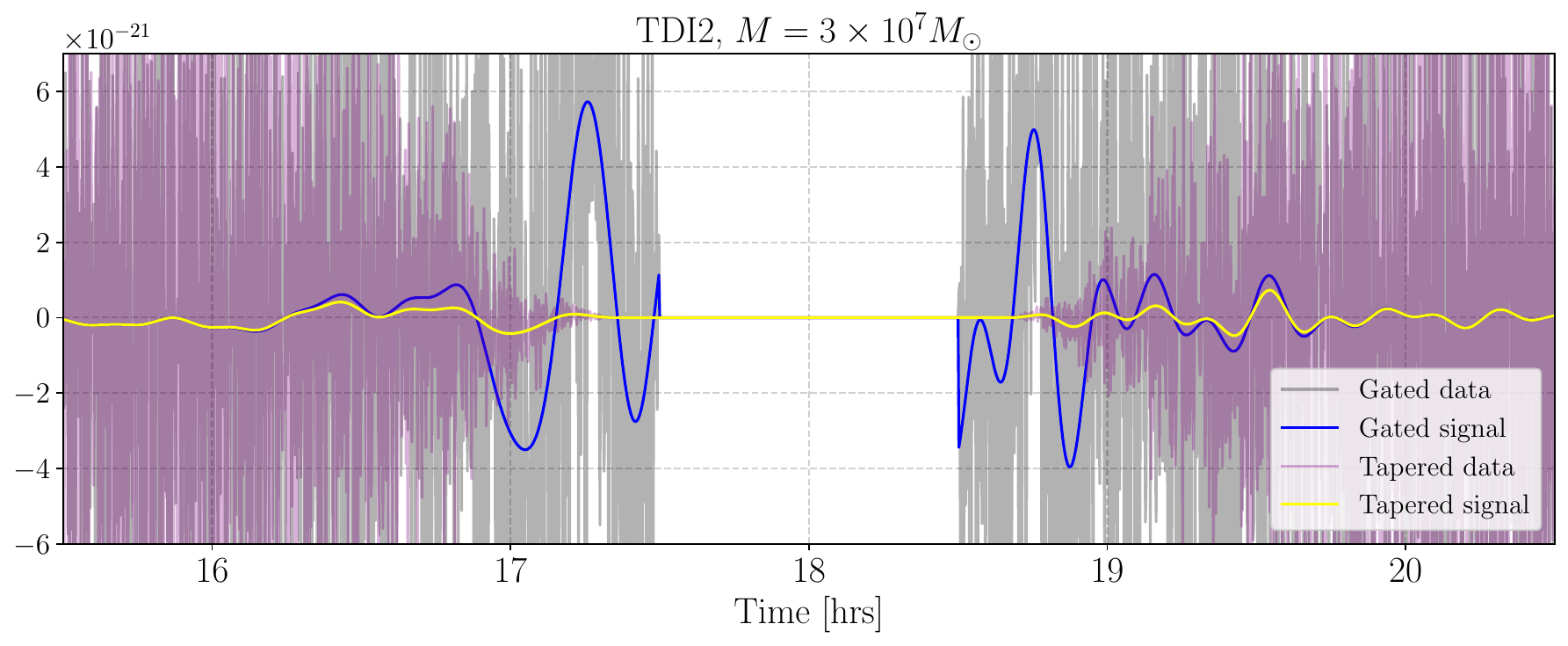}
    \caption{\textbf{(The Blue/Yellow solid curves):} The gated/tapered MBH signal for the M3e7 case with a gap at merger lasting 1 hour. \textbf{(The purple/black curves):} The gated/tapered data stream consisting of noise and the MBH signals. The SNR of the gated signal is $\sim 1397$, whereas the SNR of the tapered signal is $\sim 1393$ using a pseudo-inverse regularized according to Eq.\eqref{eq:regualrised_singular_matrix} with $\epsilon = 10^{-13}$. Assuming accurate computation of the pseudo-inverse of the noise covariance matrix, the SNR should be \emph{invariant} to the choice of tapering function with only loss of SNR occurring during the gated segment. In theory, albeit very counter-intuitive, the SNR of the orange and blue curves above are identical, provided the noise covariance includes the effect of the tapering.}
\label{fig:zoomed_in_taper_plot}
\end{figure*}
In this section, we will demonstrate the implications for the FD analysis if a smooth taper were applied to the TD data. We refer the reader to Fig.~\ref{fig:zoomed_in_taper_plot}, which illustrates our tapering scheme unique to this section. We will study a gap at merger elapsing one hour, with one window function a simple binary mask (gate) and the other a smooth taper which lobe lengths $\sim 2$ hours. The number of zero-data is identical between the two windowing cases. The blue stars in Fig.~\ref{fig:pseudo_inv_vs_SNR} are the result of FD SNR computations via Eq.\eqref{eq:SNRoptgapFD} for different choices of small $\epsilon$ when computing the pseudo-inverse using the regularization procedure outlined in Eq.~\eqref{eq:regualrised_singular_matrix}. The red horizontal dashed line is the SNR of the gated MBH signal given by the blue curve in Fig.~\ref{fig:zoomed_in_taper_plot}. The green horizontal dashed line is the SNR of the tapered MBH when using a diagonal (Whittle) based noise covariance matrix. In the limit as $\epsilon \rightarrow 0^{+}$, the FD SNR tends towards the gated SNR, implying that \emph{all} of the information is retained during the tapering scheme. Notice that the SNR assuming a Whittle-based covariance, with taper applied to the MBH, is significantly less than the SNRs computed using correct pseudo-inverses that correctly account for the windowed noise process. To put this into perspective, the SNR of the orange curve in Fig.~\ref{fig:zoomed_in_taper_plot} is \emph{equivalent} to the SNR of the blue curve in Fig.~\ref{fig:zoomed_in_taper_plot}, thanks to the appropriate incorporation of the window in the covariance itself. If the window was not incorporated into the noise covariance matrix, then this situation is equivalent to analysing a signal that tapers to zero whilst the noise does not, yielding a lower SNR. This validates the claim made around Eq.~\eqref{eq:time_domain_noise_cov_windowed}, that information is only lost during the gated segment of the data stream. Assuming the pseudo-inverse is correctly calculated, information should be retained during the smooth tapering procedure.

\begin{figure*}
    \centering
    \includegraphics[width = \textwidth]{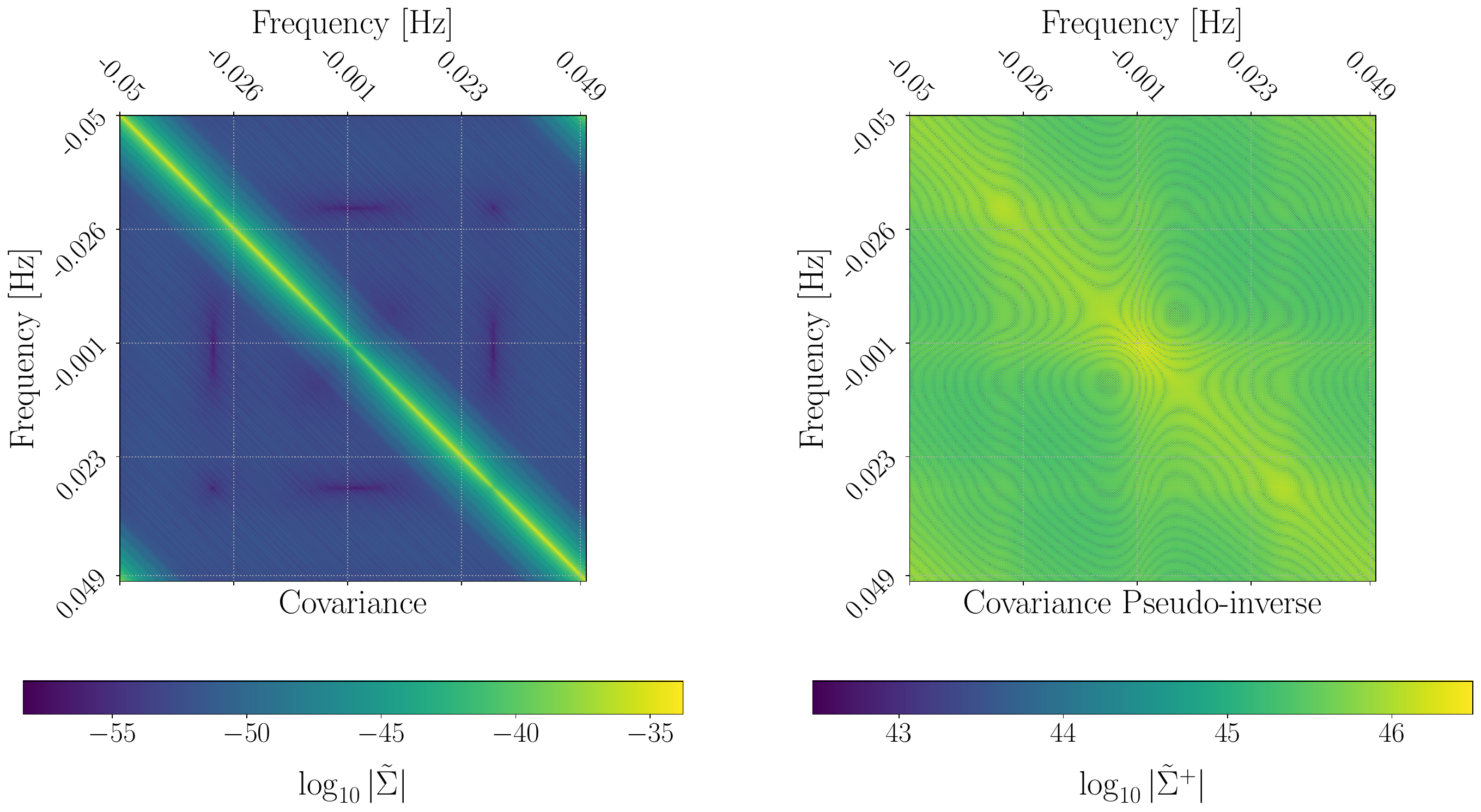}
    \caption{\textbf{(Left Panel):} The FD tapered noise covariance matrix with 1 hour gap at merger with a very smooth $\sim 2$ hours worth of lobes. Notice that the matrix appears to be band-diagonal, with all power concentrated along those specific diagonals. \textbf{(Right Panel):} The \emph{pseudo-inverse} of the matrix presented in the left panel, for a regularization choice of $\epsilon = 10^{-13}$. Aside from hermitian symmetry and psychedelic patterns, there is little exploitable structure to be had in the pseudo-inverse.}
\label{fig:cov_vs_pseudo_inv_eps}
\end{figure*}

To close this section, we plot the tapered FD covariance matrix and corresponding pseudo-inverse (with regularization $\epsilon = 10^{-13}$) in the left and right panels of Fig.~\ref{fig:cov_vs_pseudo_inv_eps}. To add to the middle and right panels of Fig.(\ref{fig:TDI0_td_fd_gap_smooth_harsh} -- \ref{fig:TDI2_td_fd_gap_smooth_harsh}), we observe that the gentler the tapering scheme the more the power is concentrated along the main diagonals. The right-most plot is a plot of the pseudo-inverse, possessing little exploitable structure aside from the fact that the matrix is Hermitian. The pseudo-inverse is a dense matrix, making FD likelihood computations expensive $\sim \mathcal{O}(N^2)$. We are hopeful that there will exist methods that can improve the cost of computing the likelihood in the FD through, perhaps, exploiting band-diagonal nature of $(\tilde{\bm{W}}\tilde{\bm{\Sigma}}\tilde{\bm{W}})$ and accelerating using parallelization techniques on Graphical Processing Units (GPUs). Performance studies of the TD and FD likelihoods are a topic we did not explore in great detail, and we will leave this matter for future work.

Since there are no obvious cost reductions (yet) for analyses conducted in the FD, we will use the TD methods described in this paper to compute quantities presented in the results Sec.~\ref{sec:results}. We have checked that calculations (involving $\overline{\Upsilon}$ and $\overline{\Xi}$ as defined in Eq. \eqref{eq:def_upsilonbar_xibar} remain unchanged whether computations are performed in either the FD or TD.

\section{Results}\label{sec:results}

The first instance of mismodeling that we will consider is the case of the improper use of the Whittle-likelihood, appropriate for Gaussian stationary noise, to analyse a data stream where missing data breaks the stationary feature of the stochastic process. 

The advantage of using the Whittle-likelihood is obvious from the computational standpoint, allowing for $\mathcal{O}(N \log_{2} N)$ likelihood evaluations. Searching for sources in LISA could be done using cheap Whittle-based likelihoods, before further refinements are made with the correct but more expensive likelihoods. We want therefore to ascertain what would be the consequences of using the Whittle-likelihood despite the presence of gaps in the data, and whether introducing a smooth taper around gaps can alleviate noise mismodeling errors.

Here, we will assume that we know exactly the PSD and the covariance $\bm{\Sigma}$ of the underlying noise process. We also assume that the noise process is coherent over the full interval, and the gap acts as a mask, causing data to be missing without any other impact on its statistics. We will come back to the case where the true noise process is independent before and after the gap in Sec.~\ref{subsec:mismodeling_independence}. We will also assume that the $\bm{\Sigma}$ covariance has a circulant structure, and not only Toeplitz, on the full segment:
\be
    \bm{\Sigma} = \bm{\Sigma}_{\rm circ}
\ee
This is not physical, as no real data stream will have this property; we make this choice so that the discrete Whittle-likelihood~\eqref{eq:DiscreteWhittle} is exact before gaps are introduced, in order to isolate other sources of mismodeling from the Toeplitz/circulant mismodeling, to which we will come back in Sec.~\ref{subsec:mismodeling_toeplitz_circulant}. 

\subsection{Mismodeling Whittle on gapped data}\label{subsec:gap_mismodeling_ratios}

\begin{table*}
\begin{tabular}{ll | cccc  cccc  cccc}
\toprule
\multicolumn{2}{c}{\shortstack{\bf Mismodeling $\overline{\Upsilon}$ \\ \bf data gaps}} & \multicolumn{4}{c}{TDI2} & \multicolumn{4}{c}{TDI1} & \multicolumn{4}{c}{TDI0} \\
\cmidrule(lr){3-6} \cmidrule(lr){7-10} \cmidrule(lr){11-14}
& & \multicolumn{2}{c}{Merger} & \multicolumn{2}{c}{Insp.} & \multicolumn{2}{c}{Merger} & \multicolumn{2}{c}{Insp.} & \multicolumn{2}{c}{Merger} & \multicolumn{2}{c}{Insp.} \\
\cmidrule(lr){3-4} \cmidrule(lr){5-6} \cmidrule(lr){7-8} \cmidrule(lr){9-10} \cmidrule(lr){11-12} \cmidrule(lr){13-14}
Data & Model & M3e7 & M3e6 & M3e7 & M3e6 & M3e7 & M3e6 & M3e7 & M3e6 & M3e7 & M3e6 & M3e7 & M3e6 \\
\midrule
Coherent, gap & Whittle, gated     & \cellcolor{red!50} 12.2 & \cellcolor{red!50} 40.5 & \cellcolor{yellow!50} 1.3 & \cellcolor{yellow!50} 1.4 & \cellcolor{yellow!50} 1.2 & \cellcolor{orange!50} 1.6 & \cellcolor{green!50} 1.0 & \cellcolor{green!50} 1.0 & \cellcolor{red!50} 39.7 & \cellcolor{red!50} 38.0 & \cellcolor{red!50} 21.9 & \cellcolor{red!50} 6.4 \\
Coherent, gap & Whittle, taper 10min    & \cellcolor{green!50} 1.0 & \cellcolor{yellow!50} 0.9 & \cellcolor{green!50} 1.0 & \cellcolor{green!50} 1.0 & \cellcolor{green!50} 1.0 & \cellcolor{yellow!50} 0.9 & \cellcolor{green!50} 1.0 & \cellcolor{green!50} 1.0 & \cellcolor{red!50} 8.6 & \cellcolor{orange!50} 3.2 & \cellcolor{red!50} 5.5 & \cellcolor{orange!50} 1.6 \\
Coherent, gap & Whittle, taper 30min    & \cellcolor{green!50} 1.0 & \cellcolor{yellow!50} 0.8 & \cellcolor{green!50} 1.0 & \cellcolor{green!50} 1.0 & \cellcolor{green!50} 1.0 & \cellcolor{yellow!50} 0.8 & \cellcolor{green!50} 1.0 & \cellcolor{green!50} 1.0 & \cellcolor{red!50} 6.1 & \cellcolor{orange!50} 2.0 & \cellcolor{orange!50} 3.0 & \cellcolor{yellow!50} 1.1 \\
\midrule
Coherent, gap & Seg. Whittle, taper 10min   & \cellcolor{yellow!50} 1.1 & \cellcolor{yellow!50} 0.9 & \cellcolor{green!50} 1.0 & \cellcolor{green!50} 1.0 & \cellcolor{green!50} 1.0 & \cellcolor{yellow!50} 0.9 & \cellcolor{green!50} 1.0 & \cellcolor{green!50} 1.0 & \cellcolor{red!50} 10.7 & \cellcolor{orange!50} 2.9 & \cellcolor{red!50} 6.5 & \cellcolor{orange!50} 1.9 \\
Coherent, gap & Seg. Whittle, taper 30min   & \cellcolor{green!50} 1.0 & \cellcolor{yellow!50} 0.8 & \cellcolor{green!50} 1.0 & \cellcolor{green!50} 1.0 & \cellcolor{green!50} 1.0 & \cellcolor{yellow!50} 0.8 & \cellcolor{green!50} 1.0 & \cellcolor{green!50} 1.0 & \cellcolor{red!50} 6.6 & \cellcolor{orange!50} 2.0 & \cellcolor{orange!50} 3.6 & \cellcolor{yellow!50} 1.4 \\
\bottomrule
\end{tabular}
\caption{\textbf{(From left to right):} The ``Data" column describes how the data stream was generated and should be regarded as the truth. The ``model" column describes the approximate likelihood used in the model. The next three main columns indicate which TDI noise model is used, TDI2/1/0 respectively. The sub-columns distinguish between a gap placement during the merger or inspiral, and between high and low-mass MBHBs with $M=3e6M_\odot$ or $M=3e7M_\odot$. We refer to Fig.~\ref{fig:signal_with_gaps_tdi2_M3e7_M3e6} for an illustration of the signal waveform and gap placement. \textbf{(From top to bottom):} ``Data" is always the same here: the data is first treated as coherent (and circulant) over the full segment, and then a gap is introduced. ``Model'':  ``Whittle" implies that the full-segment Whittle covariance is used as in~\ref{subsec:gap_whittle}, effectively ignoring the gap in the covariance, while either gating or tapering the data with a 10 or 30 minute lobe length respectively. ``Seg. Whittle'' means that a segmented Whittle likelihood is used, treating data segments before and after the gap as independent, as in Sec.~\ref{subsec:gap_segmented_whittle}; tapering is applied at both ends of each segment. \textbf{(Table entries):} individual numerical results are the scatter-to-width ratio $\overline{\Upsilon}$~\eqref{eq:def_Upsilon}, averaged over parameters as in~\eqref{eq:def_upsilonbar_xibar}. $\overline{\Upsilon}$ measures statistical inconsistency, $\overline{\Upsilon} \gg 1$ shows posteriors that are scattered from the truth much more than their typical width. The colors green/yellow/orange/red indicate the magnitude of the deviation from 1 on a scale of acceptable (green) to unacceptable (red).}
\label{tab:mismodeling_upsilon_gap}
\end{table*}

\begin{table*}
\begin{tabular}{ll | cccc  cccc  cccc}
\toprule
\multicolumn{2}{c}{\shortstack{\bf Mismodeling $\overline{\Xi}$ \\ \bf data gaps}} & \multicolumn{4}{c}{TDI2} & \multicolumn{4}{c}{TDI1} & \multicolumn{4}{c}{TDI0} \\
\cmidrule(lr){3-6} \cmidrule(lr){7-10} \cmidrule(lr){11-14}
& & \multicolumn{2}{c}{Merger} & \multicolumn{2}{c}{Insp.} & \multicolumn{2}{c}{Merger} & \multicolumn{2}{c}{Insp.} & \multicolumn{2}{c}{Merger} & \multicolumn{2}{c}{Insp.} \\
\cmidrule(lr){3-4} \cmidrule(lr){5-6} \cmidrule(lr){7-8} \cmidrule(lr){9-10} \cmidrule(lr){11-12} \cmidrule(lr){13-14}
Data & Model & M3e7 & M3e6 & M3e7 & M3e6 & M3e7 & M3e6 & M3e7 & M3e6 & M3e7 & M3e6 & M3e7 & M3e6 \\
\midrule
Coherent, gap & Whittle, gated     & \cellcolor{yellow!50} 0.8 & \cellcolor{orange!50} 0.5 & \cellcolor{green!50} 1.0 & \cellcolor{green!50} 1.0 & \cellcolor{green!50} 1.0 & \cellcolor{yellow!50} 0.9 & \cellcolor{green!50} 1.0 & \cellcolor{green!50} 1.0 & \cellcolor{yellow!50} 0.9 & \cellcolor{yellow!50} 0.9 & \cellcolor{green!50} 1.0 & \cellcolor{green!50} 1.0\\
Coherent, gap & Whittle, taper 10min    & \cellcolor{green!50} 1.0 & \cellcolor{orange!50} 1.7 & \cellcolor{green!50} 1.0 & \cellcolor{green!50} 1.0 & \cellcolor{yellow!50} 1.1 & \cellcolor{orange!50} 1.8 & \cellcolor{green!50} 1.0 & \cellcolor{green!50} 1.0 & \cellcolor{yellow!50} 1.1 & \cellcolor{orange!50} 1.7 & \cellcolor{green!50} 1.0 & \cellcolor{green!50} 1.0 \\
Coherent, gap & Whittle, taper 30min    & \cellcolor{yellow!50} 1.2 & \cellcolor{orange!50} 3.9 & \cellcolor{green!50} 1.0 & \cellcolor{green!50} 1.0 & \cellcolor{yellow!50} 1.3 & \cellcolor{orange!50} 3.9 & \cellcolor{green!50} 1.0 & \cellcolor{green!50} 1.0 & \cellcolor{yellow!50} 1.3 & \cellcolor{orange!50} 3.7 & \cellcolor{green!50} 1.0 & \cellcolor{green!50} 1.0 \\
\midrule
Coherent, gap & Seg. Whittle, taper 10min   & \cellcolor{green!50} 1.0 & \cellcolor{orange!50} 1.8 & \cellcolor{green!50} 1.0 & \cellcolor{green!50} 1.0 & \cellcolor{yellow!50} 1.1 & \cellcolor{orange!50} 1.8 & \cellcolor{green!50} 1.0 & \cellcolor{green!50} 1.0 & \cellcolor{yellow!50} 1.1 & \cellcolor{orange!50} 1.8 & \cellcolor{green!50} 1.0 & \cellcolor{green!50} 1.0 \\
Coherent, gap & Seg. Whittle, taper 30min   & \cellcolor{yellow!50} 1.2 & \cellcolor{orange!50} 4.0 & \cellcolor{green!50} 1.0 & \cellcolor{green!50} 1.0 & \cellcolor{yellow!50} 1.3 & \cellcolor{orange!50} 3.9 & \cellcolor{green!50} 1.0 & \cellcolor{green!50} 1.0 & \cellcolor{yellow!50} 1.3 & \cellcolor{orange!50} 3.8 & \cellcolor{green!50} 1.0 & \cellcolor{green!50} 1.0 \\
\bottomrule
\end{tabular}
\caption{Identical layout as Tab.\ref{tab:mismodeling_upsilon_gap} except that $\overline{\Xi}$ is shown, describing the average ratio between the true and model posterior widths, defined by Eqs.~\eqref{eq:def_Xi} and~\eqref{eq:def_upsilonbar_xibar}.}
\label{tab:mismodeling_xi_gap}
\end{table*}

In order to assess the impact of noise mismodeling in the presence of data gaps with a Whittle-based approximation, we start by computing the mismodeling ratios defined in Eq.~\eqref{eq:def_Xi} and Eq.~\eqref{eq:def_Upsilon}, for the different systems and TDI/gap configurations described in Sec.~\ref{sec:application_MBHB}. Results are reported for $\Upsilon$ in Tab.~\ref{tab:mismodeling_upsilon_gap} and for $\Xi$ in Tab.~\ref{tab:mismodeling_xi_gap} with layout of the tables described in Tab.~\ref{tab:mismodeling_upsilon_gap}. With independent codes developed specific to the FD, we have verified results found in the the top row of Tab.~\ref{tab:mismodeling_upsilon_gap} and Tab.~\ref{tab:mismodeling_xi_gap}.

The columns in those tables give the result for the different generations TDI2, TDI1 and TDI0 as introduced in Sec.~\ref{sec:application_MBHB}. The expected outcome of the L0/L1 noise reduction pipeline will be TDI2~\cite{LISA:2024hlh}, but comparing with noise processes corresponding to TDI1 and TDI0 allows us to understand the effect of the shape of the PSD, from blue-noise dominated for TDI2, to featuring both blue and red noise for TDI1, and to strongly red-noise dominated for TDI0 (as illustrated in Fig.~\ref{fig:noise_curves}). In each case we show the results for both gap configurations, during the inspiral or right at merger, and for both MBHB masses, $M=3\times 10^7 M_\odot$ (M3e7) and $M=3\times 10^6 M_\odot$ (M3e6).

The first three lines of Tab.~\ref{tab:mismodeling_upsilon_gap} and  Tab.~\ref{tab:mismodeling_xi_gap} show the result for the approach of Sec.~\ref{subsec:gap_whittle}, where we use a Whittle-likelihood for the full segment. The first line shows the result without a taper (the window is a direct gating function, 1 or 0), the second and third line for a tapering window as defined in Eq.~\eqref{eq:def_planck_window}, with lobe lengths of 10 minutes and 30 minutes respectively.

We first observe that $\overline{\Upsilon}$ is always much larger than 1 for TDI0, catastrophically so as $\Upsilon$ is counted in units of $\sigma$. Tapering helps in reducing these values significantly but does not eliminate the problem, with a slightly better performance of the longer tapering lobes. We attribute this to the fact that TDI0 is very red-noise dominated, with long-range correlations encoded in the high values of the PSD at low frequencies, which make the effect of gap mismodeling much more pronounced. The values of $\overline{\Xi}$, however, are close to 1 for the gap in the inspiral in all cases, while raising above 1 for the gap at merger when tapering is introduced.

At this point, we wish to raise a word of caution. In GW astronomy, it is common to use sky-averaged sensitivity curves~\cite{Robson:2018ifk} for SNR computations and simplified parameter estimation purposes with Fisher matrices. Sensitivity curves incorporating the transfer function into the PSDs can result in large red-noise components, worse than our TDI0 example. For such a dominant red-noise process, in early explorations we found  scatter-to-width ratios that were very severe ($\overline{\Upsilon}\gg 1$), but ratios of posteriors widths that were mild ($\overline{\Xi} \sim 1$). This indicates that exploratory studies investigating SNRs and Fisher Matrices may be fine but PE studies with simulated noise would become problematic when using sensitivity curve-based noise, with a poor representation of the impact of gaps compared to realistic TDI2 data.

Focusing on the gap at merger, we see that treating the gap through a naive gating window (with no taper) gives unfavorable $\Upsilon$ values for TDI2. The gating window case is not as bad for TDI1, with a difference that may be attributed to the stronger blue noise component of TDI2. However, introducing tapering  significantly limits the issue, with $\Upsilon$ close to 1 for the high-mass system M3e7, while the lower-mass system M3e6 has values below 1 indicating that posteriors are too broad with respect to their scatter. The values of $\Xi$ are lower than 1 for the gating window, indicating underestimated posterior widths. With tapering, the values of $\Xi$ rise above 1, particularly for the M3e6 case, indicating that tapering, if it limits statistical inconsistency by decreasing $\Upsilon$, worsens information loss by increasing $\Xi$. 

The gap in the inspiral, by contrast, induces much weaker mismodeling errors for TDI2 and TDI1, with values for both $\Upsilon$ and $\Xi$ that are close to 1 in all cases, except when using a gating window in the TDI2 case for $\Upsilon$. This is an important and positive result: the inspiral gap represents a more typical configuration, while the gap at merger represents a worst-case scenario.

The last two lines of Tab.~\ref{tab:mismodeling_upsilon_gap} and  Tab.~\ref{tab:mismodeling_xi_gap} correspond to the segmented Whittle approach of Sec.~\ref{subsec:gap_segmented_whittle}. Although they are two parts of a coherent noise process realization, the data segments before and after the gap are treated as independent, and we use a Whittle-likelihood on each of those segments, with a tapering at each end of the segments. The two lines show the results for a lobe length of 10 minutes and 30 minutes respectively.

The behavior of this segmented Whittle model is very similar to the full-data Whittle model across all configurations, with the same qualitative behavior in both $\Upsilon$ and $\Xi$ when using the same tapering lengths. This points towards a limited, or even negligible, impact of correlations across the gaps, with independence being a safe approximation to make -- we will directly test for this in Sec.~\ref{subsec:mismodeling_independence} below. 

\subsection{Parameter estimation examples}\label{subsec:Parameter_Estimation}

\begin{figure*}
    \centering
    \begin{minipage}{0.5\textwidth}
        \centering
        \includegraphics[width = \textwidth]{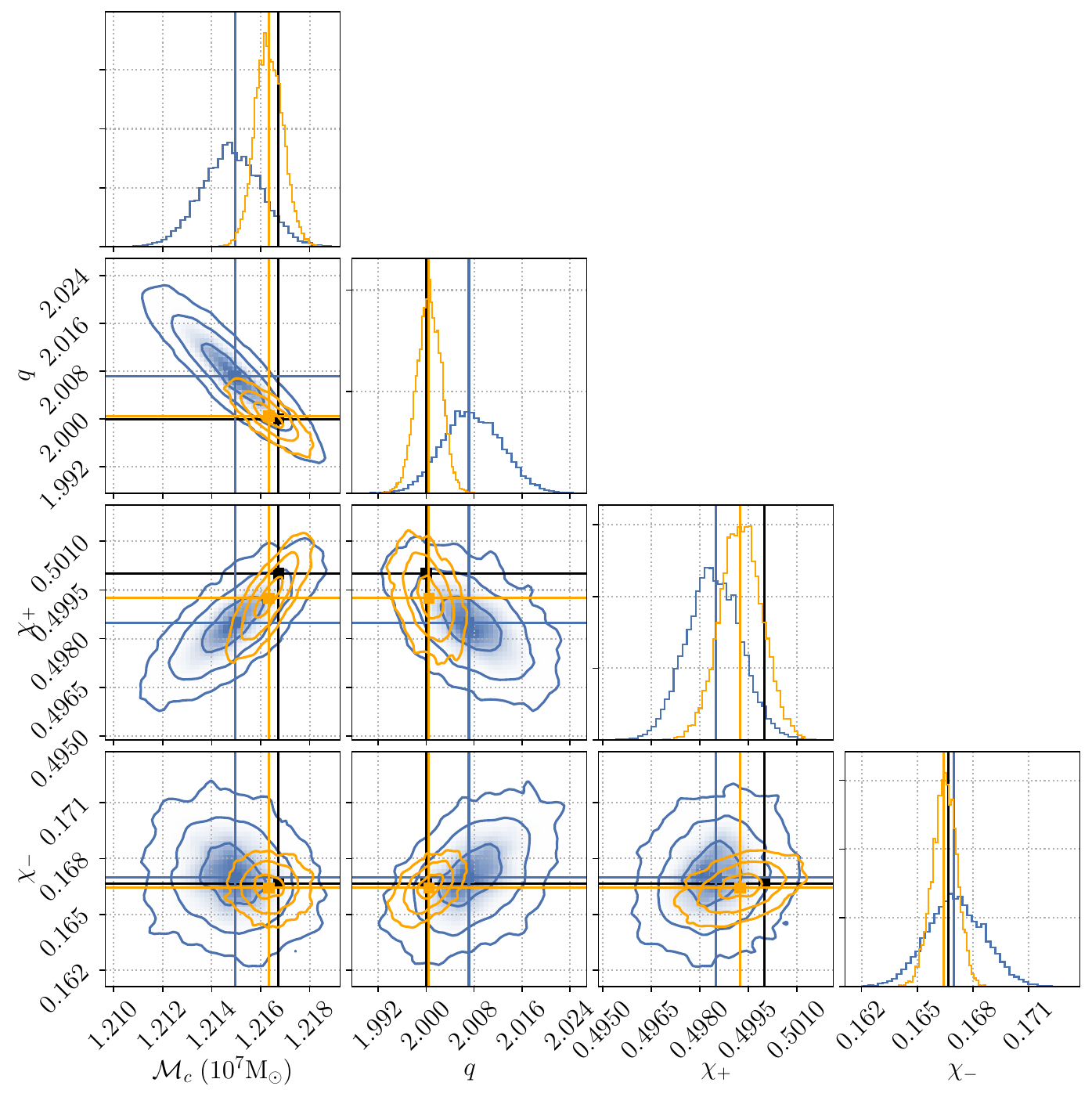}
    \end{minipage}%
    \begin{minipage}{0.5\textwidth}
        \centering
        \includegraphics[width = \textwidth]{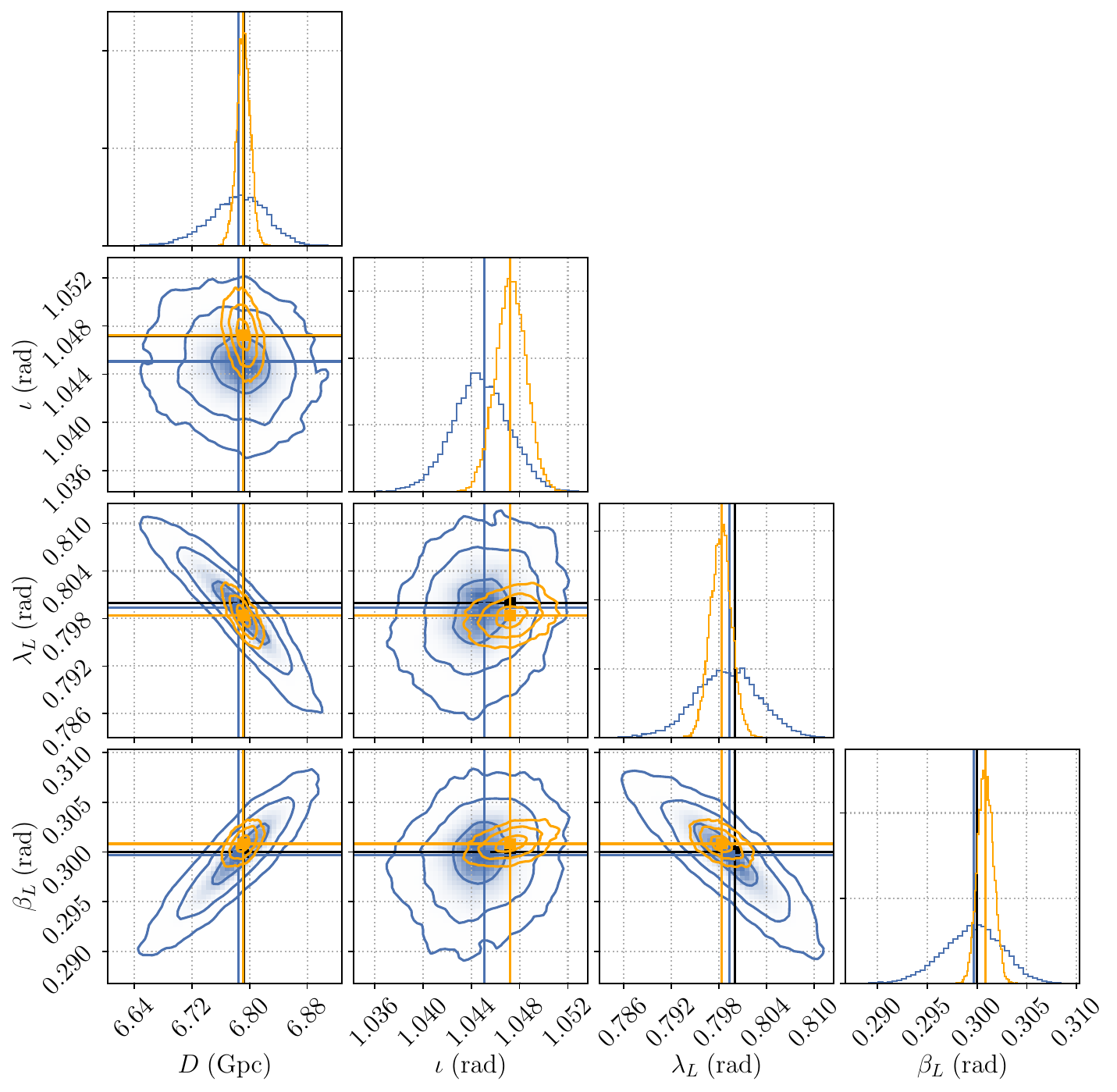}
    \end{minipage}
    \begin{minipage}{0.5\textwidth}
        \centering
        \includegraphics[width = \textwidth]{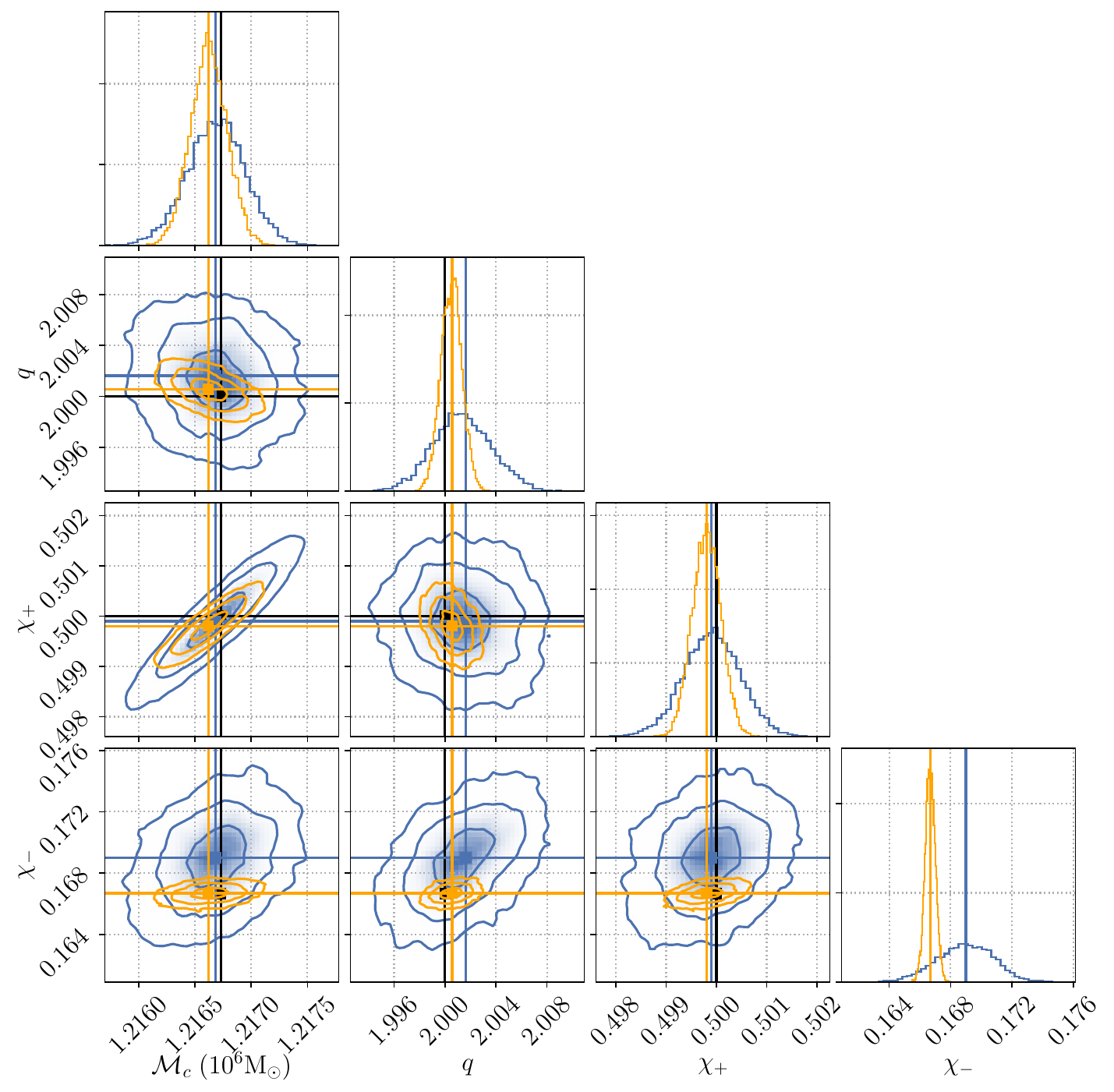}
    \end{minipage}%
    \begin{minipage}{0.5\textwidth}
        \centering
        \includegraphics[width = \textwidth]{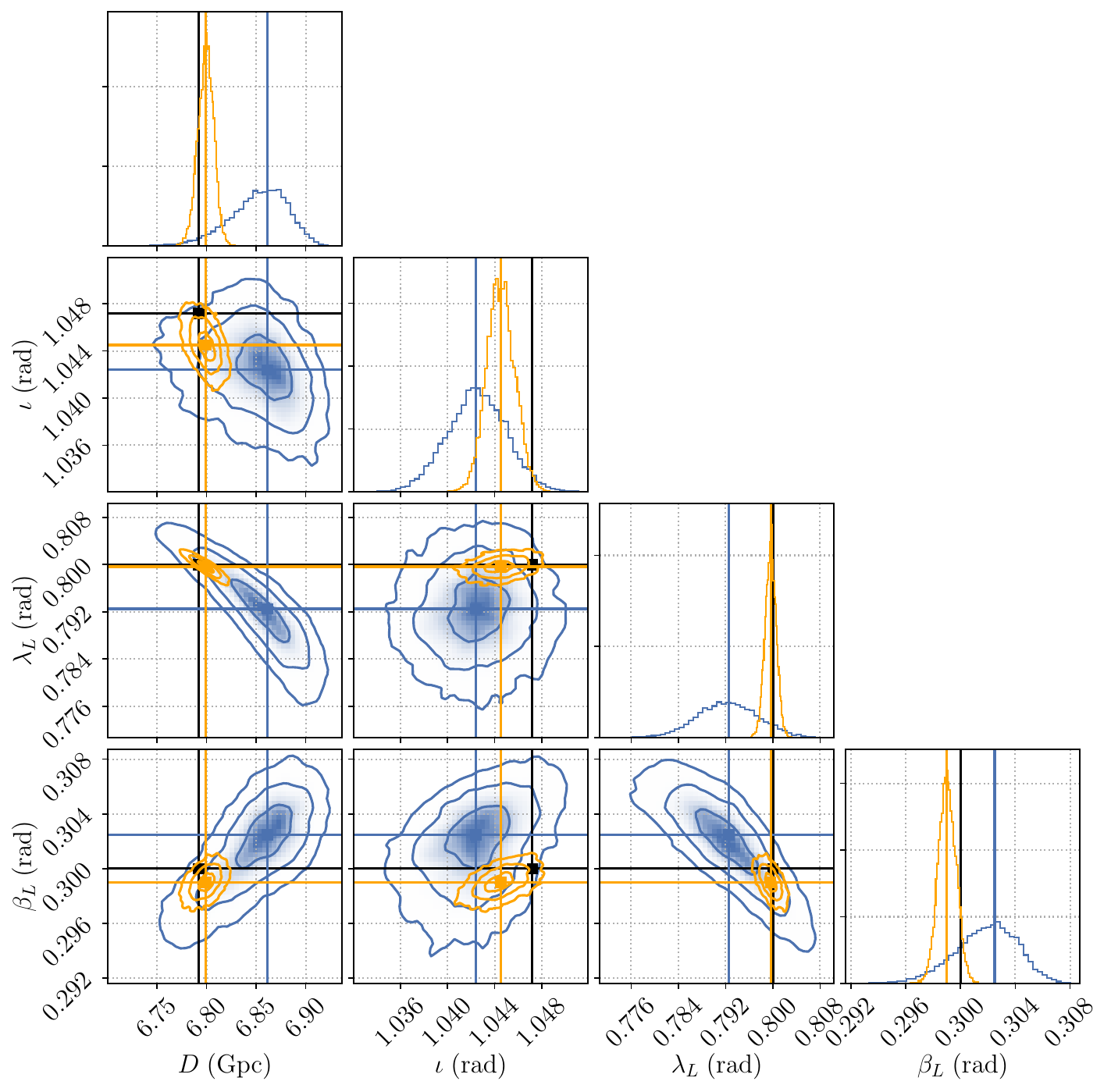}
    \end{minipage}
    \caption{\textbf{(Top Row/Bottom Row)} Parameter estimation results for the case $\text{M3e7/M3e6}$, gap at merger. The inference is done on the full parameter space but we show only sub-corner plots for the intrinsic parameters on the left and extrinsic parameters on the right. The black vertical and horizontal lines indicate the true parameters. The yellow posteriors show for reference the result of parameter estimation on a data stream without any gap using the Whittle-based likelihood (which is then the correct likelihood by assumption). The blue posteriors were generated by performing inference on a data stream with gaps during the merger phase when using a model that is statistically consistent with the data generation process, as described in Sec.~\ref{sec:methods}. The blue and yellow lines are the corresponding Fisher-based predictions of the MLE.}
    \label{fig:corner_correct_gap_vs_nogap}
\end{figure*}

\begin{figure*}
    \centering
    \begin{minipage}{0.5\textwidth}
        \centering
        \includegraphics[width = \textwidth]{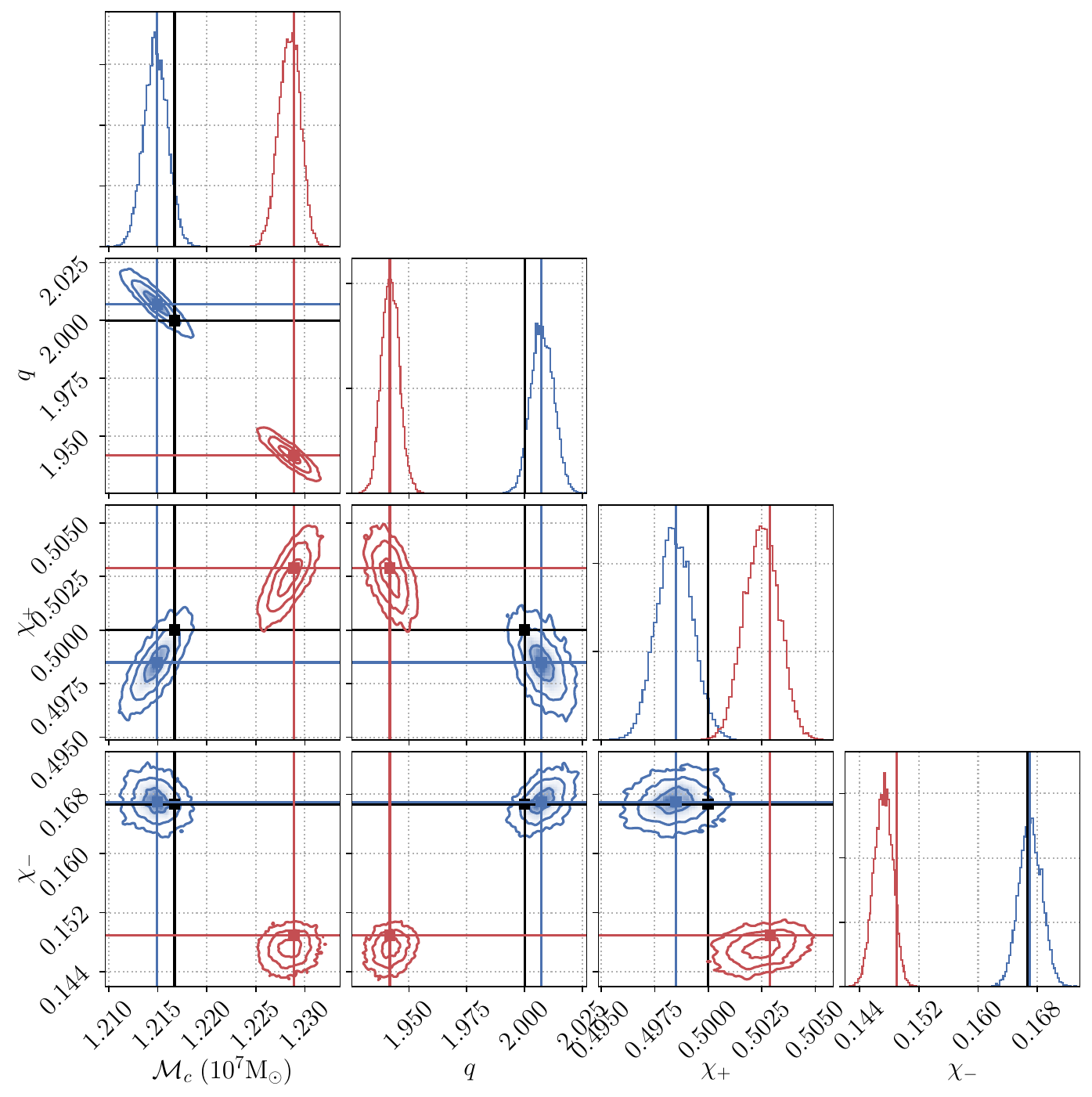}
    \end{minipage}%
    \begin{minipage}{0.5\textwidth}
        \centering
        \includegraphics[width = \textwidth]{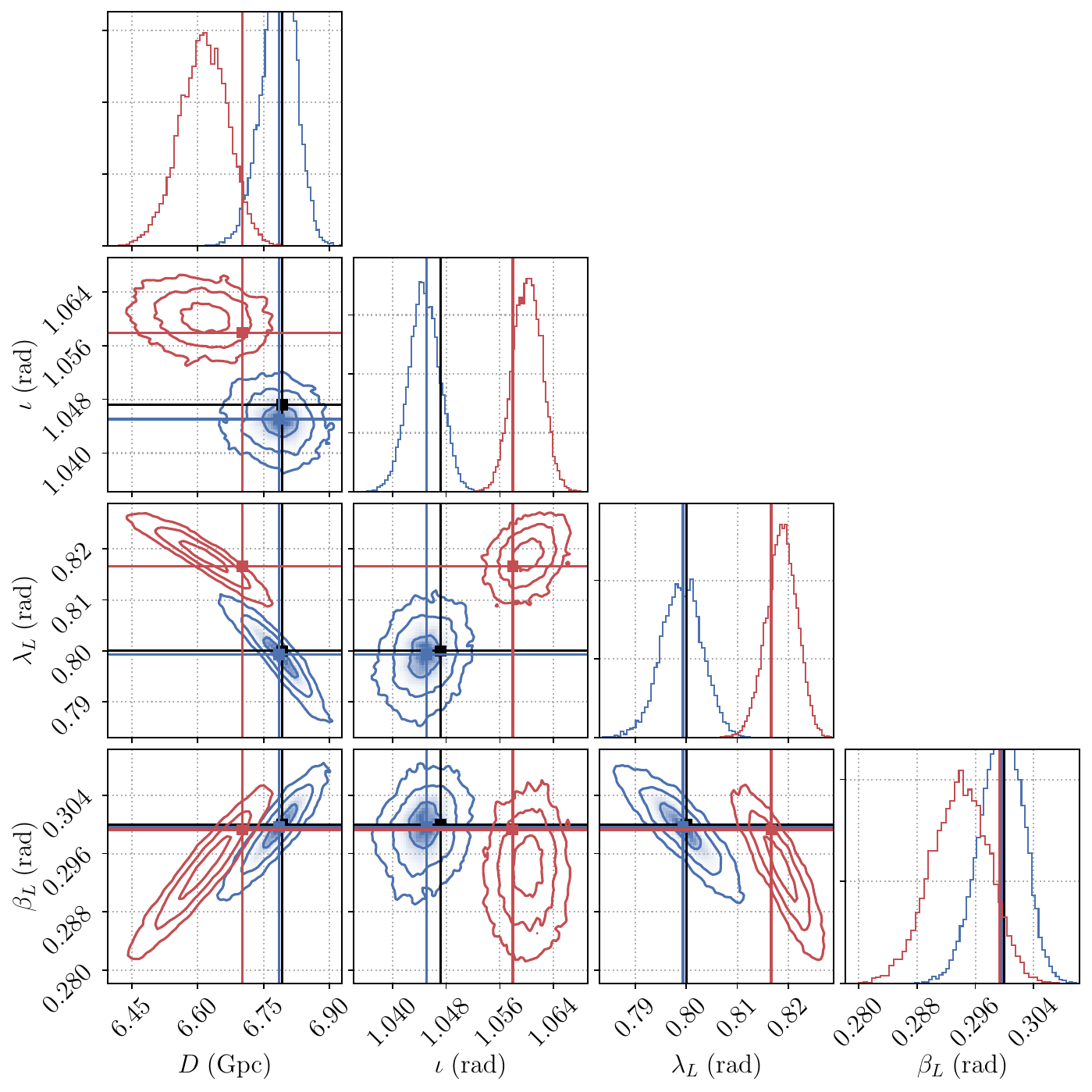}
    \end{minipage}
    \begin{minipage}{0.5\textwidth}
        \centering
        \includegraphics[width = \textwidth]{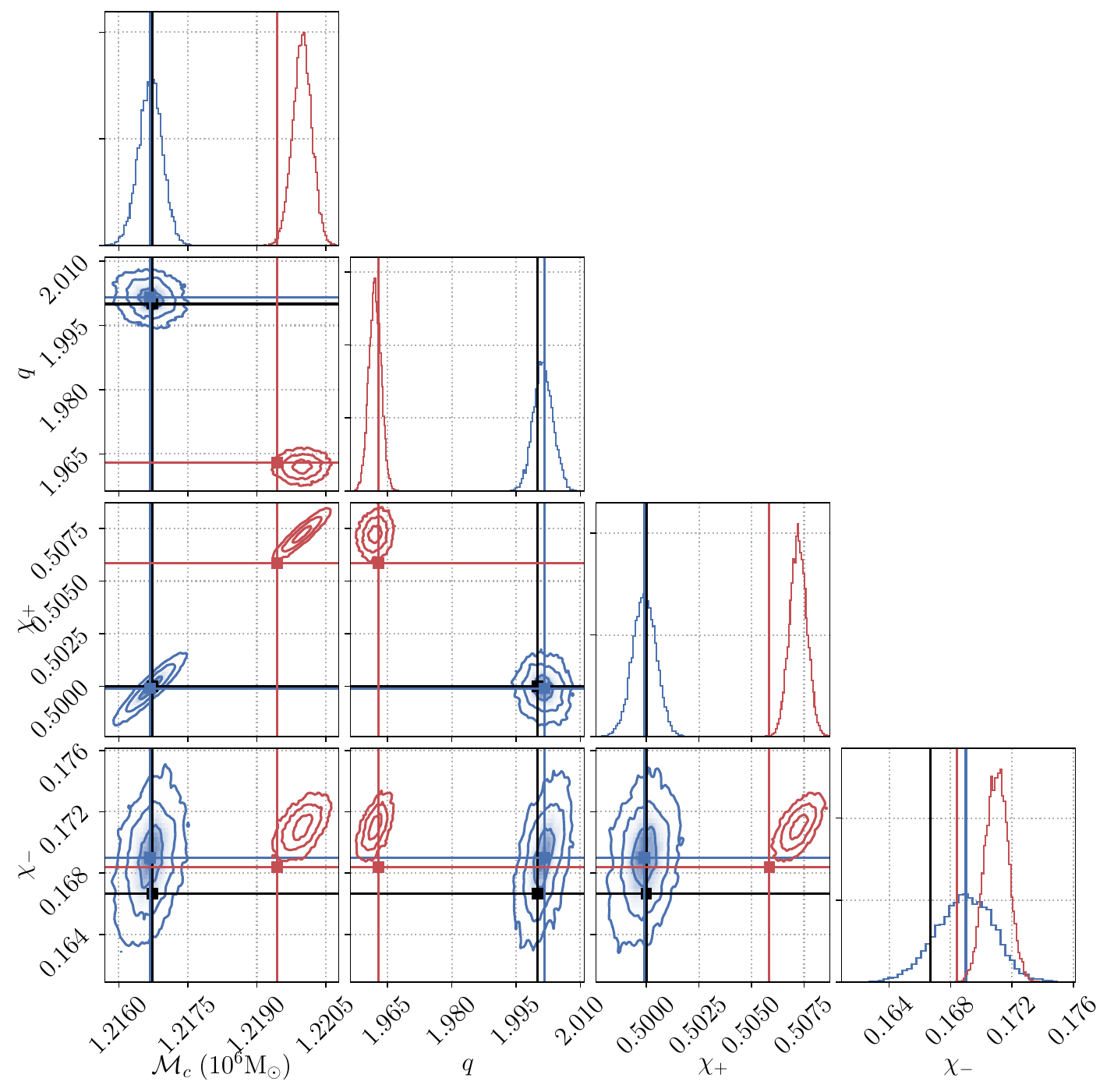}
    \end{minipage}%
    \begin{minipage}{0.5\textwidth}
        \centering
        \includegraphics[width = \textwidth]{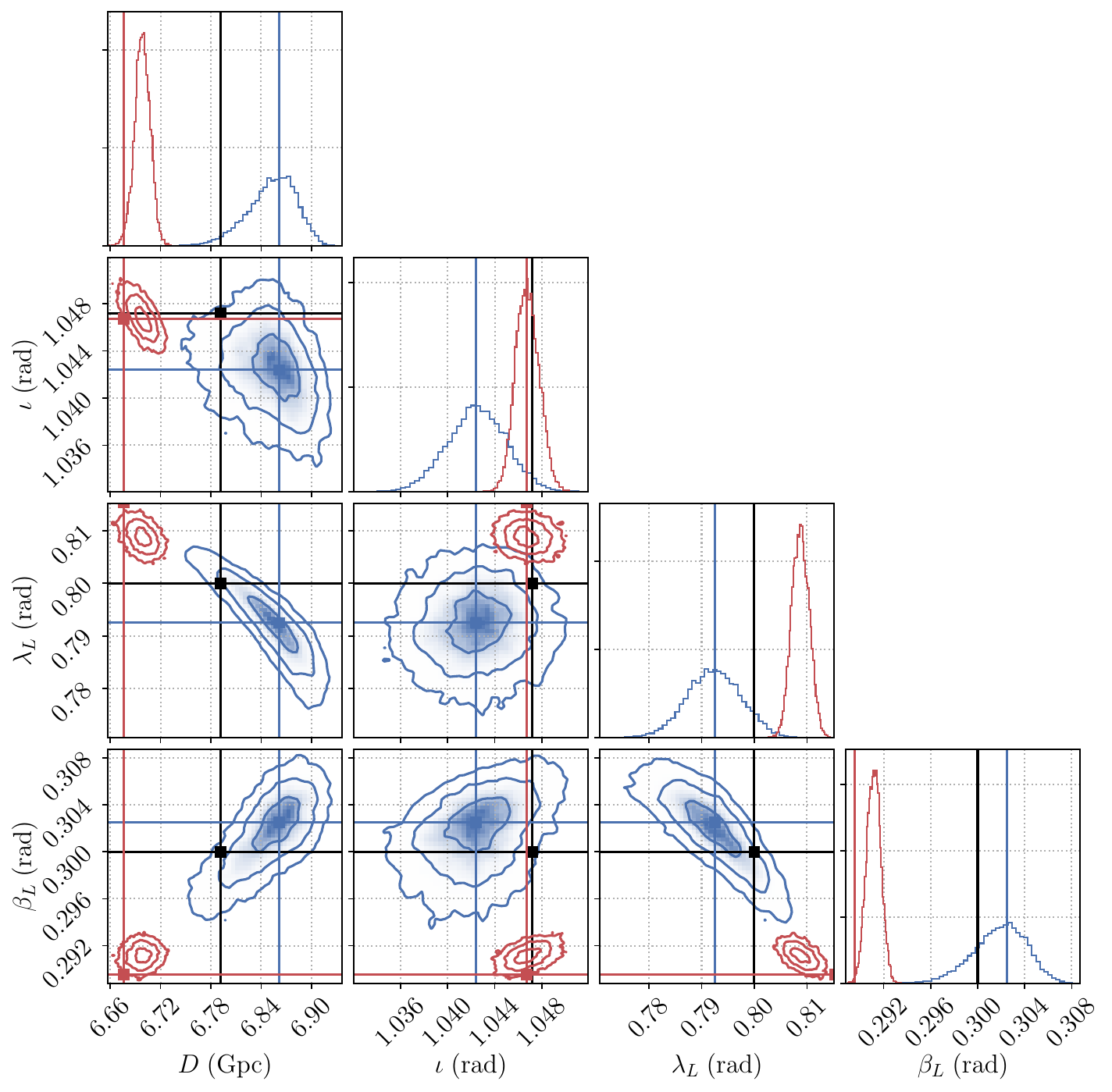}
    \end{minipage}
    \caption{A similar configuration to figure \ref{fig:corner_correct_gap_vs_nogap}. The blue posteriors are the result of parameter estimation on a data stream with gaps using the correct gated likelihood. The red posteriors were generated by performing inference on a data stream with gaps during the merger phase when gating the data (no tapering) and using the Whittle-likelihood of Sec.~\ref{subsec:gap_whittle}, which is statistically inconsistent with the data generating process.}
    \label{fig:corner_correct_vs_whittle}
\end{figure*}

\begin{figure*}
    \centering
    \begin{minipage}{0.5\textwidth}
        \centering
        \includegraphics[width = \textwidth]{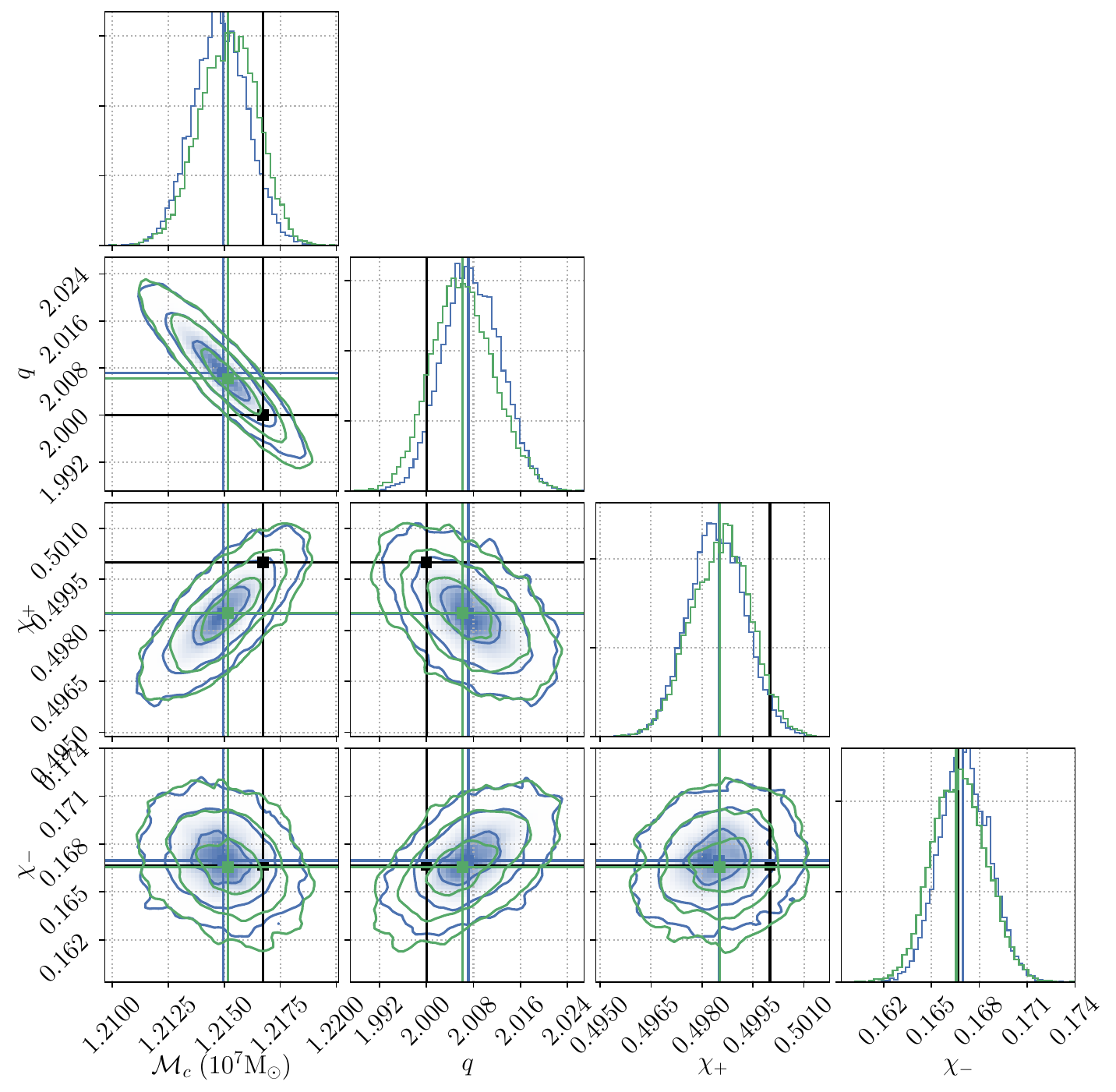}
    \end{minipage}%
    \begin{minipage}{0.5\textwidth}
        \centering
        \includegraphics[width = \textwidth]{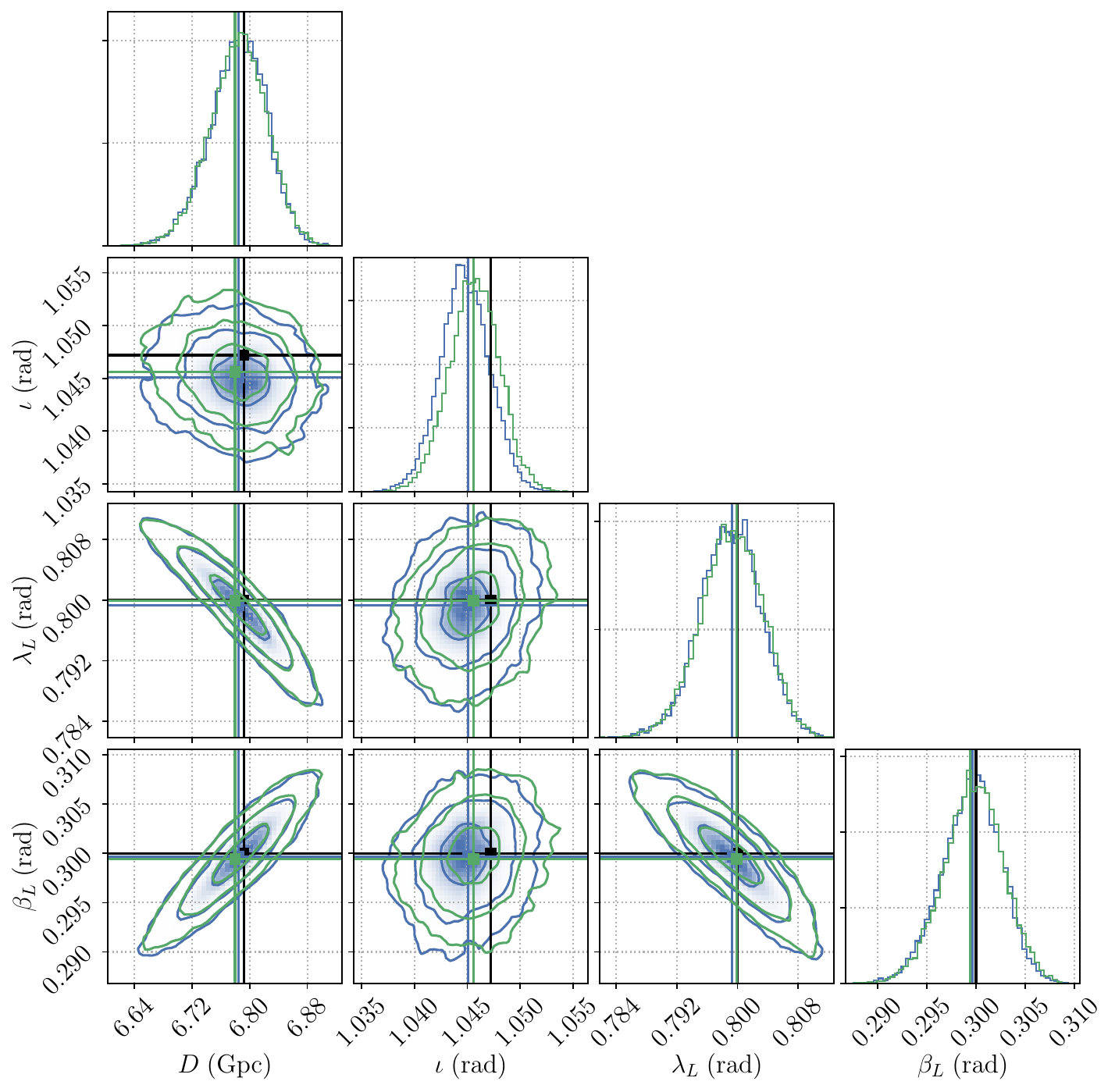}
    \end{minipage}
    \begin{minipage}{0.5\textwidth}
        \centering
        \includegraphics[width = \textwidth]{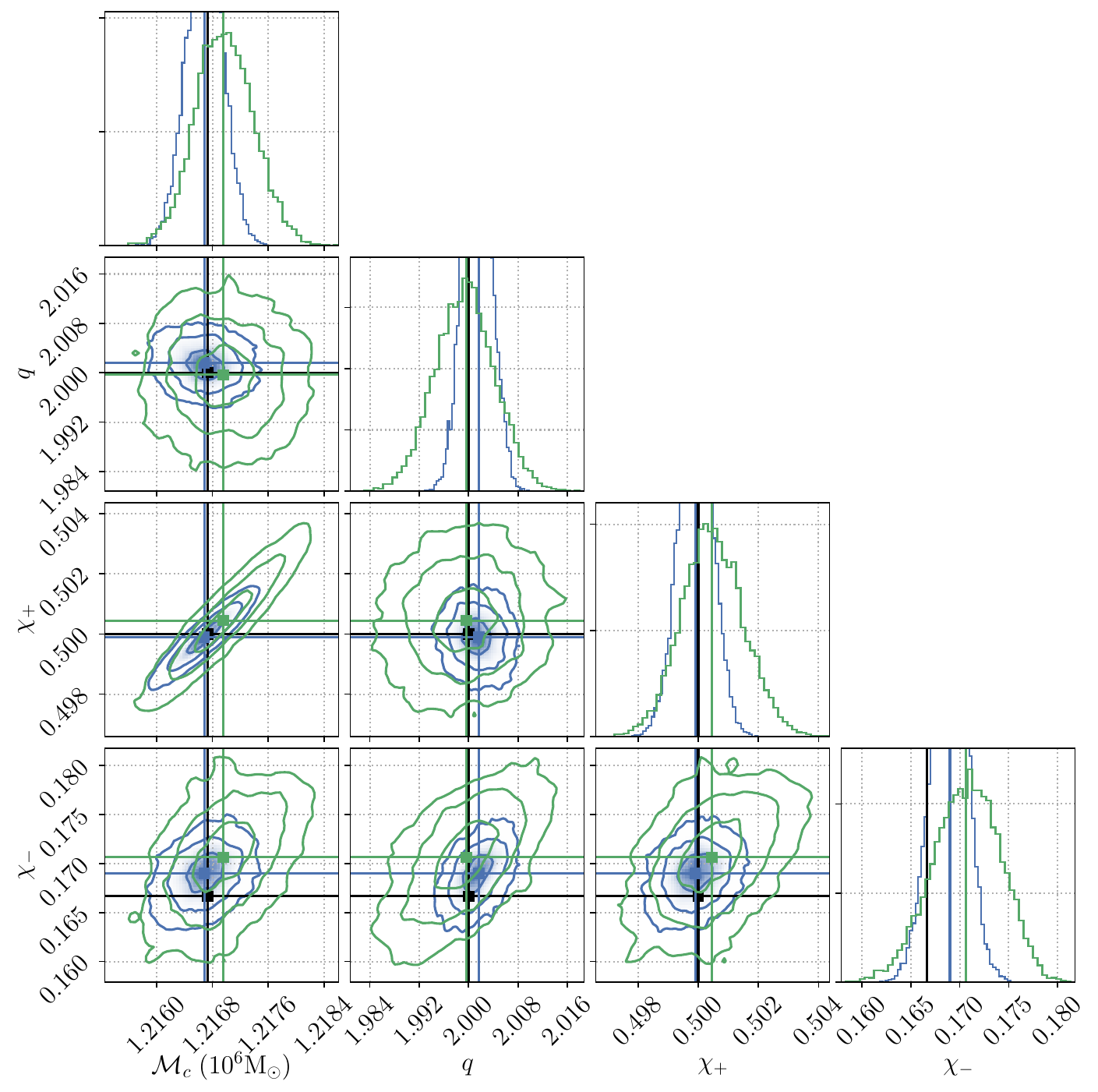}
    \end{minipage}%
    \begin{minipage}{0.5\textwidth}
        \centering
        \includegraphics[width = \textwidth]{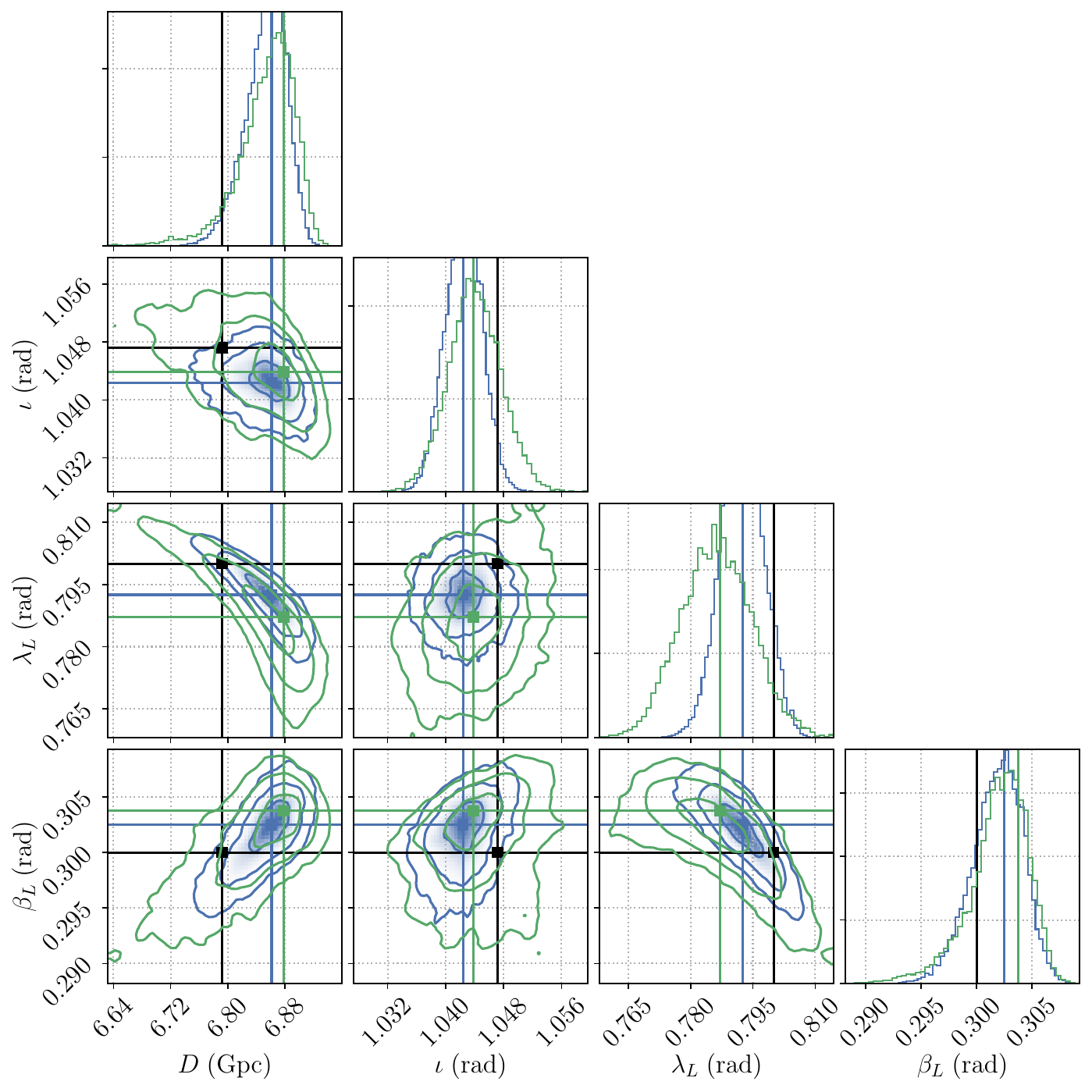}
    \end{minipage}
    \caption{The same set up as figure~\ref{fig:corner_correct_vs_whittle} but applying a 10 minute taper before and after the gap segment when using the approximate Whittle-likelihood of Sec.~\ref{subsec:gap_whittle}, represented by the green posterior. The blue posterior, as before, represents a truthful run where the gated data stream is inferred with a likelihood built from a gated covariance.}
    \label{fig:corner_correct_vs_whittle_taper10min}
\end{figure*}

\begin{figure*}
    \centering
    \begin{minipage}{0.5\textwidth}
        \centering
        \includegraphics[width = \textwidth]{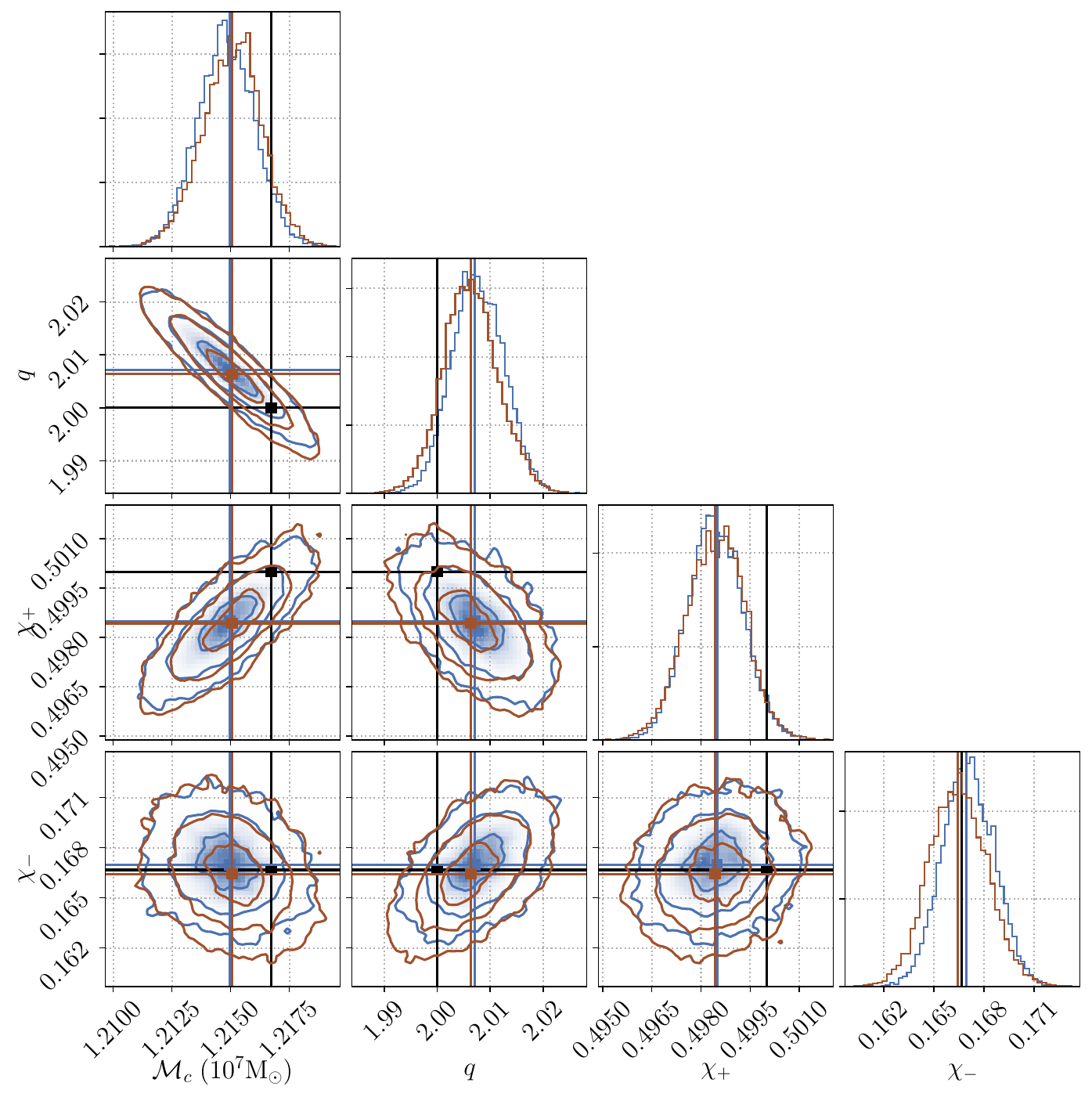}
    \end{minipage}%
    \begin{minipage}{0.5\textwidth}
        \centering
        \includegraphics[width = \textwidth]{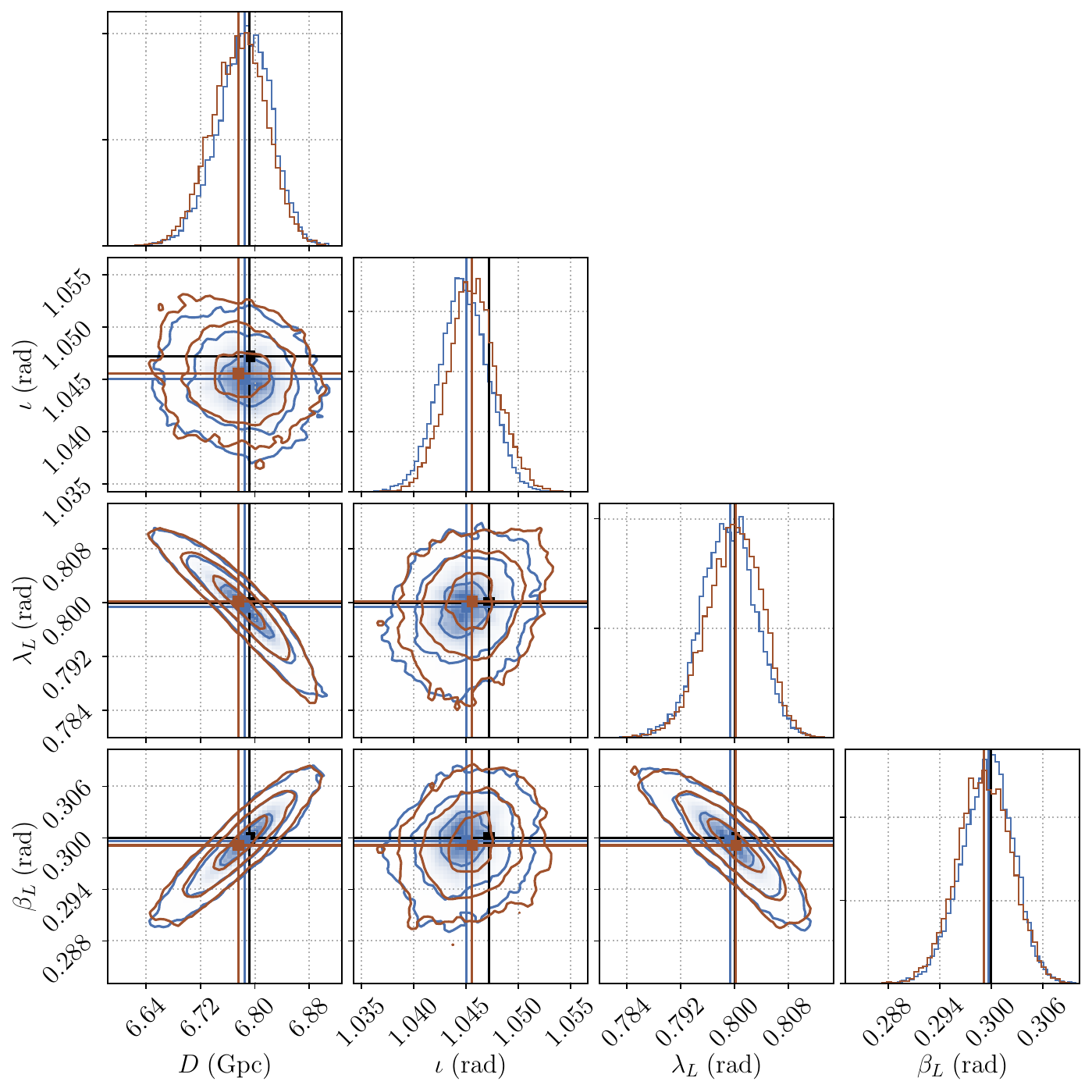}
    \end{minipage}
    \begin{minipage}{0.5\textwidth}
        \centering
        \includegraphics[width = \textwidth]{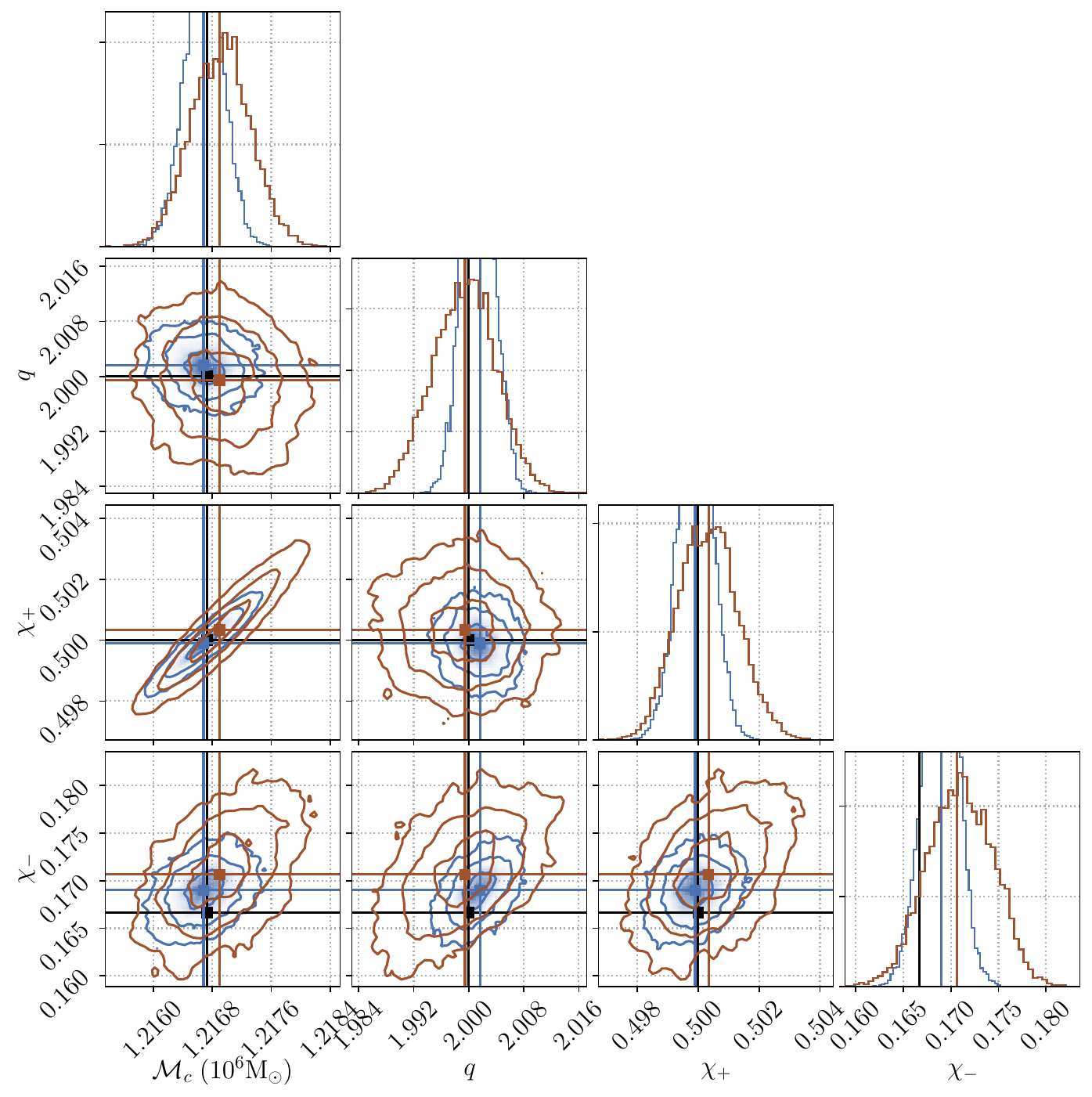}
    \end{minipage}%
    \begin{minipage}{0.5\textwidth}
        \centering
        \includegraphics[width = \textwidth]{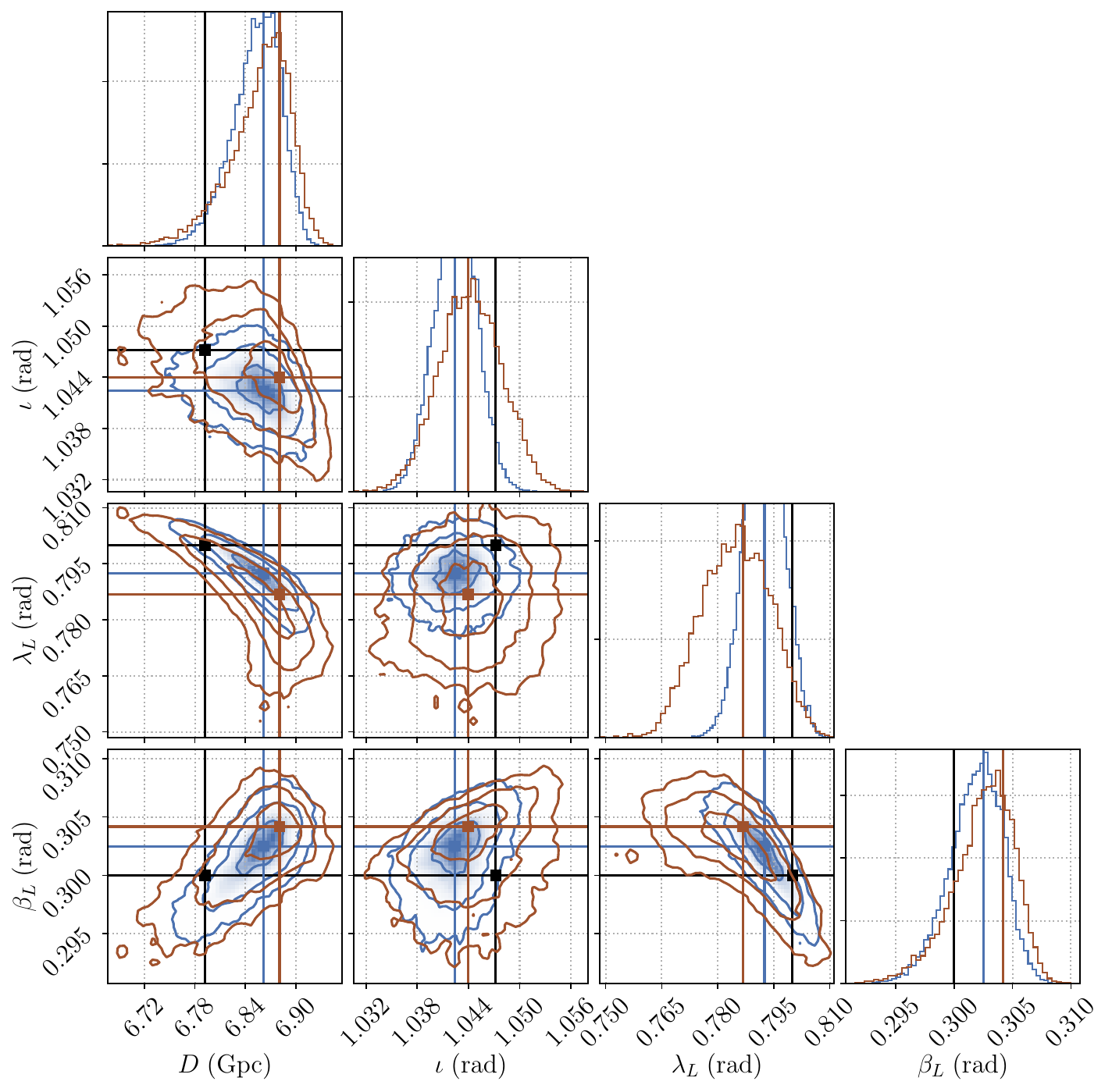}
    \end{minipage}
    \caption{The same set up as Fig.\ref{fig:corner_correct_vs_whittle_taper10min} but  where the model likelihood, represented by the brown posterior, is instead the segmented Whittle model of Sec.~\ref{subsec:gap_segmented_whittle}, and the independence between segments is assumed. A smoothing taper of 10 minutes is applied to both ends of each segment. The blue posterior, as before, represents the truthful situation where a gated likelihood is used to perform inference on a gated data stream.}
    \label{fig:corner_correct_vs_segmented_taper10min}
\end{figure*}

The mismodeling ratios $\Upsilon$ and $\Xi$ are useful to paint a qualitative picture of mismodeling errors in a synthetic manner, but they are based on the FM approach, relying on a linearized signal approximation around the true parameters. It is therefore important to validate this approach with full Bayesian PE runs, which will also give us a better illustration of the full posteriors than the Gaussian approximation with a Fisher based covariance.

The PE runs are performed with the TD implementation of the likelihoods, using the prescription outlined in Sec.~\ref{subsec:analysis_set_up} with likelihood given by Eq.\eqref{eq:windowed_likelihood_td_gate}. We will only present results for the TDI2 case, with a gap at merger, for both masses $M=3\times 10^7 M_\odot$ and $M=3\times 10^6 M_\odot$. The noise realization $\bm{n}$ is kept the same in all simulations. The posterior samples obtained span all 11 physical parameters, but to save space we will only show corner plots for two subsets: the intrinsic parameters $\{\mathcal{M}_c, q, \chi_+, \chi_-\}$, and the most important extrinsic parameters $\{ D, \iota, \lambda_L, \beta_L \}$ (see the caption of Tab.~\ref{tab:mbhb_params}). We found that the correlations in the full corner plots are comparable to those shown in these reduced plots. In all corner plots, we show the result of the linear signal approximation estimate for the max-likelihood bias obtained with each likelihood model.

We start by showing in Fig.~\ref{fig:corner_correct_gap_vs_nogap} posteriors for the case where we do not introduce any gap (where the Whittle-likelihood is valid), and for the statistically correct treatment of the merger gap described by the gated likelihood in Eq.~\eqref{eq:windowed_likelihood_td}. This comparison demonstrates how much information is erased by the gap, and we see that the introduction of the gap significantly broadens the posteriors, in line with the degradation in SNR given by Tab.~\ref{tab:mbhb_gap_snr}. The estimates for the best-fit parameters given by~\eqref{eq:cv_correct_gap} are in good agreement with the posteriors.

Next, Fig.~\ref{fig:corner_correct_vs_whittle} compares the correct gated likelihood to the Whittle-likelihood for the full segment described in Sec.~\ref{subsec:gap_whittle}, with a direct gating window. As predicted by Tab.~\ref{tab:mismodeling_upsilon_gap} and Tab.~\ref{tab:mismodeling_xi_gap}, we see that the width of the Whittle posterior is comparable to the correct one for M3e7, and significantly smaller for M3e6, while their center are moved away to the point of excluding the true parameters with great significance. Importantly, we find that the noise biases as estimated by~\eqref{eq:CV_bias_model}, while not exact, are able to accurately estimate the magnitude of the bias. This is an important check of the validity of the Fisher approach, which is intrinsically only a local approximation to the posterior distribution.

Fig.~\ref{fig:corner_correct_vs_whittle_taper10min} shows the same comparison, but this time introducing a smooth tapering window with a lobe length of 10 minutes. As predicted by Tab~\ref{tab:mismodeling_upsilon_gap} and Tab.~\ref{tab:mismodeling_xi_gap}, for the heavy system M3e7, the introduction of this taper gives a posterior that is in good agreement with the correct posterior, both in centering and width. For the M3e6 system, however, the posteriors are significantly broadened. They are still compatible with the true parameters, with slightly larger biases; in accordance with the expected value $\overline{\Upsilon}=0.9$ in Tab.~\ref{tab:mismodeling_upsilon_gap}, the broadening of the posteriors dominates over the slight change in biases, and the max-likelihood parameters are actually closer to the true parameters in units of $\sigma$-deviations. We also notice that the posteriors for the extrinsic parameters are non-Gaussian.

Finally, Fig.~\ref{fig:corner_correct_vs_segmented_taper10min} compares the correct likelihood to the segmented Whittle-likelihood as described in Sec.~\ref{subsec:gap_segmented_whittle}, with the same tapering lobe lengths equal to 10 minutes. We find comparable results to Fig.~\ref{fig:corner_correct_vs_whittle_taper10min}, in accordance to Tab.~\ref{tab:mismodeling_upsilon_gap} and Tab.~\ref{tab:mismodeling_xi_gap}, with good agreement for M3e7 and similarly broadened posteriors for M3e6. We verify again that biases seem well estimated by the formula given by Eq.~\eqref{eq:fisher_cv_bias_segmented}.

\begin{figure*}
    \centering
    \includegraphics[width = \textwidth]{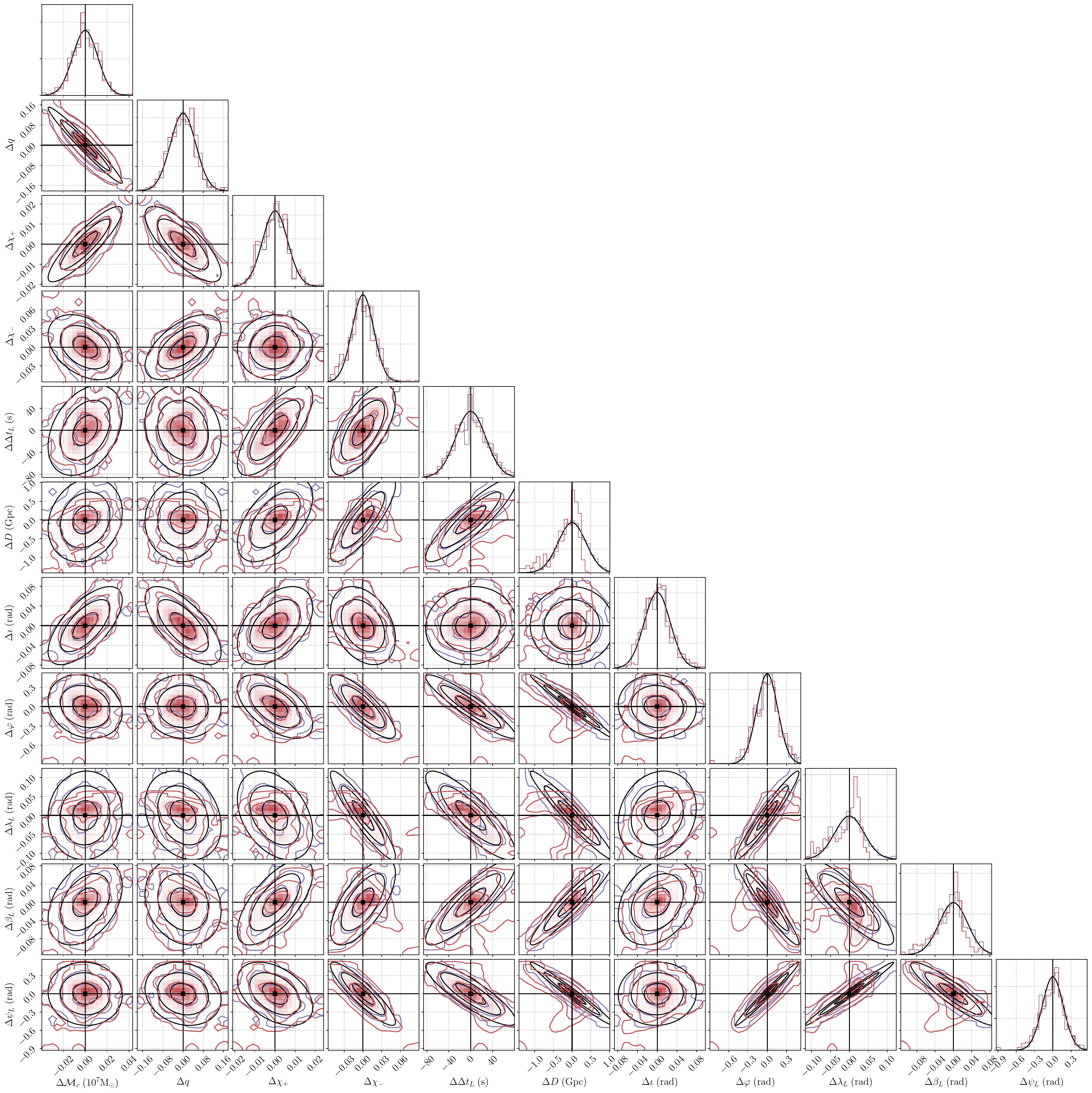}
    \caption{In this figure, which is not a corner plot for a posterior, we are checking the agreement between the Fisher-based estimation for max-likelihood estimates and max-likelihood parameters found in full PE runs. We use the TDI2 noise curve with MBH primary mass $M = 3e7$, with a gap during the merger, and consider the Whittle likelihood with gated data (the configuration of Fig.~\ref{fig:corner_correct_vs_whittle}). In red we plot the histogram of max-likelihood values obtained from full PE using 350 different noise realisations. In blue we plot the histogram of max-likelihood values calculated through Eq.\eqref{eq:fisher_cv_bias_whittle}. Finally, the black curve is a Gaussian distribution, centred on the true parameter with parameter covariance matrix equivalent to Eq.\eqref{eq:fisher_cv_bias_scatter_whittle}. Note that the scale is different from the previous corner plots: the width shown here is set by the scale of the posterior \emph{scatter} of Fig.~\ref{fig:corner_correct_vs_whittle}, which is much larger than the posterior \emph{width} for this case where $\Upsilon \gg 1$.}
    \label{fig:biascorner_incorrect_gap_pe_vs_cv}
\end{figure*}

The full PE runs that have just been discussed show a good level of agreement between the measured noise biases and~\eqref{eq:fisher_cv_bias_segmented} and the true max-likelihood parameters found in PE. In order to validate the synthetic Fisher-based error measures $\Upsilon$ and $\Xi$, we repeated this check multiple times.

For this set of PE runs, we performed 350 random draws of the noise realization $\bm{n}$, focusing on the Whittle case with a simple gating window, for TDI2 and M3e7 with a merger gap (first line, leftmost entry in Tab.~\ref{tab:mismodeling_upsilon_gap} and Tab.~\ref{tab:mismodeling_xi_gap}). This choice allows us to validate these noise bias estimates in a regime where the parameter shifts are multiple units of $\sigma$, while the Fisher approach relies on a local approximation to the likelihood. We expect the linear signal approximation bias estimates to work better in the other cases with tapering, as they show more moderate biases.

The results are shown as a corner plot in Fig.~\ref{fig:biascorner_incorrect_gap_pe_vs_cv}. Note that this is not a corner plot of posterior samples. Instead, this is showing a (multidimensional) histogram of max-likelihood values extracted from the PE runs, estimates for the max-likelihood calculated according to Eq.~\eqref{eq:fisher_cv_bias_whittle}, along with ellipses corresponding to a Gaussian distribution computed from the direct expression given by Eq.~\eqref{eq:fisher_cv_bias_scatter_whittle} for the bias covariance. The comparison between the ellipses and the contours for the biases is a pure consistency check, relying only on the Gaussian nature of the noise. The comparison between the contours for the biases and the contours for the max-likelihood results from PE is a stronger check, verifying that the Fisher results are accurate enough even though the individual posteriors might be non-Gaussian. We find an overall good agreement, with some overall noise that is due to the limited number of simulations (350). We see deviations in some parameters, for instance $D$ and $\lambda_L$, but even then the distributions are broadly consistent. 
Given that the gated Whittle has large biases and plays the role of a stress-test here, we conclude that the estimates shown in Tab.~\ref{tab:mismodeling_upsilon_gap} and Tab.~\ref{tab:mismodeling_xi_gap} are qualitatively reliable.

\subsection{Influence of the tapering length}\label{subsec:mismodeling_tapering_length}

\begin{figure*}
    \centering
    \includegraphics[width = \textwidth]{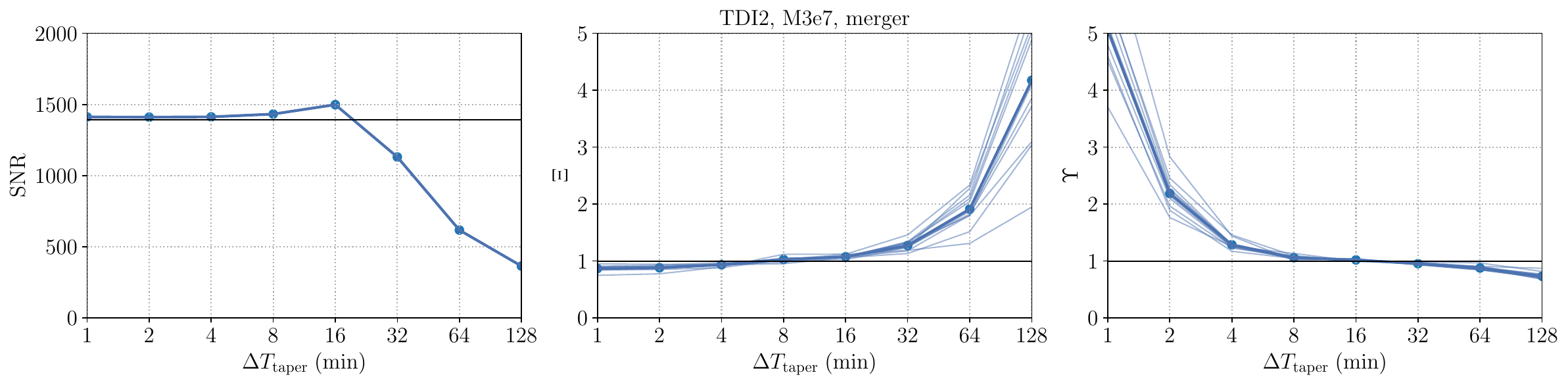}
    \includegraphics[width = \textwidth]{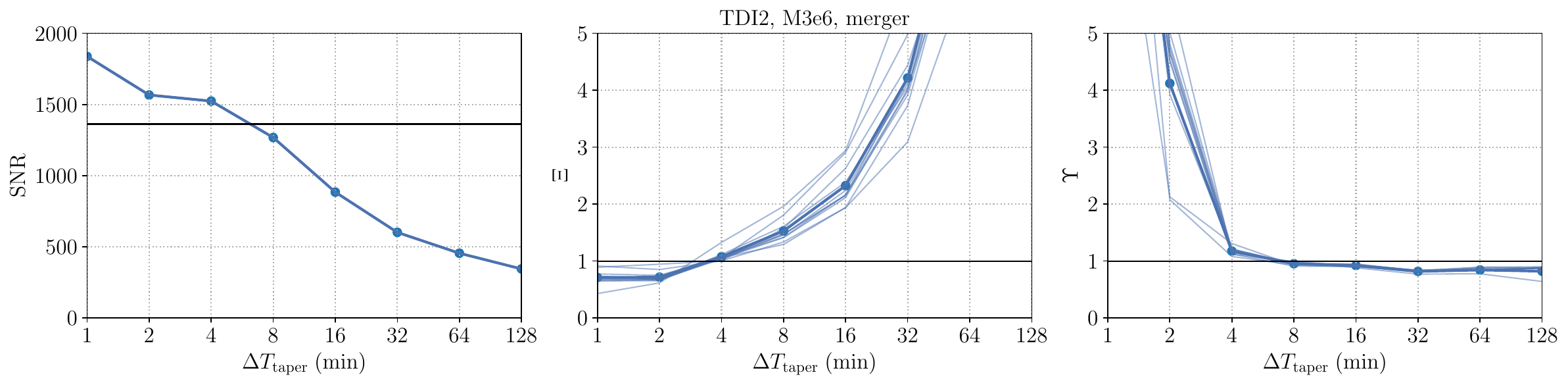}
    \caption{\textbf{(Top row from left to right):)} SNR, $\overline{\Xi}$ and $\overline{\Upsilon}$ computations as a function of the lobe-length in minutes for the $M = 3e7$ case. The horizontal black lines represent the true values with model TD noise covariance matrix consistent with the gated TD noise-covariance matrix. The blue dots are computations for SNR, $\overline{\Upsilon}$ and $\overline{\Xi}$ using a Whittle-based covariance matrix assuming dependence between segments. The fainter blue lines are the individual parameter $\Upsilon_{a}$ and $\Xi_{a}$ values. \textbf{(Bottom row from left to right):} The same set up as the top row, but for the M3e6 case.}
    \label{fig:snr_Xi_Upsilon_mismodelgap_varytaper_tdi2_merger}
\end{figure*}

\begin{figure*}
    \centering
    \includegraphics[width = \textwidth]{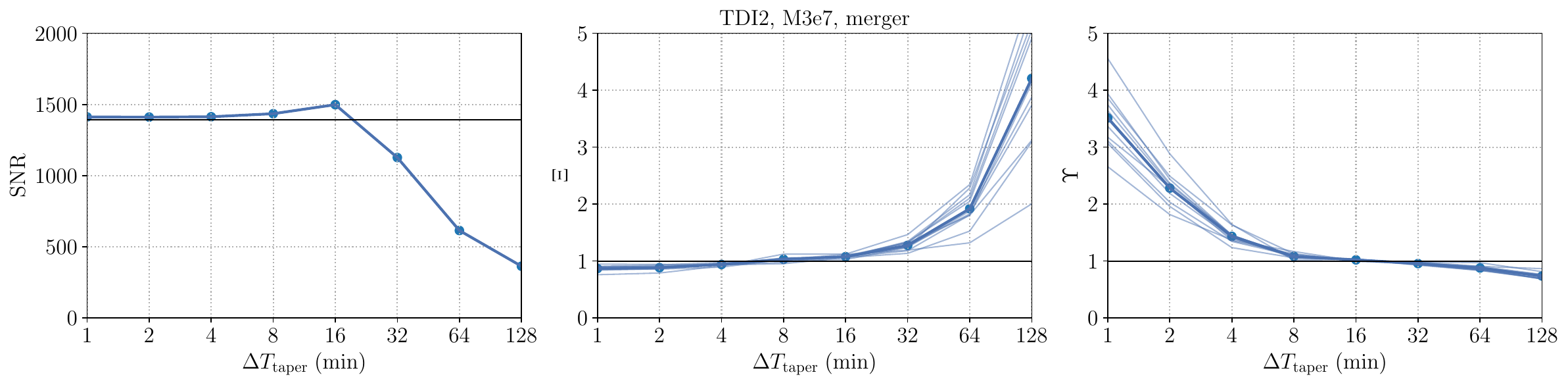}
    \includegraphics[width = \textwidth]{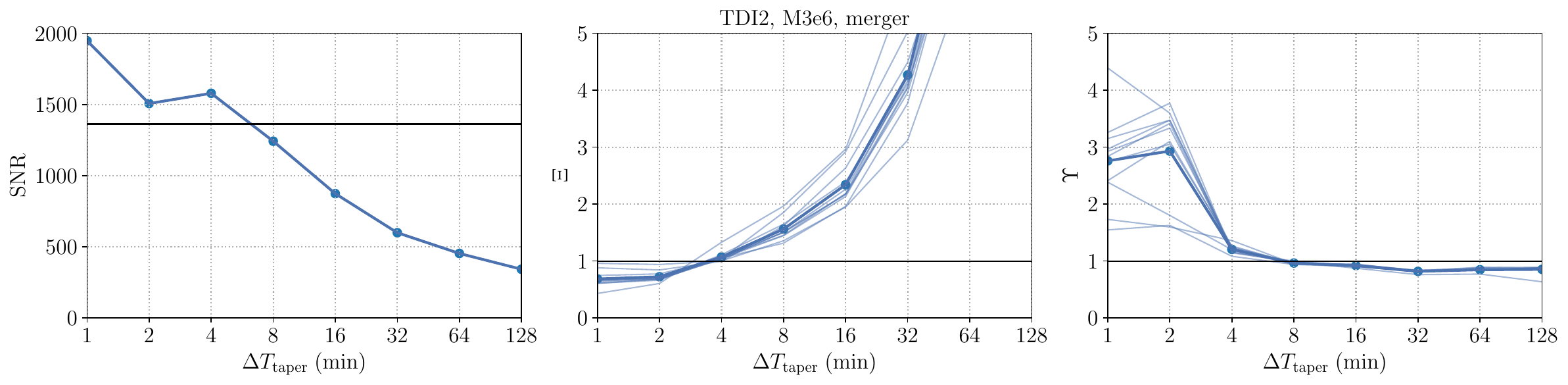}
    \caption{The same set up as figure~\ref{fig:snr_Xi_Upsilon_mismodelgap_varytaper_tdi2_merger} except with model covariance matrix replaced by the Segmented-Whittle covariance (assuming independence between gap segments).}
    \label{fig:snr_Xi_Upsilon_segmented_varytaper_tdi2_merger}
\end{figure*}

Introducing a smooth tapering window helps to reduce the mismodeling ratio $\Upsilon$ (improving the statistical consistency of posteriors) at the price of increasing the mismodeling ratio $\Xi$ (broadening the posteriors and erasing information). We found a strong dependency of the results on the chosen tapering length, and we explore this dependency in Fig.~\ref{fig:snr_Xi_Upsilon_mismodelgap_varytaper_tdi2_merger} for the full-segment Whittle-likelihood of Sec.~\ref{subsec:gap_whittle} and Fig.~\ref{fig:snr_Xi_Upsilon_segmented_varytaper_tdi2_merger} for the segmented Whittle-likelihood of Sec.~\ref{subsec:gap_segmented_whittle}

In both these figures, we show the dependency of the SNR, $\overline{\Xi}$ and $\overline{\Upsilon}$ on the tapering length, exploring lengths chosen to be powers-of-2 in minutes from 1 to 128. We focus on the TDI2 case, with a gap at merger. We recall that the gap length itself is different for the two masses, of 1 hour for M3e7 and of 15 minutes for M3e6. We also show individual curves for the different parameters $\Xi_a$, $\Upsilon_a$, illustrating that the trends are similar across parameters. We also show the reference value of the SNR obtained with the correct treatment of the gap.

Both the Whittle and the segmented Whittle-likelihood show the same qualitative trends. We see that increasing the lobe length washes away information, as expected: this is evident in both SNR and $\Xi$. There is a marked difference between the M3e7 (1-hour gap, lower-frequency signal) and M3e6 (15-minutes gap, higher-frequency signal) cases: $\Xi$ degrades much faster for increasing lobe length for M3e6, while for M3e7 there is an extended range where $\Xi \simeq 1$. For the shortest lobe lengths, in both cases, we obtain an overestimated SNR and values $\Xi < 1$. The behavior of $\Upsilon$, with an explosion towards $\Upsilon > 1$, particularly pronunced for the low-mass case M3e6, and values $\Upsilon \lesssim 1$ for longer tapering lobes.

The results for the Whittle and the segmented Whittle-likelihood are almost identical for both the SNR and $\Xi$. The behavior in $\Upsilon$ is also close, with one noticeable difference for short tapering lengths in the M3e6 case, where the segmented Whittle model seems to perform better with a more limited increase.

Overall, Fig.~\ref{fig:snr_Xi_Upsilon_mismodelgap_varytaper_tdi2_merger} and Fig.~\ref{fig:snr_Xi_Upsilon_segmented_varytaper_tdi2_merger} indicate the existence of an optimal tapering length, which is in line with intuition: the tapering window must alleviate the statistical inconsistencies concentrated near the edges of the gap, while avoiding erasing too much of the information provided by the signal there. We find indeed that $\Xi$ increases and $\Upsilon$ decreases, for an increasing lobe length. In our case, it seems that the value of 10 minutes shown in Tab.~\ref{tab:mismodeling_upsilon_gap} and Tab.~\ref{tab:mismodeling_xi_gap} is close to optimal.

Our tools allow us to look systematically for an optimal tapering length, optimizing for both $\Upsilon$ and $\Xi$ simultaneously, or possibly optimizing their product $\Xi \Upsilon$ which is bounded from below by 1 and whose dependence on tapering length has a characteristic U-shape. This could be done quite fast when working with Fisher matrices and linear signal approximation biases, without the need for full PE runs.

More explorations would be needed to understand what determines the optimal tapering length: does it depend primarily on the length of the gap ? On the morphology of the signal, in particular its high-frequency content ? How much does this optimum vary for different signals ? We leave these questions for future work.

\subsection{Influence of the cadence}\label{subsec:mismodeling_cadence}

\begin{table*}
\begin{tabular}{ll | cc  cc  cc}
\toprule
\multicolumn{2}{c}{\shortstack{\bf Mismodeling $\overline{\Upsilon}$ data gaps: \\ \bf effect of cadence -- M3e7 merger}} & \multicolumn{2}{c}{TDI2} & \multicolumn{2}{c}{TDI1} & \multicolumn{2}{c}{TDI0} \\
\cmidrule(lr){3-4} \cmidrule(lr){5-6} \cmidrule(lr){7-8}
Data & Model & $dt=10\mathrm{s}$ & $dt=100\mathrm{s}$ & $dt=10\mathrm{s}$ & $dt=100\mathrm{s}$ & $dt=10\mathrm{s}$ & $dt=100\mathrm{s}$ \\
\midrule
Coherent, gap & Whittle, gated     & \cellcolor{red!50} 12.2 & \cellcolor{yellow!50} 1.4 & \cellcolor{yellow!50} 1.2 & \cellcolor{green!50} 1.0 & \cellcolor{red!50} 39.7 & \cellcolor{red!50} 74.6 \\
\bottomrule
\end{tabular}
\caption{A similar setup to Tab.\ref{tab:mismodeling_upsilon_gap} but only focusing on the M3e7 case with gap at merger, using a model covariance matrix consistent with Whittle and gating the data. The two variables changed in our experiments here are the cadence $\Delta t \in [10,100]$ seconds and noise curve TDI2/1/0.}
\label{tab:mismodeling_upsilon_gap_cadence}
\end{table*}

We conclude this exploration of gap mismodeling with an important observation. The frequency resolution in the discrete FD is $\Delta f = 1/T$, where $T$ is the total duration of the data. On the other hand, the cadence or time step $\Delta t$ drives the frequency content, through $f_{\rm Nyquist} = 1/(2 \Delta t)$. The lowest and highest frequencies present in the data are therefore determined by $1/T$ and $1/(2 \Delta t)$ respectively. Considering the noise PSDs illustrated in Fig.~\ref{fig:noise_curves}, we can see that for a red process $T$ will affect the importance of the red noise, while for a blue process $\Delta t$ will affect the importance of the blue noise.

All our results presented above were produced with a cadence $\Delta t = 10 \mathrm{s}$ (while the L1 TDI data is rather expected to use $\Delta t = 5 \mathrm{s}$ -- this factor of 2 in the data size $N$ eases our computations). Our most massive system, with $M = 3\times 10^7 M_\odot$, is so massive that a Nyquist frequency of $5\mathrm{mHz}$ (a cadence of $\Delta t = 100\mathrm{s}$) would be sufficient to resolve it.

In Tab.~\ref{tab:mismodeling_upsilon_gap_cadence}, we contrast the mismodeling ratio $\Upsilon$ that we would obtain with a cadence of $\Delta t = 10 \mathrm{s}$ or $\Delta t = 100 \mathrm{s}$, for the gated Whittle-likelihood. We see that changing the cadence would strongly change the mismodeling ratio for TDI2, reducing it from $\sim 12$ to $\sim 1.4$. We attribute this to the fact that TDI2 is a blue process, and the smaller sampling rate reduces the relative importance of the high-frequency noise. This change would have a moderate effect in the TDI1 case, and a worsening effect in the TDI0 case, mostly likely indicative of a breakdown in the linear signal approximation, since for large values of $\Upsilon$ the best-fit parameters are many $\sigma$'s away from the truth.

We conclude that the cadence can have an effect on the gap mismodeling that must not be overlooked. Direct downsampling of the data, prior to analysis, is an operation that could be implemented easily, although only applies to low-frequency signals. Our results might change when using the more realistic cadence of $\Delta t = 5\mathrm{s}$. The effect of the total length on red noise would also need to be explored, which we leave for future work. 

\subsection{Mismodeling data segment independence}
\label{subsec:mismodeling_independence}

\begin{table*}
\begin{tabular}{ll | cccc  cccc  cccc}
\toprule
\multicolumn{2}{c}{\shortstack{\bf Mismodeling $\overline{\Upsilon}$ \\ \bf segment independence}} & \multicolumn{4}{c}{TDI2} & \multicolumn{4}{c}{TDI1} & \multicolumn{4}{c}{TDI0} \\
\cmidrule(lr){3-6} \cmidrule(lr){7-10} \cmidrule(lr){11-14}
& & \multicolumn{2}{c}{Merger} & \multicolumn{2}{c}{Insp.} & \multicolumn{2}{c}{Merger} & \multicolumn{2}{c}{Insp.} & \multicolumn{2}{c}{Merger} & \multicolumn{2}{c}{Insp.} \\
\cmidrule(lr){3-4} \cmidrule(lr){5-6} \cmidrule(lr){7-8} \cmidrule(lr){9-10} \cmidrule(lr){11-12} \cmidrule(lr){13-14}
Data & Model & M3e7 & M3e6 & M3e7 & M3e6 & M3e7 & M3e6 & M3e7 & M3e6 & M3e7 & M3e6 & M3e7 & M3e6 \\
\midrule
Coherent, gap & Incoherent, gap   & \cellcolor{green!50} 1.0 & \cellcolor{green!50} 1.0 & \cellcolor{green!50} 1.0 & \cellcolor{green!50} 1.0 & \cellcolor{green!50} 1.0 & \cellcolor{green!50} 1.0 & \cellcolor{green!50} 1.0 & \cellcolor{green!50} 1.0 & \cellcolor{green!50} 1.0 & \cellcolor{green!50} 1.0 & \cellcolor{green!50} 1.0 & \cellcolor{green!50} 1.0 \\
Incoherent, gap & Coherent, gap   & \cellcolor{yellow!50} 1.2 & \cellcolor{yellow!50} 1.1 & \cellcolor{yellow!50} 1.1 & \cellcolor{orange!50} 1.5 & \cellcolor{green!50} 1.0 & \cellcolor{green!50} 1.0 & \cellcolor{green!50} 1.0 & \cellcolor{green!50} 1.0 & \cellcolor{red!50} 26.2 & \cellcolor{red!50} 8.1 & \cellcolor{red!50} 13.4 & \cellcolor{red!50} 14.2 \\
\midrule
Coherent, split & Incoherent, split   & \cellcolor{green!50} 1.0 & \cellcolor{green!50} 1.0 & \cellcolor{green!50} 1.0 & \cellcolor{green!50} 1.0 & \cellcolor{green!50} 1.0 & \cellcolor{green!50} 1.0 & \cellcolor{green!50} 1.0 & \cellcolor{green!50} 1.0 & \cellcolor{green!50} 1.0 & \cellcolor{green!50} 1.0 & \cellcolor{green!50} 1.0 & \cellcolor{green!50} 1.0 \\
Incoherent, split & Coherent, split   & \cellcolor{red!50} 5.6 & \cellcolor{orange!50} 2.1 & \cellcolor{yellow!50} 1.3 & \cellcolor{yellow!50} 1.5 & \cellcolor{yellow!50} 1.3 & \cellcolor{orange!50} 1.6 & \cellcolor{green!50} 1.0 & \cellcolor{green!50} 1.0 & \cellcolor{red!50} 27.3 & \cellcolor{red!50} 41.0 & \cellcolor{red!50} 11.8 & \cellcolor{red!50} 11.9 \\
\bottomrule
\end{tabular}
\caption{The same set up as Tab.\ref{tab:mismodeling_upsilon_gap} with different assumptions. The data generation process ``Data" and the ``Model" used for the likelihood all start from the same underlying circulant full-segment covariance and apply the correct treatment of the gap from Sec.~\ref{sec:methods}. For ``Coherent" the covariance blocks describing correlations between before/after gap segments are kept, for ``Incoherent'' they are ignored (see Eqs.~\eqref{eq:cov_data_coherent_model_incoherent} and~\eqref{eq:cov_data_incoherent_model_coherent}). In the second part of the table, ``Split'' indicates that the data stream is directly split into two parts, with no gap in-between.}
\label{tab:mismodeling_upsilon_independence}
\end{table*}

\begin{table*}
\begin{tabular}{ll | cccc  cccc  cccc}
\toprule
\multicolumn{2}{c}{\shortstack{\bf Mismodeling $\overline{\Xi}$ \\ \bf segment independence}} & \multicolumn{4}{c}{TDI2} & \multicolumn{4}{c}{TDI1} & \multicolumn{4}{c}{TDI0} \\
\cmidrule(lr){3-6} \cmidrule(lr){7-10} \cmidrule(lr){11-14}
& & \multicolumn{2}{c}{Merger} & \multicolumn{2}{c}{Insp.} & \multicolumn{2}{c}{Merger} & \multicolumn{2}{c}{Insp.} & \multicolumn{2}{c}{Merger} & \multicolumn{2}{c}{Insp.} \\
\cmidrule(lr){3-4} \cmidrule(lr){5-6} \cmidrule(lr){7-8} \cmidrule(lr){9-10} \cmidrule(lr){11-12} \cmidrule(lr){13-14}
Data & Model & M3e7 & M3e6 & M3e7 & M3e6 & M3e7 & M3e6 & M3e7 & M3e6 & M3e7 & M3e6 & M3e7 & M3e6 \\
\midrule
Coherent, gap & Incoherent, gap   & \cellcolor{green!50} 1.0 & \cellcolor{green!50} 1.0 & \cellcolor{green!50} 1.0 & \cellcolor{green!50} 1.0 & \cellcolor{green!50} 1.0 & \cellcolor{green!50} 1.0 & \cellcolor{green!50} 1.0 & \cellcolor{green!50} 1.0 & \cellcolor{green!50} 1.0 & \cellcolor{green!50} 1.0 & \cellcolor{green!50} 1.0 & \cellcolor{green!50} 1.0 \\
Incoherent, gap & Coherent, gap   & \cellcolor{green!50} 1.0 & \cellcolor{green!50} 1.0 & \cellcolor{green!50} 1.0 & \cellcolor{green!50} 1.0 & \cellcolor{green!50} 1.0 & \cellcolor{green!50} 1.0 & \cellcolor{green!50} 1.0 & \cellcolor{green!50} 1.0 & \cellcolor{green!50} 1.0 & \cellcolor{green!50} 1.0 & \cellcolor{green!50} 1.0 & \cellcolor{green!50} 1.0 \\
\midrule
Coherent, split & Incoherent, split   & \cellcolor{green!50} 1.0 & \cellcolor{green!50} 1.0 & \cellcolor{green!50} 1.0 & \cellcolor{green!50} 1.0 & \cellcolor{green!50} 1.0 & \cellcolor{green!50} 1.0 & \cellcolor{green!50} 1.0 & \cellcolor{green!50} 1.0 & \cellcolor{green!50} 1.0 & \cellcolor{yellow!50} 1.1 & \cellcolor{green!50} 1.0 & \cellcolor{green!50} 1.0 \\
Incoherent, split & Coherent, split   & \cellcolor{green!50} 1.0 & \cellcolor{green!50} 1.0 & \cellcolor{green!50} 1.0 & \cellcolor{green!50} 1.0 & \cellcolor{green!50} 1.0 & \cellcolor{green!50} 1.0 & \cellcolor{green!50} 1.0 & \cellcolor{green!50} 1.0 & \cellcolor{green!50} 1.0 & \cellcolor{yellow!50} 0.9 & \cellcolor{green!50} 1.0 & \cellcolor{green!50} 1.0 \\
\bottomrule
\end{tabular}
\caption{The same set up as Tab.\ref{tab:mismodeling_upsilon_independence}, but computing $\overline{\Xi}$ instead.}
\label{tab:mismodeling_Xi_independence}
\end{table*}

Data segment independence might be an important question for LISA data analysis, as interruptions in the data stream might be of different nature. At one extreme, for missing data, the data gap acts as a mask and the underlying unmasked noise process $\bm{n}$ does not lose coherence and is not affected by the gap at all. At another extreme, the data gap corresponds to a major disturbance of the instrument, and the data before and after the gap are best represented as pertaining to two separate, statistically independent experiments. Note that the latter case cannot be treated via data imputation methods~\cite{Baghi:2019eqo, Wang:2024ovi, Mao:2024jad, Blelly:2021oim}, as they cannot invent data joining two independent processes in a consistent way. For a real instrument, the truth might lie somewhere in-between these two extremes.  

More generally, we might want to cut the data stream into different independent pieces for computational reasons, in order to keep the length of each data sub-segment low. By construction, under the independence assumption, the covariance matrix becomes block-diagonal, with blocks that are much more manageable than the full $N\times N$ matrix. Another motivation can be to improve the stationary nature on each sub-segment by considering a different PSD on each, better resolving PSD drifts and slow variations in the GB foreground.

Therefore, we investigate the mismodeling errors induced by both scenarios: assuming that the data segments are independent when they are in fact segments from a coherent stochastic process; and assuming that the data segments are coherent when they are in fact independent. Here we want to isolate the effect of segment independence from the effect of gap mismodeling. To this end, we do not use the Whittle-likelihood but use the correct gap treatment of Sec.~\ref{subsec:windowing_procedure}.

In the first scenario, where the data is coherent but the model independent, we have for the data and model respectively
\begin{align}\label{eq:cov_data_coherent_model_incoherent}
    \langle (\bm{W}\bm{n}) (\bm{W}\bm{n})^T \rangle &= \bm{W}\bm{\Sigma}\bm{W} = \begin{pmatrix}
    \bm{\Sigma}_{00} & \bm{0} & \bm{\Sigma}_{02} \\
    \bm{0}  & \bm{0}  & \bm{0}  \\
    \bm{\Sigma}_{02}^{T} & \bm{0} & \bm{\Sigma}_{22}
    \end{pmatrix} \,, \nonumber\\
    (\bm{\Sigma}^{\prime})^+ &= \begin{pmatrix}
    \bm{\Sigma}_{00}^{-1} & \bm{0} & \bm{0} \\
    \bm{0}  & \bm{0}  & \bm{0}  \\
    \bm{0} & \bm{0} & \bm{\Sigma}_{22}^{-1} 
    \end{pmatrix} \,.
\end{align}

We remark here that the configurations presented in Eq.~\eqref{eq:cov_data_coherent_model_incoherent}  are subtly different from the model specified by Eq.~\eqref{eq:block_circulant_assume_indep}, in Sec.~\ref{subsec:gap_segmented_whittle}. Notice that the individual blocks in $\bm{\Sigma}_{00}$ and $\bm{\Sigma}_{22}$ are Toeplitz and, in the model assuming incoherence, we are using block inverses of those specific Toeplitz matrices. This isolates two sources of mismodelling that were present in Eq.~\eqref{eq:cov_data_coherent_model_incoherent}, where one assumes the underlying process is \emph{circulant} and the data segments between gaps are \emph{independent}. In this case, we are mismodelling independence but correctly modeling the underlying process as Toeplitz. The model likelihood here $\bm{\Sigma}^{\prime}$, although modeling stationary noise, is \emph{not} Whittle-based.

In the second scenario, where the data is independent but the model is coherent, we exchange the role of the two matrices for the data and model:
\begin{align}\label{eq:cov_data_incoherent_model_coherent}
    \langle (\bm{W}\bm{n}) (\bm{W}\bm{n})^T \rangle &= \bm{W}\bm{\Sigma}\bm{W} = \begin{pmatrix}
    \bm{\Sigma}_{00} & \bm{0} & \bm{0} \\
    \bm{0}  & \bm{0}  & \bm{0}  \\
    \bm{0} & \bm{0} & \bm{\Sigma}_{22}
    \end{pmatrix} \,, \nonumber\\
    (\bm{\Sigma}^{\prime})^+ &= \begin{pmatrix}
    \bm{\Sigma}_{00} & \bm{0} & \bm{\Sigma}_{02} \\
    \bm{0}  & \bm{0}  & \bm{0}  \\
    \bm{\Sigma}_{02}^{T} & \bm{0} & \bm{\Sigma}_{22}
    \end{pmatrix}^+ \,.
\end{align}

We extend the analysis by considering a data split, without any missing data between the two data segments. This situation is closest to the voluntary segmentation that one might implement for long signals for computational reasons~\cite{Gair:2004iv, Amaro-Seoane:2007osp, Messenger:2011rg, Bandopadhyay:2023gkb, Bandopadhyay:2024lwv}. It represents a useful check: the longer the gap is, the more justified it will be to treat the data before/after the gap as independent, since correlations will be dominated by short timescales (the near-diagonal elements of the covariance matrix); a zero-length gap is therefore a worst-case scenario for ignoring correlations between segments present in the data. In this scenario, the gated covariance and the pseudo-inverse model covariance matrices are the same as in Eq.~\eqref{eq:cov_data_coherent_model_incoherent} and Eq.~\eqref{eq:cov_data_incoherent_model_coherent}, but eliminating the middle line and column since there is no gap present.

We report the resulting mismodeling ratios in Tab.~\ref{tab:mismodeling_upsilon_independence} and Tab.~\ref{tab:mismodeling_Xi_independence}. The results for the case where the data is coherent (dependent) but the model incoherent (independent) show very little mismodeling in both $\Upsilon$ and $\Xi$, whether the gap is present or whether we simply split the data in two segments. We find small losses of SNR and deviation of $\Xi$ from 1, but they are limited to less than 1\%. Our intuition would tell us that ignoring cross-block terms in the covariance matrix means that we choose to neglect some information by ignoring correlations. This test shows that the amount of information erased is negligible. By contrast, in the case where the data is incoherent but the model is coherent, while we still find $\Xi \simeq 1$, we also find $\Upsilon > 1$ in most cases. These values are very large for TDI0, but also deviate significantly from 1 for TDI2, and are worse for the case where the data is simply split in two. Here, our intuition would say that we are analyzing the data while inventing correlations that are not here in reality. The presence of the gaps (putting some distance between the two segments) alleviates but does not eliminate the issue.

The near-absence of mismodeling errors when enforcing segment independence, even without gaps, at the expense of a negligible loss of information, is a validation of methods based on data segmentation. It is also consistent with the findings of Sec.~\ref{subsec:gap_mismodeling_ratios} where the full-segment Whittle model and the segmented Whittle model were giving comparable results. On the other hand, the presence of mismodeling errors when enforcing data coherence on independent segments is a cautionary tale for treating instrument disturbances as missing data in a coherent process. In short, wrongly assuming coherence can be dangerous, but assuming independence appears to be safe.

We did not explore the impact of tapering windows in this experiment. Although the correct gap treatment does not require any tapering, in the present case the independence represents another source of mismodeling, and it is possible that tapering would alleviate the mismodeling in the case where the data is incoherent and the model coherent (at the expense of erasing some information, as always).

\subsection{Mismodeling the noise PSD at low frequencies}
\label{subsec:mismodeling_psd_lowf}

\begin{figure}
    \centering
    \includegraphics[width = 0.49\textwidth]{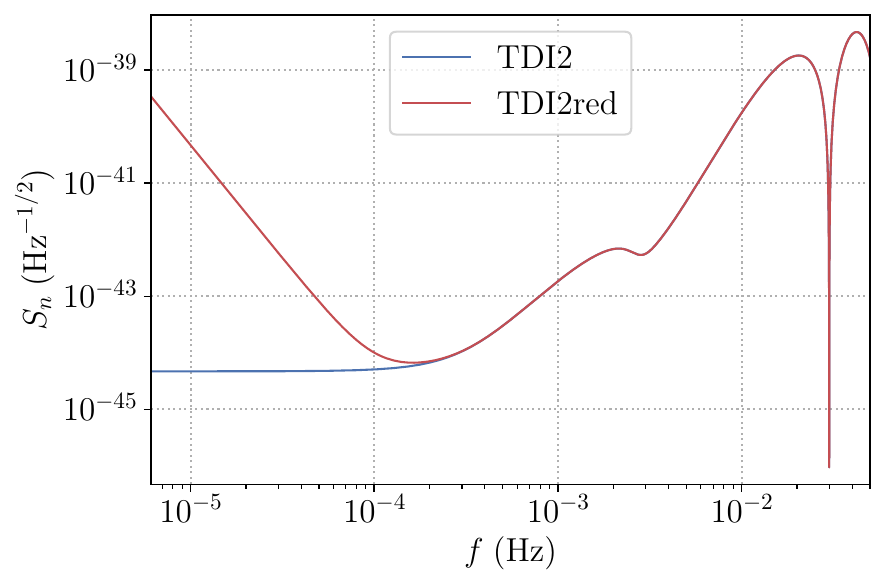}
    \caption{\textbf{(In blue):} The usual noise curve for TDI2 given by Eq.~\eqref{eq:noise_curve_TDI2}. \textbf{(In red):} The noise curve for TDI2 but with a $f^{-4}$ red noise degradation at low frequencies, as given in~\eqref{eq:def_TDI2_red_PSD}.}
    \label{fig:noise_psd_tdi2_tdi2red}
\end{figure}

\begin{table*}
\begin{tabular}{ll | cccc}
\toprule
\multicolumn{2}{c}{\shortstack{\bf Mismodeling $\overline{\Upsilon}$ \\ \bf low-$f$ PSD}} & \multicolumn{4}{c}{TDI2} \\
\cmidrule(lr){3-6}
& & \multicolumn{2}{c}{Merger} & \multicolumn{2}{c}{Insp.} \\
\cmidrule(lr){3-4} \cmidrule(lr){5-6}
Data & Model & M3e7 & M3e6 & M3e7 & M3e6 \\
\midrule
PSD TDI2${}_{\rm red}$, no gap & PSD TDI2, no gap     & \cellcolor{green!50} 1.0 & \cellcolor{green!50} 1.0 & \cellcolor{green!50} 1.0 & \cellcolor{green!50} 1.0 \\
PSD TDI2${}_{\rm red}$, gap & PSD TDI2, gap     & \cellcolor{orange!50} 2.2 & \cellcolor{yellow!50} 1.3 & \cellcolor{yellow!50} 1.2 & \cellcolor{yellow!50} 1.1 \\
PSD TDI2${}_{\rm red}$, gap & PSD TDI2, Whittle, gated     & \cellcolor{red!50} 13.4 & \cellcolor{red!50} 40.9 & \cellcolor{orange!50} 1.6 & \cellcolor{yellow!50} 1.4 \\
PSD TDI2${}_{\rm red}$, gap & PSD TDI2, Whittle, taper 10min    & \cellcolor{orange!50} 3.6 & \cellcolor{yellow!50} 1.2 & \cellcolor{yellow!50} 1.3 & \cellcolor{yellow!50} 1.1 \\
PSD TDI2${}_{\rm red}$, gap & PSD TDI2, Whittle, taper 30min    & \cellcolor{orange!50} 2.2 & \cellcolor{green!50} 1.0 & \cellcolor{yellow!50} 1.2 & \cellcolor{yellow!50} 1.1 \\
\midrule
PSD TDI2${}_{\rm red}$, gap & PSD TDI2${}_{\rm red}$, Whittle, gated     & \cellcolor{red!50} 9.6 & \cellcolor{red!50} 39.0 & \cellcolor{yellow!50} 1.2 & \cellcolor{yellow!50} 1.2 \\
PSD TDI2${}_{\rm red}$, gap & PSD TDI2${}_{\rm red}$, Whittle, taper 10min    & \cellcolor{orange!50} 1.6 & \cellcolor{yellow!50} 1.3 & \cellcolor{green!50} 1.0 & \cellcolor{green!50} 1.0 \\
PSD TDI2${}_{\rm red}$, gap & PSD TDI2${}_{\rm red}$, Whittle, taper 30min    & \cellcolor{yellow!50} 1.2 & \cellcolor{yellow!50} 1.1 & \cellcolor{green!50} 1.0 & \cellcolor{green!50} 1.0 \\
\bottomrule
\end{tabular}
\caption{The same set up as Tab.~\ref{tab:mismodeling_upsilon_gap}, except that we test for the mismodeling of noise processes with unknown large red noise components. Here $\text{TDI2}_{\text{red}}$ is the TDI2 noise curve with an $f^{-4}$ explosion at low frequencies given in Eq.~\eqref{eq:def_TDI2_red_PSD} and shown in Fig.~\ref{fig:noise_psd_tdi2_tdi2red}, while $\text{TDI2}$ is the usual noise curve that approaches a constant as $f \rightarrow 0^{+}$. In the first line, no gap is present and the only mismodeling is that of the PSD.
In all other cases, the data features a gap. In the second line, the model likelihood incorporates the gap correctly according to Sec.~\ref{sec:methods}, while using the wrong PSD. In the Whittle case (lines 3-5), either gated or tapered, the same models are used as in Tab.~\ref{tab:mismodeling_upsilon_gap}, but this time the PSD is another source of mismodeling. We also show in the last three lines the result for a Whittle likelihood model where the PSD $\text{TDI2}_{\text{red}}$ is correctly modeled.}
\label{tab:mismodeling_upsilon_lowfnoise}
\end{table*}

\begin{table*}
\begin{tabular}{ll | cccc}
\toprule
\multicolumn{2}{c}{\shortstack{\bf Mismodeling $\overline{\Xi}$ \\ \bf low-$f$ PSD}} & \multicolumn{4}{c}{TDI2} \\
\cmidrule(lr){3-6}
& & \multicolumn{2}{c}{Merger} & \multicolumn{2}{c}{Insp.} \\
\cmidrule(lr){3-4} \cmidrule(lr){5-6}
Data & Model & M3e7 & M3e6 & M3e7 & M3e6 \\
\midrule
PSD TDI2${}_{\rm red}$, no gap & PSD TDI2, no gap     & \cellcolor{green!50} 1.0 & \cellcolor{green!50} 1.0 & \cellcolor{green!50} 1.0 & \cellcolor{green!50} 1.0 \\
PSD TDI2${}_{\rm red}$, gap & PSD TDI2, gap     & \cellcolor{green!50} 1.0 & \cellcolor{green!50} 1.0 & \cellcolor{green!50} 1.0 & \cellcolor{green!50} 1.0 \\
PSD TDI2${}_{\rm red}$, gap & PSD TDI2, Whittle, gated     & 
\cellcolor{yellow!50} 0.8 & \cellcolor{orange!50} 0.5 & \cellcolor{green!50} 1.0 & \cellcolor{green!50} 1.0 \\
PSD TDI2${}_{\rm red}$, gap & PSD TDI2, Whittle, taper 10min    & \cellcolor{green!50} 1.0 & \cellcolor{orange!50} 1.7 & \cellcolor{green!50} 1.0 & \cellcolor{green!50} 1.0 \\
PSD TDI2${}_{\rm red}$, gap & PSD TDI2, Whittle, taper 30min    & \cellcolor{yellow!50} 1.2 & \cellcolor{orange!50} 3.9 & \cellcolor{green!50} 1.0 & \cellcolor{green!50} 1.0 \\
\midrule
PSD TDI2${}_{\rm red}$, gap & PSD TDI2${}_{\rm red}$, Whittle, gated     & \cellcolor{yellow!50} 0.9 & \cellcolor{yellow!50} 0.5 & \cellcolor{green!50} 1.0 & \cellcolor{green!50} 1.0 \\
PSD TDI2${}_{\rm red}$, gap & PSD TDI2${}_{\rm red}$, Whittle, taper 10min    & 
\cellcolor{yellow!50} 1.1 & \cellcolor{orange!50} 1.7 & \cellcolor{green!50} 1.0 & \cellcolor{green!50} 1.0 \\
PSD TDI2${}_{\rm red}$, gap & PSD TDI2${}_{\rm red}$, Whittle, taper 30min    & \cellcolor{yellow!50} 1.2 & \cellcolor{orange!50} 3.9 & \cellcolor{green!50} 1.0 & \cellcolor{green!50} 1.0 \\
\bottomrule
\end{tabular}
\caption{The same set up as Tab.\ref{tab:mismodeling_upsilon_lowfnoise}, but computing $\overline{\Xi}$ instead.}
\label{tab:mismodeling_xi_lowfnoise}
\end{table*}

We have seen in previous sections that the frequency content of the noise process, from a red process like TDI0 to a blue process like TDI2, can strongly affect the importance of mismodeling errors. Another important point is the role of mismodeling the low-frequency noise.

In the standard setting of gravitational-wave data analysis, the stationarity justifies independence in the FD, and the bucket shape of the instrument's sensitivity justifies ignoring the low and high frequency regions where all signals would be drowned in noise. The lower frequencies, in particular, are a difficult limit to build a statistical understanding of any instrument: longer data stretches are needed to resolve lower frequencies, and stationarity makes less sense as an assumption. This is true of the LIGO-Virgo instruments and also of LISA Pathfinder~\cite{Armano:2016bkm,Armano:2018kix,Armano:2020zyx,LISAPathfinder:2022awx}. PSD estimation itself is challenging at low frequencies, with the lack of data posing issues for statistical averaging procedures (such as Welch's) and resulting in a large uncertainty~\cite{Biscoveanu:2020kat,Talbot:2020auc,Edwards:2019mwh,Baghi:2016myd}.

As the presence of gaps breaks the stationary feature of the noise, by introducing correlations across the frequency band, a natural question arises: do we need to model accurately the PSD at low frequencies to avoid introducing noise mismodeling errors~?

We will address this question by focusing on the TDI2 case. As shown in Fig.~\ref{fig:noise_psd_tdi2_tdi2red}, a simple noise model for TDI2, dominated by point-mass noise at low frequencies, gives white noise below $\sim 10^{-4} \, \mathrm{Hz}$. However, it is possible that other unmodeled sources of noise would start dominating there; and it is likely that the instrument models that we build on the ground would become more inaccurate in that range. We will consider an alternative PSD with a $f^{-4}$ red noise explosion below $\sim 10^{-4} \mathrm{Hz}$, following (see~\eqref{eq:acc})
\be\label{eq:def_TDI2_red_PSD}
    P_{\rm acc} (f) \rightarrow P_{\rm acc} (f) \left[ 1 +\left(\frac{f_0}{f} \right)^\alpha \right] \,,
\ee
with the values $f_0=10^{-4}\mathrm{Hz}$, $\alpha=4$. The corresponding PSD, that we will call $\mathrm{TDI2}_{\rm red}$, is shown in Fig~\ref{fig:noise_psd_tdi2_tdi2red}. We stress that this model is \emph{ad hoc} and not based on any detailed instrumental considerations. In the following, it will serve as a toy model to investigate the effect of mismodeling this type of additional red noise.

Assuming that the true PSD of the noise process is $\mathrm{TDI2}_{\rm red}$ while the PSD $\mathrm{TDI2}$ is used in the model, we compute mismodeling ratios $\Upsilon$ and $\Xi$ in different gap scenarios and report the results in Tab.~\ref{tab:mismodeling_upsilon_lowfnoise} and Tab.~\ref{tab:mismodeling_xi_lowfnoise}.

In the first scenario, we do not introduce any gap in the data stream. The two processes are stationary (and circulant) so the covariance is diagonal in the FD. We are still mismodeling the PSD at low frequencies, but even with the underestimation of the noise there the signal is drowned in noise, neutralizing mismodeling errors. We find that $\Upsilon=1$ and $\Xi=1$ in that case.

In the second scenario, we introduce a gap that is correctly modeled, while mismodeling the low-$f$ PSD. That is to say, the model likelihood respects the construction of Sec.~\ref{subsec:windowing_procedure}, but starting with a PSD that is $\mathrm{TDI2}$ instead of $\mathrm{TDI2}_{\rm red}$. While we obtain $\Xi \simeq 1$, we get values $\Upsilon > 1$ in all cases, the worst being the M3e7 case with the gap at merger. This demonstrates that, in the presence of gaps, the low-$f$ PSD mismodeling contaminates the analysis, even though the gaps themselves are properly accounted for.

In the third scenario (lines 3-5) we investigate the case where we both: (i) mismodel the PSD at low-$f$ and (ii) mismodel the gap by using the Whittle-likelihood, as in Sec.~\ref{subsec:mismodeling_gaps}, with different tapering windows. With a simple gating window, we find large values $\Upsilon >1$ and also values of $\Xi <1$. Both issues are mitigated but not eliminated by introducing a tapering window. In particular, we find some mismodeling $\Upsilon > 1$ (but $\Xi \simeq 1$) even for the inspiral gap,  in contrast to Tab.~\ref{tab:mismodeling_upsilon_gap}.

These results indicate that we should be wary of the impact of mismodeled low-frequency noise, even when correctly taking the gaps into account. However, we have not explored possible simple solutions here. In the stationary case without gaps , applying a simple high-pass filter to the data eliminates the low-frequency content. One could expect to construct a filtering procedure, properly generalized to work in the presence of gaps. We also note that the large uncertainty on the low-$f$ part of the PSD should be marginalized over as part of the LISA global fit. We leave these questions for future work.

\subsection{Mismodeling the stationary process as circulant}
\label{subsec:mismodeling_toeplitz_circulant}

\begin{table*}
\begin{tabular}{ll | cc  cc  cc}
\toprule
\multicolumn{2}{c}{\shortstack{\bf Mismodeling $\overline{\Upsilon}$ \\ \bf Toeplitz/circulant}} & \multicolumn{2}{c}{TDI2} & \multicolumn{2}{c}{TDI1} & \multicolumn{2}{c}{TDI0} \\
\cmidrule(lr){3-4} \cmidrule(lr){5-6} \cmidrule(lr){7-8}
Data & Model & M3e7 & M3e6 & M3e7 & M3e6 & M3e7 & M3e6 \\
\midrule
Toeplitz, no gap & Circulant, no gap, no taper  & \cellcolor{yellow!50} 1.1 & \cellcolor{yellow!50} 1.4 & \cellcolor{green!50} 1.0 & \cellcolor{green!50} 1.0 & \cellcolor{red!50} 37.3 & \cellcolor{red!50} 37.3 \\
Toeplitz, no gap & Circulant, no gap, taper 1hr  & \cellcolor{green!50} 1.0 & \cellcolor{green!50} 1.0 & \cellcolor{green!50} 1.0 & \cellcolor{green!50} 1.0 & \cellcolor{orange!50} 2.1 & \cellcolor{yellow!50} 1.3 \\
\bottomrule
\end{tabular}
\caption{The same set up as Tab.\ref{tab:mismodeling_upsilon_gap}, except computing mismodeling errors $\overline{\Upsilon}$ under the assumption that the underlying noise process is circulant when in reality it is Toeplitz. We either keep the data as it is or apply a taper of 1hr at both ends of the data segment. We do not incorporate data gaps in this specific study.}
\label{tab:mismodeling_upsilon_toeplitz}
\end{table*}

We will finisg our mismodeling investigations with a limited investigation of the effect of mismodeling a Toeplitz covariance matrix as a circulant matrix. As explained in Sec.~\ref{subsec:mismodeling_gaps}, the true covariance of the stationary noise process will have a Toeplitz structure. But the FD covariance is only diagonal when the covariance is circulant. In our previous tests, we have taken a circulant covariance over the full segment to be the true covariance, because we wanted to isolate the different sources of mismodeling. For a reminder, we refer the reader to figures \ref{fig:circulant_td_all_matrices} and \ref{fig:toeplitz_td_all_matrices} for a visual illustration of the differences between circulant and Toeplitz matrices respectively. It may also be worth reading the caption and the surrounding text of Fig.(\ref{fig:toeplitz_vs_PSD}) of the potential impact of mismodeling a Toeplitz process as circulant.

We now investigate the Toeplitz-circulant mismodeling errors more rigorously by considering a data stream with no gaps, computing mismodeling ratios when the true covariance is Toeplitz while the model likelihood is circulant (Whittle). The results for $\Upsilon$ are reported in Tab.~\ref{tab:mismodeling_upsilon_toeplitz}, comparing two scenarios: in one, we use the raw data; in the other, we taper the data at each end of the data segment.

We can see that we find $\Upsilon > 1$ for TDI2, large values of $\Upsilon$ for TDI0, while $\Upsilon \simeq 1$ TDI1. Introducing a tapering window gives $\Upsilon \simeq 1$ for TDI2, and reduces but does not fully eliminate the mismodeling for TDI0. We do not present a table for $\overline{\Xi}$ since in all cases we find $\Xi \simeq 1$. The tapering introduced here is at the edges of the data segment, away from the merger that dominates the SNR.

Note also that, when using a segmented Whittle-likelihood, the model covariance is circulant on each segment while the true covariance, taken to be circulant over the full range of the data, is not circulant on the segments. Yet, in Tab.~\ref{tab:mismodeling_upsilon_gap} and Tab.~\ref{tab:mismodeling_xi_gap}, the segmented Whittle and full Whittle (i.e., coherent) give similar results, pointing towards a limited impact of this mismodeling.

% The cow is watching

We conclude that tapering the ends of the data seems to prevent mismodeling errors due to the use of a circulant structure, except for the red-noise dominated process TDI0. Note that when Whittle is used in practice, it is generally with a tapering window. Transient signals that are confined to the inside of the data interval will not be affected much by such a window tapering the end points of the data.

\section{Conclusions and outlook}\label{sec:conclusions_outlook}

\subsection{Conclusions}\label{subsec:conclusions}

The primary goal of this work is to understand the impact data gaps have on PE studies for GW signals in LISA data. Data gaps, whether planned or unplanned, are unavoidable and will have to be included in future LISA data analysis pipelines. In our work, limiting ourselves to a stationary Gaussian underlying noise process, we have explored how the presence of gaps breaks the stationary feature of the observed data.

Working in the TD, in Sec.~\ref{sec:methods} we presented a formalism describing the likelihood of the observed data obtained by marginalizing over unobserved data values, then translated the result to another point of view where the gap is represented by a gating window applied to the original data. The use of the Moore-Penrose pseudo-inverse is crucial, allowing for a direct translation of all results to the discrete FD. The formalism allows for the introduction of smooth windows tapering each gap edge. This approach relies on using $N \times N$ covariance matrices and is limited to short data stretches, however it represents an exact treatment of gaps and can be used to assess mismodeling errors induced by other approximate treatments. The concepts within this paper are robust to any family of data gaps LISA may expect -- be them scheduled or unplanned gaps of any duration.

We developed a formalism to describe these mismodeling errors in Sec.~\ref{sec:mismodeling}, based on the FM approach. Mismodeling the noise process will lead to biases in the PE that, although zero-mean, will have the wrong statistics. For a generic mismodeled Gaussian likelihood, We introduce two complementary error measures: $\Xi$ in Eq.\eqref{eq:def_Xi} that measures the broadening of the posteriors, describing how much information is erased (notably, by the use of tapering windows), and $\Upsilon$ in Eq.\eqref{eq:def_Upsilon}, which measures the bias covariance (scatter of the max-likelihood) in units of the $\sigma$ of the model posterior, describing the statistical inconsistency induced by the mismodeling. These error measures can be computed for any Gaussian model likelihood, as long as the true process can be simulated, and are much faster to evaluate than PP plots requiring full Bayesian PE. 

We focused on MBHB signals in LISA, as described in Sec.~\ref{sec:application_MBHB}. We only considered a short data stretch of less than one day, which is enough to represent the merger of MBHB signals if their masses are large enough, and a limited array of cases, for a gap at merger or in the late inpiral, and for a heavier of lighter black hole. We investigate different TDI generations (TDI2, TDI1, TDI0), to explore the effect of the frequency content of the noise, from blue noise at high frequencies to red noise at low frequencies. In Sec.~\ref{sec:analysis_noise_cov} we compared covariance matrices for both circulant and Toeplitz structured covariance matrices, showing qualitatively their differences in both the TD and FD. We also demonstrated the impact of gaps with/without tapering on the covariance matrices, showing that hard cut-offs in the window function used to mimic gaps introduce highly non-trivial leakage in the FD noise-covariance matrix, giving both dense and degenerate matrices. Tapering concentrates the power more along the main diagonals, but degeneracy is still present. We then demonstrated a practical method to compute the pseudo-inverse of both the TD and FD noise covariance matrices by identifying the singular values corresponding to the gaps and excluding them. 

We then applied our error measure formalism to different mismodeling cases in Sec~\ref{sec:results}, exploring the effects of mismodeling the gaps, data segment independence, and the low-frequency PSD, as well as a limited exploration of the mismodeling of the true Toeplitz covariance with a circulant covariance. We also ran a set of full PE runs validating our approach. A summary of our findings is: 
\begin{itemize}
    \item Assuming the correct pseudo-inverse for the gated covariance matrix has been calculated, the likelihood is invariant to whether the model templates are gated or not. See Eq.~\eqref{eq:windowed_likelihood_td_gate} and Eq.~\eqref{eq:windowed_likelihood_fd_alt}.
    \item There is \emph{no} loss of information concerning the tapered signal assuming that the tapering scheme has been correctly incorporated into the pseudo-inverse of the windowed noise covariance matrix.
    \item There is a strong contrast between blue and red-noise processes; red noise is associated with long-lived correlations, and gap mismodeling is the worst in that case, but a naive gating window with no smooth taper also gives large errors for the blue process TDI2. This is demonstrated in Tab.~\ref{subsec:mismodeling_gaps}.
    \item A gap during the inspiral gives much smaller errors than a worst-case gap happening at merger for MBHs. This is consistent with the analysis presented in~\cite{Dey:2021dem,Wang:2024ovi,Mao:2024jad,Castelli:2024sdb}.
    \item Introducing tapering windows reduces statistical inconsistency errors significantly, at the expense of broadened posteriors.
    \item Adjusting the tapering length makes a significant difference, with an exchange between statistical inconsistency and information erasure; our methods can be used to cheaply find an optimal tapering length. 
    \item A segmented Whittle-likelihood, treating the data before/after gap as independent, gives results that are similar to a full-segment Whittle-likelihood.
    \item Treating data segments as independent when they are in reality correlated, even in the case of a data split (no in-between gap), gives little to no errors.
    \item Treating data segments as correlated when they are in reality independent leads to significant errors.
    \item mismodeling the low-$f$ PSD by underestimating the red noise leads to significant errors, even when properly taking the gaps into account. Tapering mildly alleviates this issue, but does not completely remove it. 
    \item Treating the data as circulant when it is in fact Toeplitz appears reasonable provided smooth tapers are applied to the data stream.
\end{itemize}

We believe that the methodology discussed in this paper is a suitable procedure for assessing noise mismodeling errors in the presence of data gaps and may be essential in validating LISA global fit pipelines. When in doubt, we strongly suggest that you smoothly taper your data.

\subsection{Future work and outlook}\label{subsec:outlook}

The present paper presents a preliminary exploration of gap mismodeling, calling for many extensions, some of which could be achieved with the same formalism. One strong limitation of the results presented is that we only considered a short stretch of data, appropriate only for merger-dominated MBHB signals. This choice was made to allow for a complete exploration and thorough validation of the method, comparing with the exact treatment of the gaps requiring to work with $N \times N$ matrices. However, assessing the mismodeling error $\Upsilon$, representing statistical inconsistency, is in fact only as expensive as working with the fast approximate likelihood, provided the true noise can be simulated (note that this is not true for $\Xi$, which compares the model Fisher covariance with the true model covariance). Having built confidence in the methods, they could be applied to longer signals: GBs, EMRIs, SBHBs. In particular, the segmented likelihood seems a promising avenue to explore and our method can be used to assess the systematics introduced.

We also limited ourselves to a few examples, and a more systematic study of the role of the gap length and signal characteristics would be valuable. Our method applies to any data plan, for instance for short repeated gaps.

Some of our formalism translates directly to other contexts, where instead of a gap we have an abrupt termination of the data. In ringdown analyses~\cite{Veitch:2014wba,siegel2024analyzingblackholeringdownsii,Isi:2021iql,Kastha:2021chr,Gennari:2023gmx} the analyst purposely selects data that starts shortly after the amplitude peak, right where the signal is strongest. In pre-merger analyses of MBHBs~\cite{CabournDavies:2024hea}, the observed data can end right where the signal is the loudest.

While we saw that the low-frequency noise can be responsible for noise mismodeling errors, we did not explore how filtering or whitening methods, that can readily eliminate low-$f$ noise in the stationary case, can be adapted to the presence of gaps.

We have seen that mismodeling the coherence of the noise over the gap can lead to biased results. In particular, we found that assuming a ``missing data'' model in which the noise is coherent across the gap leads to biases if there is a change in the properties of the noise between either side of the gap. This  should be a note of caution as the missing data model is being widely used in simulations of LISA data. As data gaps are typically associated with disturbances to one or more of the satellites, a data gap will certainly not be just missing data. The two cases we considered here represent two extremes - one in which the noise is completely coherent and one in which it is completely independent. The reality will be somewhere in between, as certain environmental noise sources might well be coherent, while instrumental noise sources probably won't be. This should be further explored in the future. Ideally, realistic gapped data would be produced so that the robustness of the independent segment model in more realistic data generation scenarios can be assessed.

Beyond data gaps, the same mismodeling formalism could be used to assess other approximations and errors made in the covariance matrix. The mismodeling ratios that we introduced allow for a much faster computation than building PP plots, which require repeating full PE runs for many noise realizations. The approximations that we could explore include non stationarity in the true noise process, either in the form of instrumental PSD drifts or in the form of annual modulations of the GB confusion foreground. Our formalism only requires that the model likelihood is Gaussian, and that the true noise can be simulated; as such, it could also be used to explore noise mismodeling errors caused by non-Gaussian features, such as distributions with heavier tails~\cite{Karnesis:2024pxh, Sasli:2023mxr, Rover:2011qd} or glitches~\cite{Houba:2024tyn, sauter2025maximumlikelihooddetectioninstrumental, Baghi:2021tfd}.

Time-frequency methods using the short-time-fourier-transforms (STFTs)~\cite{Tenorio:2025yca, Gair:2008ec, Kawahara:2018xjl} or Wilson-Daubechies-Meyer (WDM) basis functions~\cite{Necula:2012zz} appear to be growing in popularity. Within the WDM formalism, the window functions in both time and frequency are local, which have proved advantageous on data streams with mild non-stationary features (e.g., drifts to the PSD)~\cite{Digman:2022igm,Digman:2022jmp,Cornish:2020odn}. We have demonstrated in our TD / FD analysis that it may be reasonable to assume independence between segments if the underlying data stream is dependent. The same could be said for analyses conducted in the time-frequency domain. However, we are certain that working in time-frequency would still be negatively impacted by noise processes with significant red noise. Similarly, time-frequency analyses are not invincible to artefacts arising from edge-effects due to working with time-series of shorter length, as discussed in~\cite{Cornish:2020odn}.
This rich set of investigations will be left to future work.

\section{Acknowledgements}
O.Burke and S.Marsat made equal contributions in the preparation of this work. They also acknowledge support from the French space agency CNES in the framework of LISA. O.Burke and S.Marsat thank John Baker and Christian Chapman-Bird for their careful reading of our manuscript. O.Burke shows immense appreciation towards collaborator and friend Lorenzo Speri for pushing him to complete this work. We express our deepest thanks towards Quentin Baghi, Avi Vajpeyi, Giorgio Mentasti, Matthew Edwards, Vasco Gennari, Natalia Korsakova, Michael J. Williams, Peter Wolf, Jean Baptiste-Bayle and Olaf Hartwig, all for insightful discussions during the preparation of this work. 
This work has made use of \texttt{pastamarkers}~\cite{2024arXiv240320314P}.

\appendix

\section{Discrete Fourier transform of the time-domain covariance}

\subsection{The circulant case}
\label{app:diagonal_cov_matrix_derivation_circulant}

We will now derive a FD equivalent of eq.~\eqref{eq:defTDcovariance}, for the matrix $\langle \tilde{\bm{n}} \tilde{\bm{n}}^{\dagger} \rangle$. In our notations,
\begin{subequations}
\begin{align}
	 \tilde{\bm{n}} &= \Delta t \sqrt{N} \bm{P} \bm{n} \,,\\
	 \tilde{\bm{n}}^{\dagger} &= \Delta t \sqrt{N} \bm{n}^{T}\bm{P}^\dagger \,,
\end{align}
\end{subequations}
so that the FD analog of the covariance reads
\be
	\tilde{\bm{\Sigma}} \equiv \langle \tilde{\bm{n}} \tilde{\bm{n}}^{\dagger} \rangle = 
    N \Delta t^{2} \bm{P} \langle \bm{n} \bm{n}^{T}\rangle \bm{P}^{\dagger} = N \Delta t^{2} \bm{P} \bm{\Sigma} \bm{P}^{\dagger} \,.
	\label{eq:defSigmatilde}
\ee
We can compute explicitly (with implicit labels on sums extending from $0$ to $N-1$):
\begin{widetext}
\begin{subequations}
\begin{align}
	 (\bm{P} \bm{\Sigma} \bm{P}^{\dagger})_{ij} &= \frac{1}{N} \sum_{k} \sum_{l} \omega^{-ik} \bm{\Sigma}_{kl} \omega^{lj} \,,\\
	  &= \frac{1}{N} \sum_{k} \sum_{l} \omega^{-ik} \omega^{lj} \bm{\sigma}_{l-k} \,, \\
	  &= \frac{1}{N} \sum_{k} \sum_{l=0}^{k-1} \omega^{-ik} \omega^{lj} \bm{\sigma}_{k-l} + \frac{1}{N} \sum_{k} \sum_{l=k}^{N-1} \omega^{-ik} \omega^{lj} \bm{\sigma}_{l-k} \,, \\
	  &= \frac{1}{N} \sum_{k} \sum_{m=1}^{k} \omega^{k(j-i)} \omega^{-mj} \bm{\sigma}_{m} + \frac{1}{N} \sum_{k} \sum_{m=0}^{N-k-1} \omega^{k(j-i)} \omega^{mj} \bm{\sigma}_{m} \,,
\end{align}
\end{subequations}
where we used the Toeplitz structure, the symmetry of $\bm{\sigma}$ for negative indices, and changed variables to $m=k-l$ and $m=k+l$. Using the circulant structure, we can rewrite the first term as:
\be
	\frac{1}{N} \sum_{k} \sum_{m=1}^{k} \omega^{k(j-i)} \omega^{-mj} \bm{\sigma}_{m} = \frac{1}{N} \sum_{k} \sum_{m=1}^{k} \omega^{k(j-i)} \omega^{-mj} \bm{\sigma}_{N-m} = \frac{1}{N} \sum_{k} \sum_{m=N-k}^{N-1} \omega^{k(j-i)} \omega^{mj} \bm{\sigma}_{m} \,,
	\label{eq:circulantrewrite}
\ee
where we used $\omega^{N}=1$. We can now gather terms as
\be
	(\bm{P} \bm{\Sigma} \bm{P}^{\dagger})_{ij} = \left( \frac{1}{N} \sum_{k} \omega^{k(j-i)}\right) \sum_{m=0}^{N-1} \omega^{mj} \bm{\sigma}_{m} = \delta_{ij} \sum_{m=0}^{N-1} \omega^{-mj} \bm{\sigma}_{m} = \delta_{ij} \sqrt{N} \left( \bm{P} \bm{\sigma} \right)_{i} \,,
	\label{eq:diagonalrewrite}
\ee
where we used the circulant condition again in the last equality to change the sign of the exponent.
\end{widetext}

We can also do a similar computation with the transpose of $\tilde{N}$ instead of its Hermitian conjugate, which will allow us to complete the description of the statistics of the real and imaginary parts of $\tilde{N}$. This is a repetition of the computation of $P\Sigma P^{\dagger}$, changing simply the sign of the exponent of $\omega$. Without detailing each step, we obtain (noting the special case $i=j=0$)
\begin{widetext}
\be
	(\bm{P} \bm{\Sigma} \bm{P}^{T})_{ij} = \left( \frac{1}{N} \sum_{k} \omega^{-k(i+j)}\right) \sum_{m=0}^{N-1} \omega^{-mj} \bm{\sigma}_{m} = \left(\delta_{i,0}\delta_{j,0} + \delta_{N,i+j} \right) \sum_{m=0}^{N-1} \omega^{-mj} \bm{\sigma}_{m} \,.
\ee
\end{widetext}
We arrive at (note that the last index is $j$ and not $i$)\be\label{eq:PSigmaPT_circ}
	(\bm{P} \bm{\Sigma} \bm{P}^{T})_{ij} = \sqrt{N} \left(\delta_{i,0}\delta_{j,0} + \delta_{N,i+j} \right) (\bm{P}\bm{\sigma})_{j} \,.
\ee
Thus, considering the $N\times N$ matrix form with four quadrants for positive and negative frequency entries, $\bm{P} \bm{\Sigma} \bm{P}^{\dagger}$ has a diagonal structure through the positive-positive and negative-negative quadrants, while $\bm{P} \bm{\Sigma} \bm{P}^{T}$ has an anti-diagonal-like structure $i+j=N$ (and not $i+j=N-1$, as it would be for the main anti-diagonal) extending through the positive-negative and negative-positive frequency quadrants, plus an extra component at $i=j=0$.

\subsection{The Toeplitz case}\label{app:The_Toeplitz_Case_Derivation}

We now investigate the case where the TD covariance $\Sigma$ has a Toeplitz but not circulant structure.

First, we note that, if we build the aucovariance as an IFT of the noise PSD, we will automatically get a circulant structure: if $\tilde{\bm{\sigma}}$ is built according to~\eqref{eq:sigmatildePSD} and if $\bm{\sigma} = P^{\dagger}\tilde{\bm{\sigma}} / (\Delta t \sqrt{N})$, then we have explicitly
\be
	\bm{\sigma}_{i} = \frac{1}{2 N \Delta t} \left( S_{n}^{0} - S_{n}^{N/2} + \sum_{k=1}^{N/2-1} \left( \omega^{ik} + \omega^{-ik} \right) S_{n}^{k}\right) \,,
\ee
so that $\bm{\sigma}_{i} \in \mathbb{R}$ and $\bm{\sigma}_{i} = \bm{\sigma}_{N-i}$.

This means that a non-circulant autocorrelation requires another generation procedure; one such way is to generate a circulant covariance of size $2N\times 2N$ with the IFT, then extract its first quadrant to obtain an $N\times N$ Toeplitz matrix.

Coming back to the computation of $\langle \tilde{\bm{n}} \tilde{\bm{n}}^{\dagger}\rangle$, we have to remember that we used the circulant condition twice. We will get an extra contribution from $\bm{\sigma}_{N-m} \neq \bm{\sigma}_{m}$ in~\eqref{eq:circulantrewrite}, and we have to stop before the last inequality in~\eqref{eq:diagonalrewrite}. As a result,
\be
	(P\Sigma P^{\dagger})_{ij} = \delta_{ij} \sum_{m}\omega^{im}\bm{\sigma}_{m} + \Delta_{ij} \,,
\ee
where
\begin{widetext}
\be
	\Delta_{ij} = \frac{1}{N} \sum_{k}  \sum_{m=1}^{k}  \omega^{k(j-i)} \omega^{-mj} \left(\bm{\sigma}_{m} - \bm{\sigma}_{N-m}\right) = \frac{1}{N} \sum_{m=1}^{N-1}  \left(\sum_{k=m}^{N-1}  \omega^{k(j-i)} \right)\omega^{-mj} \left(\bm{\sigma}_{m} - \bm{\sigma}_{N-m}\right) \,.
\ee
For $i=j$, the sum in parentheses is $N-m$, and we have
\be
	\Delta_{ii} = \frac{1}{N} \sum_{m=1}^{N-1}  (N-m)\omega^{-mi} \left(\bm{\sigma}_{m} - \bm{\sigma}_{N-m}\right) = \sum_{m=1}^{N-1} \omega^{-mi} \bm{\sigma}_{m} - \frac{1}{N} \sum_{m=1}^{N-1}  m\left(\omega^{mi} + \omega^{-mi} \right) \bm{\sigma}_{m} 
\ee
\end{widetext}
For $i \neq j$,
\be
	\sum_{k=m}^{N-1}  \omega^{k(j-i)} = - \frac{1 - \omega^{m(j-i)}}{1 - \omega^{j-i}}\;.
\ee
So, for $i\neq j$ we have
\begin{subequations}
\begin{align}
	 \Delta_{ij} &= -\frac{1}{N} \sum_{m=1}^{N-1}  \frac{\omega^{-mj} - \omega^{-mi}}{1 - \omega^{j-i}} \left(\bm{\sigma}_{m} - \bm{\sigma}_{N-m}\right) \,, \\
	  &= \frac{1}{1 - \omega^{j-i}} \frac{1}{N} \sum_{m=0}^{N-1} \left( \omega^{-mi} - \omega^{-mj} - \omega^{mi} + \omega^{mj} \right) \bm{\sigma}_{m} \,,
\end{align}
\end{subequations}
which we can rewrite using $\bm{P}\bm{\sigma}$. Gathering the diagonal and off-diagonal terms, we obtain
\begin{subequations}
\label{eq:Toeplitz_Matrix}
\begin{align}
	 (\bm{P}\bm{\Sigma} \bm{P}^{\dagger})_{ij} &= \delta_{ij} \bm{\Lambda}_{i} + \bm{\Omega}_{ij} \,, \\
	 \bm{\Lambda}_{i} &= \bm{\sigma}_0 + \frac{1}{N} \sum_{m=1}^{N-1} (N - m) \left(\omega^{mi} + \omega^{-mi} \right) \bm{\sigma}_{m} \,, \\
	 \bm{\Omega}_{ij} &= \frac{1}{\sqrt{N}} \frac{1}{1 - \omega^{j-i}} \left[ (\bm{P}\bm{\sigma})_{i} - (\bm{P}\bm{\sigma})_{i}^{*} \right.\nonumber \\
     &\hspace{2cm} \left. - (\bm{P}\bm{\sigma})_{j} + (\bm{P}\bm{\sigma})_{j}^{*} \right] \; \text{for} \; i\neq j \,.
\end{align}
\end{subequations}
In this rewriting, we can check that $\Lambda_{i} \in \mathbb{R}$ and $\Omega$ is Hermitian, $\Omega^{\dagger} = \Omega$.

This calculation can be repeated, with few changes, for the computation of $\bm{P} \bm{\Sigma} \bm{P}^{T}$. This is required if one needs a complete control on both real and imaginary parts in the FD; however, $\bm{P}\bm{\Sigma} \bm{P}^{\dagger}$ is enough to express the likelihood. The result reads
\begin{widetext}
\begin{subequations}
\label{eq:PSigmaPT_toeplitz}
\begin{align}
	 (\bm{P} \bm{\Sigma} \bm{P}^{T})_{00} &= \bm{\sigma}_0 + \frac{2}{N} \sum_{m=1}^{N-1} (N - m) \bm{\sigma}_{m} \,, \\
	 (\bm{P} \bm{\Sigma} \bm{P}^{T})_{i,N-i} &= \bm{\sigma}_0 + \frac{1}{N} \sum_{m=1}^{N-1} (N - m) \left(\omega^{mi} + \omega^{-mi} \right) \bm{\sigma}_{m} \quad \text{for} \; i=1,\dots,N-1 \,, \\
	 (\bm{P} \bm{\Sigma} \bm{P}^{T})_{ij} &= \frac{1}{\sqrt{N}} \frac{1}{1 - \omega^{-i-j}} \left[ (\bm{P}\bm{\sigma})_{i} - (\bm{P}\bm{\sigma})_{i}^{*} + (\bm{P}\bm{\sigma})_{j} - (\bm{P}\bm{\sigma})_{j}^{*} \right] \quad \text{for} \; i+j \neq 0,N \,.
\end{align}
\end{subequations}

\end{widetext}

\section{Fourier-domain noise generation in the circulant case}\label{app:draw_circ_noise_fd}

We give here a detailed description of noise generation in the discrete FD, in the circulant case. This will constitute a discrete equivalent of the better-known functional relations~\eqref{eq:wiener_khintchine}.   

We wish to draw a FD realization of the noise, $\tilde{\bm{n}}$, given a PSD $S_n(f)$. We first recall that, since $\bm{n} \in \mathbb{R}^N$, $\tilde{\bm{n}}$ is entirely represented by its $N/2+1$ components $\tilde{\bm{n}}_j$ for $j=0,\dots,N/2$ inclusive, with the other components being determined from $\tilde{\bm{n}}_{N-j} = \tilde{\bm{n}}_j^*$ for $j=1,\dots,N/2-1$. Moreover, $\tilde{\bm{n}}_0,\tilde{\bm{n}}_{N/2} \in \mathbb{R}$, so there are the expected $N$ degrees of freedom.

We will introduce random variables for the real and imaginary parts as $\tilde{\bm{n}}_{j} = A_{j} + i B_{j}$ for $j=0,\dots,N/2$ inclusive (we will come back later to $B_0$ and $B_N/2$). By assumption, $A_{j}$, $B_{j}$ are Gaussian (as linear combinations of the Gaussian vector $\bm{n}$). Considering~\eqref{eq:FDcovarianceWhittle} and~\eqref{eq:PSigmaPT_circ} in the quadrant $\llbracket 0,N/2 \rrbracket^2$, we see that $\langle \tilde{\bm{n}}\tilde{\bm{n}}^\dagger\rangle$ has support on the diagonal, while $\langle \tilde{\bm{n}}\tilde{\bm{n}}^T\rangle$ only has support at $i=j=0$ and $i=j=N/2$. Off-diagonal, we get
\be
	 \langle (A_{j} + i B_{j}) (A_{k} \pm i B_{k}) \rangle = 0 \quad (j\neq k)\,,
\ee
from which we deduce that $A_{j}$, $B_{j}$ are independent from $A_{k}$, $B_{k}$ when $j \neq k$. In turn, along the diagonal, for $j=1,\dots,N/2-1$, we get
\begin{subequations}
\begin{align}
	 \langle (A_{j} + i B_{j}) (A_{j} - i B_{j}) \rangle &= \frac{S_{n}^{j}}{2\Delta f} \,, \\
	 \langle (A_{j} + i B_{j}) (A_{j} + i B_{j}) \rangle &= 0 \,, \quad (j=1,\dots,N/2-1)
\end{align}
\end{subequations}
The real and imaginary part of the second relation give that $\langle A_{j}^{2} \rangle = \langle B_{j}^{2} \rangle$ and $\langle A_{j} B_{j} \rangle = 0$, so the Gaussian variables $A_{j}$, $B_{j}$ are independent with the same variance. The first relation gives $\langle A_{j}^{2} + B_{j}^{2}\rangle = 2 \langle A_{j}^{2}\rangle = 2 \langle B_{j}^{2}\rangle = S_{n}^{j} / (2\Delta f)$, and we obtain that
\be
	\mathrm{Re} \, \tilde{\bm{n}}_{j}, \; \mathrm{Im} \, \tilde{\bm{n}}_{j} \sim \mathcal{N}\left(0, \frac{S_{n}(j \Delta f)}{4 \Delta f} \right) \quad (j=1,\dots,N/2-1)\,,
\ee
and are independent both from each other and across frequency bins.

Coming back to the special cases $j=0$ and $j=N/2$ with a non-zero component for $\langle \tilde{\bm{n}}\tilde{\bm{n}}^T\rangle$, we get
\begin{subequations}
\begin{align}
	 \langle (A_{j} + i B_{j}) (A_{j} - i B_{j}) \rangle &= \frac{S_{n}^{j}}{2\Delta f} \,, \\
	 \langle (A_{j} + i B_{j}) (A_{j} + i B_{j}) \rangle &= \frac{S_{n}^{j}}{2\Delta f} \,, \quad (j=0,N/2)
\end{align}
\end{subequations}
The second equation gives $\langle A_j B_j \rangle = 0$ as before. We then get $\langle A_{j}^{2} + B_{j}^{2}\rangle = \langle A_{j}^{2} - B_{j}^{2}\rangle$, which shows that $B_j$ is identically zero, as it should be for $\tilde{\bm{n}}$ to be the DFT of a real vector. We arrive at
\be
	\mathrm{Re} \, \tilde{\bm{n}}_{j} \sim \mathcal{N}\left(0, \frac{S_{n}(j \Delta f)}{2 \Delta f} \right) \,, \; \mathrm{Im} \, \tilde{\bm{n}}_{j} = 0 \quad (j=0,N/2)\,,
\ee
which completes the result. Thanks to independence, we can draw FD circulant noise with individual Gaussian draws.

Once noise is generated in the FD efficiently thanks to the independence property, it can be translated back to the TD via an IFFT. One efficient strategy to generate noise in the Toeplitz case is to artificially extend the Toeplitz covariance as a circulant covariance over a segment of length $2N$, draw an extended circulant noise realization in the FD, and perform an IFFT to obtain TD noise of length $2N$; the first half of this vector then gives an $N$-sized TD realization of the desired noise process with a Toeplitz covariance.

\section{Likelihood with gaps using a permutation and masks}\label{app:likelihood_using_masks_permutations}

We can treat the general case of an arbitrary data plan, with multiple gaps of varying lengths,with the help of Boolean masks. We can define a mask as a boldface index $\bm{g}$ such that $\bm{x}_{\bm{g}}$ is the subvector of values inside gap segments, of length $N_g$. Similarly, for matrices $\bm{\Sigma}_{\neg \bm{g}, \neg \bm{g}}$ is the $(N-N_g) \times (N-N_g)$ submatrix of values outside of the gap segments. The marginalization over elements of a Gaussian vector,
\begin{equation}\label{eq:marginalLike_masks}
	p(\bm{n}_{\neg \bm{g}}) = \int d\bm{n}_{\bm{g}} \, p(\bm{n}_{\bm{g}}, \bm{n}_{\neg \bm{g}}) \,,
\end{equation}
gives $p(\bm{n}_{\neg \bm{g}}) \sim \mathcal{N} \left( \bm{0}, \bm{\Sigma}_{\neg \bm{g}, \neg \bm{g}} \right)$. This gives the log-likelihood
\begin{align}
    \ln \mathcal{L}_{\rm gap} (\bm{d}_{\neg \bm{g}} | \bm{\theta}) = &-\frac{1}{2} \bm{x}_{\neg \bm{g}}^{T} \bm{\Sigma}_{\neg \bm{g}, \neg \bm{g}}^{-1} \bm{x}_{\neg \bm{g}} - \frac{1}{2}\det \bm{\Sigma}_{\neg \bm{g}, \neg \bm{g}} \nonumber\\
    & - \frac{1}{2}(N - N_g)\ln 2\pi \,,
	\label{eq:gated_likelihood_td_mask}
\end{align}

We can reorder indices to put all indices pertaining to the gaps at the end, by introducing a permutation\footnote{Not to be confused with the auto-covariance vector.} $\sigma$, such that
\begin{equation}\label{eq:gating_perm}
    \begin{cases}
\sigma(i) \leq N - N_g \quad \text{for} \ i \not\in \text{gap segments} \\
\sigma(i) > N - N_g \quad \text{for} \ i\in \text{gap segments} \,.
\end{cases}
\end{equation}
The precise ordering within the gap and no-gap segments will not matter, for a simple connection with the mask notation we can assume that we keep the relative ordering on each sub-segment. In our linear algebra notations, this permutation can be represented by a matrix $\bm{R}$ such that
\be\label{eq:def_Rperm}
    \bm{R}_{ij} = \delta_{i \sigma(j)} \,.
\ee
A permutation matrix is automatically orthogonal (it preserves the scalar product), $\bm{R}^T \bm{R} = \mathds{1}$. If $\bm{W} = \mathrm{Diag}(w_g(i\Delta t), i=0,\dots,N-1)$ with $w_g$ a gating window, we will define the \emph{reordered gated covariance} as
\be
    \mathring{\bm{\Sigma}} \equiv \bm{R} \bm{W} \bm{\Sigma} \bm{W} \bm{R}^T \,, 
\ee
and we can check explicitly, using\eqref{eq:def_Rperm}, that
\begin{align}
    \mathring{\bm{\Sigma}}_{ij} &= \sum_k \sum_l \bm{R}_{ik} \bm{w}_k \bm{\Sigma}_{kl} \bm{w}_{l} \bm{R}_{lj}\nonumber\\
    &= \sum_k \sum_l \delta_{i \sigma(k)} \bm{w}_k \bm{\Sigma}_{kl} \bm{w}_{l} \delta_{l\sigma{j}}\nonumber\\
    &= \bm{w}_{\sigma^{-1}(i)} \bm{\Sigma}_{\sigma^{-1}(i) \sigma^{-1}(j)} \bm{w}_{\sigma^{-1}(j)}\nonumber\\
    &= \begin{cases}
        \bm{\Sigma}_{\sigma^{-1}(i) \sigma^{-1}(j)} \quad \text{if} \ i,j \leq N-N_g \\
        0 \quad \text{else} \ i\in \text{gap segments} \,,
       \end{cases}
\end{align}
so that the reordered covariance has the block structure
\be
    \mathring{\bm{\Sigma}} = \begin{pmatrix}
					\bm{\Sigma}_{\neg \bm{g}, \neg \bm{g}} & \bm{0} \\
                    \bm{0}  & \bm{0}
                    \end{pmatrix} \,,\label{app:eq_reordered_covariance}
\ee
with pseudo-inverse 
\be
    \mathring{\bm{\Sigma}}^{+} = \begin{pmatrix}
					(\bm{\Sigma}_{\neg \bm{g}, \neg \bm{g}})^{-1} & \bm{0} \\
                    \bm{0}  & \bm{0}
                    \end{pmatrix} \,.
\ee
Similarly, if we define reordered gated vectors
\be
    \mathring{\bm{x}} \equiv \bm{R} \bm{W} \bm{x} = \begin{pmatrix}
					\bm{x}_{\neg \bm{g}} \\
                    \bm{0}
                    \end{pmatrix} \,, 
\ee
the block structure shows directly that the marginalized likelihood can be written in terms of our reordered matrices and vectors as (ignoring constants)
\be
    \ln \mathcal{L}_{\rm gap} (\bm{d}_{\neg \bm{g}} | \bm{\theta}) = -\frac{1}{2} \mathring{\bm{x}}^T \mathring{\bm{\Sigma}}^+ \mathring{\bm{x}} \,.
\ee
Now, we can use the properties of the pseudo-inverse and the fact that $\bm{R}$ is orthogonal to rewrite
\begin{align}
    \mathring{\bm{x}}^T \mathring{\bm{\Sigma}}^+ \mathring{\bm{x}} &= \bm{x}^T \bm{W} \bm{R}^T \left( \bm{R} \bm{W} \bm{\Sigma} \bm{W} \bm{R}^T\right)^+ \bm{R} \bm{W} \bm{x} \nonumber\\
    &= \bm{x}^T \bm{W} \bm{R}^T \bm{R} \left( \bm{W} \bm{\Sigma} \bm{W} \right)^+  \bm{R}^T \bm{R} \bm{W} \bm{x} \nonumber\\
    &= (\bm{W}\bm{x})^T  \left( \bm{W} \bm{\Sigma} \bm{W} \right)^+ \bm{W} \bm{x} \,.
\end{align}
Thus, we arrive at our expression~\eqref{eq:windowed_likelihood_td} for the likelihood in the presence of gaps.

Note that we could also have used the re-ordered but not gated data vectors 
\be
    \overset{\bullet}{\bm{x}} \equiv \bm{R} \bm{x} = \begin{pmatrix}
					\bm{x}_{\neg \bm{g}} \\
                    \bm{x}_{\bm{g}}
                    \end{pmatrix} \,, 
\ee
and the likelihood would retain the same expression thanks to the structure of $\mathring{\bm{\Sigma}}^{+}$: 
\be
    \ln \mathcal{L}_{\rm gap} (\bm{d}_{\neg \bm{g}} | \bm{\theta}) = -\frac{1}{2} \overset{\bullet}{\bm{x}}^T \mathring{\bm{\Sigma}}^+ \overset{\bullet}{\bm{x}} \,.
\ee
Similarly to the previous case, we have
\be
    \overset{\bullet}{\bm{x}}^T \mathring{\bm{\Sigma}}^+ \overset{\bullet}{\bm{x}} = \bm{x}^T  \left( \bm{W} \bm{\Sigma} \bm{W} \right)^+ \bm{x} \,,
\ee
in which we recognize the property that the likelihood is insensitive to the content of the residual vector inside the gap, so gating the residual is facultative. 

\section{Degenerate gated covariance matrices in time and frequency domain}\label{app:proof_eigenvalues_matrices}
Let $\bm{W}\bm{\Sigma}\bm{W} = \bm{\Sigma}^{\text{gap}}\in\mathbb{R}^{N\times N}$. From the rank--nullity theorem, we must have that $\text{rank}(\bm{\Sigma}^{\text{gap}}) + \text{nullity}(\bm{\Sigma}^{\text{gap}}) = N$. Making reference to App.\ref{app:likelihood_using_masks_permutations}, Eq.~\eqref{app:eq_reordered_covariance} there are an equivalent number of linearly dependent rows and columns in $\bm{\Sigma}^{\text{gap}}$, implying that the dimension of the column space $\text{dim}(\text{Col}(\bm{\Sigma}^{\text{gap}})) =\text{rank}(\bm{\Sigma}^{\text{gap}}) = N - N_{g}$, for $N_{g}$ the number of zero columns of the matrix $\bm{\Sigma}^{\text{gap}}$. We remark here that the swapping of rows and columns will not change the dimension of the column space. By the rank-nullity theorem, the dimension of the null space, $\text{nullity}(\bm{\Sigma}^{\text{gap}}) = N_{g}$, meaning that there are $N_{g}$ vectors $\bm{v}\in\text{null}(\bm{\Sigma}^{\text{gap}})$ such that $\bm{\Sigma}^{\text{gap}}\bm{v} = \lambda \bm{v} = 0$. Hence there must exist $N_{g}$ zero eigenvalues of the matrix $\bm{\Sigma}^{\text{gap}}$. We can translate this into the FD via the unitary DFT matrix $\bm{P}$
\begin{align*}
\bm{\Sigma}^{\text{gap}}\bm{v} &= \bm{0} \\
\bm{P}[(\bm{W}\bm{\Sigma}\bm{W})] \bm{v}\bm{P}^{\dagger} &= \bm{0} \\
(\tilde{\bm{W}}\tilde{\bm{\Sigma}}\tilde{\bm{W}})\tilde{\bm{u}} &= \bm{0} \\
\tilde{\bm{\Sigma}}^{\text{gap}}\tilde{\bm{u}} & = 0
\end{align*}
implying that $\bm{P}\bm{v}\bm{P}^{\dagger} = \tilde{\bm{u}}\in\text{null}(\tilde{\bm{\Sigma}}^{\text{gap}})$. Finally, noting that $\text{rank}(\bm{\Sigma}^{\text{gap}}) = \text{rank}(\bm{P}^{\dagger}\tilde{\bm{\Sigma}}^{\text{gap}}\bm{P}) = \text{rank}(\tilde{\bm{\Sigma}}^{\text{gap}})$, which implies that $\text{nullity}(\bm{\Sigma}^{\text{gap}}) = \text{nullity}(\tilde{\bm{\Sigma}}^{\text{gap}})$, there must be $N_{g}$ zero-eigenvalues in the FD gated covariance matrix $\tilde{\bm{\Sigma}}^{\text{gap}}$. Finally, since $\tilde{\bm{u}} = \bm{P}\bm{v}\bm{P}^{\dagger}$, there exists a 1-1 correspondence between the eigenvectors with zero-eigenvalues of the TD and FD covariance matrices. This means that the eigenvectors with zero-eigenvalues present due to the gated segment in the TD matrix correspond to the transformation of the eigenvectors with zero-eigenvalues of the FD matrix. Finally, Since the matrix $\tilde{\bm{\Sigma}}^{\text{gap}}$ is hermitian symmetric, the Singular Values (SVs) following a SVD are equivalent to the absolute values of the eigenvalues, i.e., $\bm{S}(\bm{\tilde{\Sigma}}^{\text{gap}})_{i} = |\bm{\lambda}(\bm{\tilde{\Sigma}}^{\text{gap}})_{i}|$. In conclusion, there are $N_{g}$ many singular values of the FD gated noise covariance matrix.

\section{Details on the tapering window}\label{app:window}

When smoothly tapering signal at the edges of a data gap or at the ends of the data segment, we will use a Planck window, as described in~\cite{McKechan:2010kp}. For a window $w(t)$ that is zero outside of $[t_i, t_f]$, smoothly transitions from 0 to 1 on $[t_i, t_i + \Delta t_i]$ and back from 1 to 0 on $[t_f - \Delta t_f, t_f]$, the expression is
\begin{align}\label{eq:def_planck_window}
    w(t) = \begin{cases} 0 \; \text{for} \; t \leq t_i \\
                         \frac{1}{1 + \mathrm{exp}\left[ \frac{\Delta t_i}{t - t_i} + \frac{\Delta t_i}{t - (t_i + \Delta t_i)} \right]} \; \text{for} \; t_i < t < t_i + \Delta t_i \\
                         1 \; \text{for} \; t_i + \Delta t_i \leq t \leq t_f - \Delta t_f \\
                         \frac{1}{1 + \mathrm{exp} \left[ -\frac{\Delta t_f}{t - t_f} - \frac{\Delta t_f}{t - (t_f - \Delta t_f)} \right]} \; \text{for} \; t_f - \Delta t_f < t < t_f \\
                         0 \; \text{for} \; t \geq t_f \\
           \end{cases}
\end{align}
For instance, multiplying the data with $1-w$ would implement a gap on $[t_i + \Delta t_i, t_f - \Delta t_f]$ with tapering lobes $\Delta t_i$, $\Delta t_f$.

\bibliographystyle{apsrev4-2}
\bibliography{Data_Gaps_arXiv_Sub_2}% Produces the bibliography via BibTeX.

\end{document}